\documentclass[12pt,preprint]{aastex}
\usepackage{natbib}
\usepackage{color}
\usepackage{soul}
\usepackage{ulem}
\usepackage{amsmath}
\usepackage[nolists,nomarkers]{endfloat}
\usepackage{booktabs}
\usepackage{afterpage}
\usepackage{multirow}

\setul{-0.8ex}{0.40ex}
\setulcolor{red}

\newcommand{\frca}{\mathfrak{a}}
\newcommand{\frcA}{\mathfrak{A}}

\shorttitle{Estimation of High-Degree p-Mode Parameters}

\shortauthors{Reiter et al.}

\begin{document}


\title{A Method for the Estimation of p-Mode Parameters from
Averaged Solar Oscillation Power Spectra}


\author{J. Reiter}
\affil{Zentrum Mathematik, M17, Technische Universit\"at M\"unchen,
       \linebreak
       D-85748 Garching bei M\"unchen, Germany}
\email{jreiter@lrz.tum.de}

\author{E. J. Rhodes, Jr.\altaffilmark{1}}
\affil{Department of Physics and Astronomy, University of Southern
       California, \linebreak Los Angeles, CA 90089-1342, U.S.A}
\email{erhodes@solar.stanford.edu}

\author{A. G. Kosovichev\altaffilmark{2}, J. Schou\altaffilmark{3}, 
        P. H. Scherrer, and T. P. Larson}
\affil{W. W. Hansen Experimental Physics Laboratory,
       Stanford University, \linebreak
       Stanford, CA 94305-4085, U.S.A}
\email{sasha@bbso.njit.edu, schou@mps.mpg.de,\\
       pscherrer@solar.stanford.edu, tplarson@sun.stanford.edu}

\altaffiltext{1}{Astrophysics and Space Sciences Section, Jet Propulsion
                 Laboratory, California Institute of Technology, 
                 4800 Oak Grove Dr., Pasadena, CA 91109-8099}
\altaffiltext{2}{New Jersey Institute of Technology, Newark, NJ 07102, U.S.A.}
\altaffiltext{3}{Max-Planck-Institut f\"ur Sonnensystemforschung,
                 Justus-von-Liebig-Weg 3, D-37077 G\"ottingen, Germany}


%%\altaffiltext{1}{Visiting Astronomer, Cerro Tololo Inter-American Observatory.
%%CTIO is operated by AURA, Inc.\ under contract to the National Science
%%Foundation.}


\begin{abstract} 
A new fitting methodology is presented which is equally well suited for the
estimation of low-, medium-, and high-degree mode parameters from $m$-averaged
solar oscillation power spectra of widely differing spectral resolution. This
method, which we call the ``Windowed, MuLTiple-Peak, averaged spectrum", or WMLTP Method,
constructs a theoretical profile by convolving the weighted sum of the profiles
of the modes appearing in the fitting box with the power spectrum of the window
function of the observing run using weights from a leakage matrix that takes
into account both observational and physical effects, such as the distortion of
modes by solar latitudinal differential rotation. We demonstrate that the WMLTP
Method makes substantial improvements in the inferences of the properties of the solar
oscillations in comparison with a previous method that employed a single
profile to represent each spectral peak. We also present an inversion for the
internal solar structure which is based upon 6,366 modes that we have computed
using the WMLTP method on the 66-day long 2010 SOHO/MDI Dynamics Run. To
improve both the numerical stability and reliability of the inversion we
developed a new procedure for the identification and correction of outliers in
a frequency data set. We present evidence for a pronounced departure of the
sound speed in the outer half of the solar convection zone and in the
subsurface shear layer from the radial sound speed profile contained in Model~S
of Christensen-Dalsgaard and his collaborators that existed in the rising phase
of Solar Cycle~24 during mid-2010.
\end{abstract}


%% subject header.  Objects should be in the appropriate "individual"
%% headers (e.g. quasars: individual, stars: individual, etc.) with the
%% additional provision that the total number of headers, including each
%% individual object, not exceed six.  The \objectname{} macro, and its
%% alias \object{}, is used to mark each object.  The macro takes the object
%% name as its primary argument.  This name will appear in the paper
%% and serve as the link's anchor in the electronic edition if the name
%% is recognized by the data centers.  The macro also takes an optional
%% argument in parentheses in cases where the data center identification
%% differs from what is to be printed in the paper.



\section{Introduction\label{Sec1}}

Helioseismology provides a unique opportunity to investigate in great detail
the internal structure and rotation of the Sun. The starting point for the
study of the solar interior using helioseismology can be identified with the
observational confirmation by \cite{Deu75}, and independently by \cite{Rho77},
of the standing-wave nature of the solar five-minute oscillations that were
observed in the solar photosphere as proposed by \cite{Ulr70}, and
independently by \cite{Lei71}. The remarkable qualitative agreement of the
observations of \cite{Deu75} and of \cite{Rho77} with the predictions of the
theoretical studies meant that the five-minute oscillations could be regarded
as a superposition of acoustic normal modes that are trapped in the interior of
the Sun. However, the frequencies of the observed ridges of power in the
dispersion plane were systematically lower by about 5\,\% than the theoretical
predictions. This discrepancy allowed \cite{Rho76b}, and independently
\cite{Gou77} to provide observational estimates of the depth of the solar convection zone.
These estimates are now believed to be the first helioseismic inferences of
solar internal structure.

The observed modes are predominantly $p$-modes, for which pressure provides the
dominant restoring force. Also observed is the $f$-mode, which at high
spherical harmonic degree $l$ has the character of a surface gravity wave.
Both the \cite{Deu75} and \cite{Rho77} studies employed intermediate- and
high-degree $f$- and $p$-modes, while \cite{Cla79} employed low-degree $p$-modes that
penetrated into the solar core. Today, the line of demarcation between low- and
intermediate-degree modes is at $l=4$ 
(the highest degree that can be observed in integrated light), while that
between intermediate- and high-degree modes is where individual modes are no
longer resolved, or at about $l=300$ for the $f$-mode, between $l=200$ and
$l=150$ for the $n=1$ through $n=4$ ridges (and at lower values of $l$ for the
higher radial orders $n$). 

The observations by \cite{Deu75}, \cite{Rho77}, and \cite{Cla79} raised
considerable debate over the proper characterization of the oscillation modes
that had been observed. \cite{Deu77} cited a pre-publication version of
\cite{Deu79} to claim that the solar $p$-mode oscillations were truly global in
nature, but subsequently, \cite{Ulr79} disagreed with this conclusion and
argued that ``...the modes of greatest interest are not globally coherent
because of the effect of convective motions associated with the
supergranulation''. Furthermore, \cite{HHill80} and \cite{Gou80} pointed out
that the coherence times of up to nine hours that were cited by \cite{Deu79}
and by \cite{Cla80} were not sufficient to demonstrate the global nature of
these modes. Today, the term ``global helioseismology'' refers either to
studies that employ the low- and intermediate-degree $p$-modes whose lifetimes
are truly long enough for them to be globally-coherent or to studies that
employ spherical harmonic decompositions that are computed from nearly the
entire visible solar hemisphere. Studies that do not use modes which have such
long lifetimes or which are computed from observations that cover much smaller
portions of the visible hemisphere are considered to employ the tools of
``local helioseismology''.

\begin{figure}
\epsscale{0.80}
\plotone{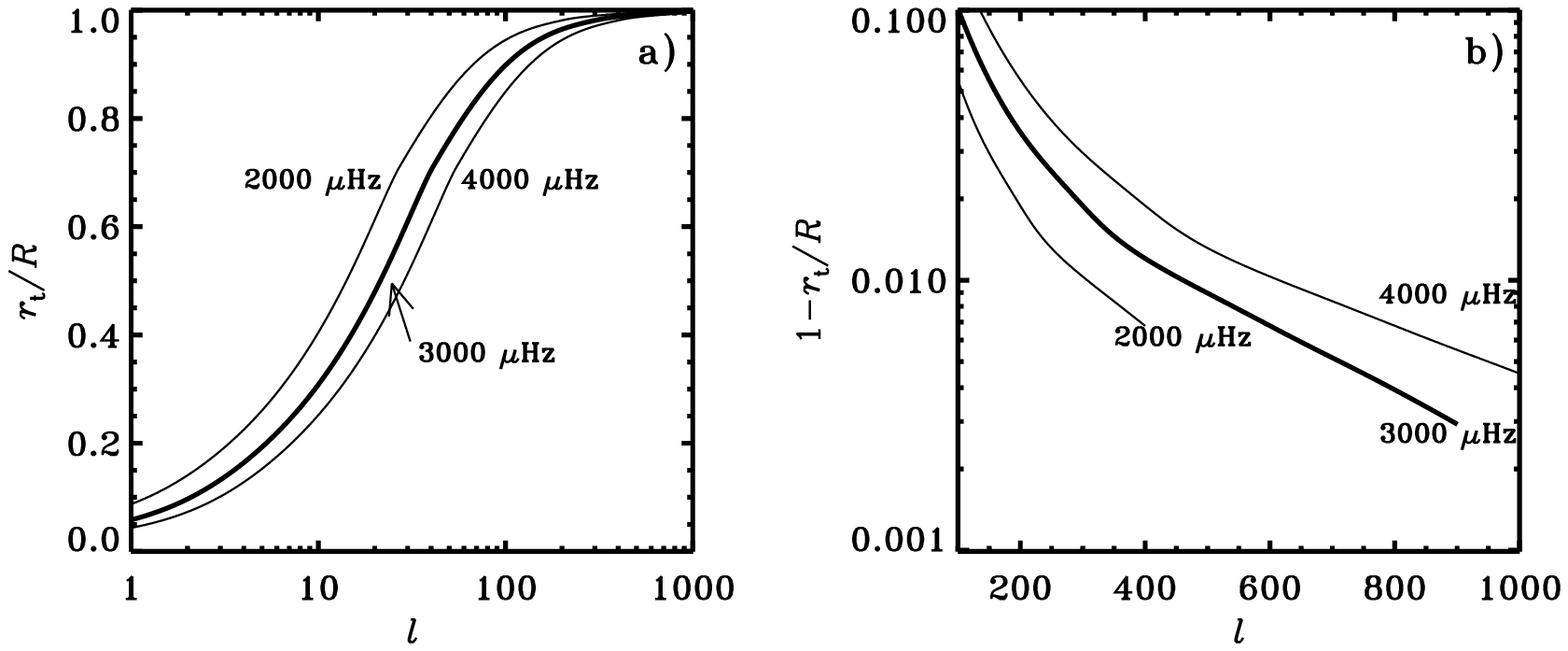}
\caption{Location $r_{\rm t}$ of the inner turning point (a), and depth of
penetration $R-r_{\rm t}$ (b), in units of the solar radius
$R$, for $p$-modes in a standard solar model. The results are shown as
functions of degree $l$, for three typical frequencies. Adapted from
\cite{jcd03}.
\label{itpr}}
\end{figure}

Depending on the frequency and degree $l$, the modes propagate within different
acoustic cavities inside the Sun between two turning points. Outside the
acoustic cavity in which a mode is propagating, the mode is evanescent.
Therefore, the mode characteristics (e.g., frequency) are relatively insensitive
to conditions outside the associated acoustic cavity, particularly far from the
turning points. Moreover, modes which propagate in the direction of solar
rotation have higher frequencies than modes with the same resonant properties
propagating in the opposite direction. This effect is called ``rotational
frequency splitting''. The amount of splitting depends on the rotation rate
inside the acoustic cavity in which the mode is propagating as well as on the
azimuthal order $m$ of the mode. Utilizing the differential penetration and the
frequency splitting of the modes allows the internal structure and rotation of
the Sun to be inferred, as a function of position \citep[cf.][]{jcd02a}. This
possibility of carrying out inversions of the observed frequencies and
frequency splittings is a key issue in the applications of helioseismology. So
far, however, the vast majority of inversions performed have only included
frequencies and frequency splittings of the low- and the intermediate-degree
oscillations \citep[see, e.g.,][]{Gou96,Thom96,Antc98,Kos98,Schou98}. On the
other hand, high-degree $f$- and $p$-modes have an immense potential in the helioseismic
probing of the sub-surface layers of the Sun. This is demonstrated here in
Figure~\ref{itpr}, where the dependence of the inner turning-point radius
$r_{\rm t}$ on spherical harmonic degree $l$ is shown for three different
frequencies, spanning the range of the observations. For small $l$, the
inner turning point is rather close to the center of the Sun, whereas for higher degrees
it moves closer to the surface. In particular, for $l > 150$ the modes are
essentially trapped in the outer 65 Mm below the solar surface. Therefore,
accurate measurements of high-degree mode frequencies and splittings allow us
to improve our inferences regarding the large-scale structure and dynamics of
the sub-surface layers \citep[see, e.g.,][]{Rab00,Rab08,Dim02}.

%Fig. 1

Due to the use of a modal concept in global helioseismology, the diagnostic
potential of the data is necessarily limited. Specifically, to first order the
standing acoustic modes sense only the longitudinally averaged, north-south
symmetric average of the internal stratification of the Sun. Moreover, in
contrast to solar differential rotation, flows in meridional planes (meridional
circulation) have only a tiny effect on global oscillation frequencies
\citep{Woo00,Rth08,Schad11,Vor11,Schad13}, which severely hampers any attempt to
detect such flows by global mode frequency analysis. Such limitations are
avoided in local helioseismology, which is based upon the assumption that the
solar oscillations locally behave as propagating acoustic waves that are
scattered and absorbed by local inhomogeneities and advected by local flow
fields. \cite{Bra87} were the first to demonstrate the utility of this approach
by showing that propagating acoustic waves could be absorbed by the strong
magnetic field associated with sunspots, thus potentially providing information
about the magnetic field itself. Subsequently, three methods of analyzing
propagating acoustic waves in a localized area on the solar surface were
devised: ring-diagram analysis \citep{Hil88,GH08}, time-distance analysis
\citep{Duv93,Kos96a,Giz05,Zha04,Zha08}, and acoustic holography
\citep{Lin00,Lin04}. The application of these methods has led to spectacular
results, as has been demonstrated by, e.g., \cite{Gil97}, \cite{Kos00},
\cite{Bra01}, \cite{Beck02}, \cite{Hab02}. For a review of local
helioseismology we refer the reader to \cite{Giz10}.

While in recent years the progress made in local helioseismology has been
substantial, it has been much slower in global helioseismology. This is in part
a consequence of the difficulties inherent in the generation of reliable
high-degree mode parameters, but misconceptions as to the roles of local and
global helioseismology may have contributed as well. One such misconception is
that the global mode measurements can be replaced with measurements from local
analyses. As \cite{Rei07} has demonstrated, at large-scales the local
measurements are much less precise than the global measurements and, therefore,
there is a strong complementarity between the local and global techniques.

The estimation of high-degree mode parameters is made difficult due to the fact
that high-degree modes cannot be observed as sharp, isolated peaks but only as
ridges of power comprised of overlapping modes. Because of the asymmetrical
distribution of the amplitudes of the modes that blend together, the central
frequency of each ridge deviates from the frequency of the target mode. Hence,
to recover the underlying mode frequency from fitting the ridge, an accurate
model of the ridge power as a function of frequency is required. With such an
accurate model the global analysis provides the most robust estimates of the
mean structure and rotation of the Sun which are important for testing theories
of stellar structure, evolution, and differential rotation.

We began to delve into the fitting of solar oscillation spectra in the
late-1980s with the development of our first-generation, or Single-Peak,
Averaged-Spectrum, fitting method, which we will refer to in the following as
either Method~1 or the SPAS method, and which is briefly described here in
Appendix~\ref{secm1}. Because of insurmountable problems with the determination
of unbiased ridge-fit frequencies we had to abandon this method, however. We
therefore began to develop our second-generation, or Windowed, MuLTiple-Peak,
averaged-spec\-trum, fitting method, which we will refer to in the following as
either Method~2 or the WMLTP method. This method is equally well suited for the
unbiased estimation of low-, medium-, and high-degree mode parameters from
$m$-averaged solar oscillation power spectra. A detailed description of this
method will be presented in Section~\ref{secm2}, after we have given an outline
of the data analysis generally employed in global helioseismology in
Section~\ref{dagh}, and after we have addressed the problems inherent in the
analysis of high-degree power spectra in Section~\ref{pahds}. The issue of the
sensitivity of Method~2 in terms of both the $m$-averaging procedure and the
effective leakage matrix is addressed in Section~\ref{sfmp}. Sample results
from Method~2 are presented in Section~\ref{SR}. In Section~\ref{SR} we will
also demonstrate the substantial improvements that Method~2 makes in the
frequencies, linewidths, and amplitudes that we generated with this method by
comparing them with corresponding quantities that we generated using our
Method~1. Finally, in Section~\ref{hinv} we firstly describe a new procedure
for the identification and correction of outliers in frequency data sets that
are to be used for solar structure inversions, before we present a new
structural inversion from a set of frequencies computed from 66-day long
spectra obtained with the Michelson Doppler Imager (MDI) \citep{Scher95} on
board the SOlar and Heliospheric Observatory (SOHO) \citep{Dom95} at the
beginning of solar cycle~24 in 2010, followed by our concluding remarks in
Section~\ref{concl}.

At this point we note that the current paper is the first of a series of three
papers. While the focus of the paper at hand is on Method~2, we will present in
the second paper our third-generation, or Multiple-Peak, Tesseral-Spectrum,
fitting method, which we will refer to in the following as either Method~3 or
the MPTS method. This method directly fits the tesseral, zonal, and sectoral
spectra at each degree rather than resorting to $m$-averaged spectra as is the
case with both Method~1 and 2. In that paper we will also intercompare the
results obtained from Method~2 and 3, and we will investigate the systematic
effects introduced in Method~2 by the $m$-averaging procedure. The purpose of
the third paper will be the intercomparison of our results obtained from both
Method~2 and 3 with results from ``established'' fitting methodologies at low,
intermediate, and high degrees.

\section{Data analysis in global helioseismology\label{dagh}}

In global helioseismic studies the data reduction typically includes the
following major steps. First, the observed Dopplergrams $V(\theta,\phi,t)$ are
spatially decomposed into spherical harmonic coefficients $\frca_{l,m}(t)$, i.e.,
\begin{equation}
\frca_{l,m}(t) = \int_{\mathcal{D}} W(\theta,\phi) V(\theta,\phi,t) 
    Y_{l}^{m}(\theta,\phi) \,\mathrm{d}\sigma,
\label{shc}
\end{equation}
where $\theta$ is co-latitude, $\phi$ is longitude, $Y_{l}^{m}(\theta,\phi)$ is
a spherical harmonic function of degree $l$ and azimuthal order $m$,
$W(\theta,\phi)$ is an apodization chosen to reduce the contribution from the
noise close to the solar limb, $\mathcal{D}$ is the visible hemisphere of the
Sun or a portion thereof, and $t$ is time. In this step of the data reduction
spatial side-lobes are introduced into the target spectrum $(l,m)$ because the
spherical harmonic functions $Y_{l}^{m}(\theta,\phi)$ are not orthogonal on
$\mathcal{D}$. However, even if the entire surface of the Sun could be observed
spatial side-lobes would be present because of the distortion of the mode
eigenfunctions by solar latitudinal differential rotation, and also because of
velocity projection effects. The integral in equation (\ref{shc}) can be very
expensive to compute. A typical approach is to use interpolation to remap, for
given time $t$, the product $W(\theta,\phi) V(\theta,\phi,t)$ onto some
coordinate system in which the integration over $\phi$ may be represented as a
Fourier transform, so that Fast Fourier Transform (FFT) techniques may be
applied. In the second step a spectral analysis is carried out for each
spherical harmonic coefficient $\frca_{l,m}(t)$, i.e.,
\begin{equation}
\frcA_{l,m}(\nu) = \int_{-\infty}^{+\infty} \frca_{l,m}(t)
    \exp(2\pi i\nu t) \,\mathrm{d} t,
\label{FT}
\end{equation}
where $\nu=\omega/2\pi$ is cyclic frequency. In this step temporal side-lobes
are introduced into the target spectrum $(l,m)$ if periodic gaps are present in
the time series of the spherical harmonic coefficients $\frca_{l,m}(t)$ due to the
day-night cycle, say. In practice, the integral in equation (\ref{FT}) is first
approximated by a discrete Fourier transform over a finite interval of time,
which then is efficiently computed using a FFT technique. In the third step the
power spectrum $\Phi_{l,m}(\nu) = |\frcA_{l,m}(\nu)|^2$ is calculated for each
$(l,m)$. The final step in the data reduction consists in the peak fitting of
the power spectra $\Phi_{l,m}(\nu)$. Alternatively, the peaks in the complex
spectra $\frcA_{l,m}(\nu)$ can be fitted as well. For example, such approach is
employed in the fitting methodology of \cite{Schou92}.

\subsection{Generation of un-averaged power spectra\label{gunavgspc}}

The results presented in this investigation are based upon four different sets
of un-averaged power spectra $\Phi_{l,m}(\nu)$ that were created from
observations obtained with the MDI instrument during 1996, 2001 and 2010. The
MDI was operated on the SOHO spacecraft between April 1996 and April 2011. The
MDI observations which we have employed were all obtained as part of the MDI
Full-Disk Program \citep{Scher95}, and are listed in Table~\ref{tab1},
where we also indicate the naming convention we use in this paper to refer to
each observing run.

Time series of Dopplergrams that resulted from each of the four observing runs
listed in Table~\ref{tab1} were converted into $l+1$ complex time series
(purely real for $m=0$) which were gap-filled using an auto-regressive gap
filling procedure based upon the approach of \cite{Fah82}, using a reduction
pipeline that was developed at Stanford University for the processing of the
MDI data.  The resulting gap-filled duty cycles are listed in the last column
of Table~\ref{tab1}. Using standard FFT techniques the gap-filled $l+1$ complex
time series were converted into a group of $2l+1$ zonal, tesseral, and sectoral
power spectra for $0\leq l\leq 1000$ for each of the four observing runs. In
doing so, the positive frequency part is identified with $m < 0$, while the
negative frequency part is identified with $m\geq 0$.

Within each of these four groups of un-averaged power spectra, each target
spectrum $(l,m)$ contains a number of frequency bins equal to one-half of the
number of samples that is listed in the corresponding row of column~3 in
Table~\ref{tab1}. For the MDI instrument, these frequency bins span the
frequency range of zero to the temporal Nyquist frequency of $8333\,\,\mu$Hz.
Within each group of spectra the zonal (i.e., $m=0$) spectrum for a given
degree, $l$, contains a variable number of sets of isolated peaks (at low- and
intermediate degrees) or a set of ridges (at higher degrees) of power that
correspond to a collection of $f$- and $p$-modes. For the cases in which the
peaks are isolated, each set of peaks consists of a peak for the target mode
$(n,l)$, the set of temporal sidelobes, and a set of spatial sidelobes that
have leaked into the target spectrum from nearby spectra. We will refer to the
entire collection of $2l+1$ target peaks and their spatial and temporal
sidelobes that share a common $n$-value as the $(n,l)$ multiplet. When we refer
to the mode $(n,l)$, we are actually referring to the $m$-average of the $2l+1$
modes that share the same values of $n$ and $l$.

For the tesseral and sectoral spectra the corresponding peaks in each spectrum
are shifted to lower frequencies by solar rotation for the spectra having
$m<0$, while they are shifted to higher frequencies for the spectra having
$m>0$ \citep[cf.][]{jcd00}.

\subsection{Procedures for the generation of \textit{\textbf{m}}-averaged power
spectra \label{gmavg}}

For a given degree, $l$, the $m$-averaged power spectrum
$\Phi_{l}^{\mathrm{mav}}(\nu)$ is defined as
\begin{equation}
\Phi_{l}^{\mathrm{mav}}(\nu) = 
    \langle \Phi_{l,m}(\nu) \rangle_{m},
\label{eqspcmav}
\end{equation}
where the symbol $\langle\,\rangle_{m}$ means averaging over $m$, is computed
in a three-step procedure. In the first step of this procedure, the frequency
shift 
resulting from the effect of solar rotation and asphericity is calculated, for
each of the $2l+1$ un-averaged spectra, $\Phi_{l,m}(\nu)$, using an iterative
cross-correlation method \citep{Bro85,Tom88,Kor90}.
In this method the frequencies within a multiplet $(n,l)$
are approximated with a polynomial expansion similar to that
of \cite{Duv86}, i.e.,
\begin{equation}
\nu_{n,l,m}=\nu_{n,l} + L\sum_{k=1}^{6} a^{(n,l)}_{k} P_{k}(m/L).
\label{splitc}
\end{equation}
Here, $\nu_{n,l}$ is the frequency of the multiplet $(n,l)$, $L^2=l(l+1)$,
$a^{(n,l)}_{k}$ are the so-called frequency-splitting coefficients, and $P_{k}$
is the Legendre polynomial of degree $k$. The splitting coefficients
$a^{(n,l)}_{k}$ with odd $k$ arise from solar internal rotation, while the
coefficients with even $k$ are caused by departures from spherical symmetry in
solar structure, or from effects of magnetic fields. Most of the $m$-averaged
power spectra that we fit for this manuscript were generated using only the
three lowest, odd-$k$ frequency-splitting coefficients (i.e., $a_1$, $a_3$, and
$a_5$). In the second step of the process in which we computed the $m$-averaged
spectra, we shifted each tesseral and sectoral spectrum by the calculated
frequency shift for that $m$-value. In the third step in this procedure, we
averaged all of these shifted spectra together with the un-shifted zonal
spectrum, $\Phi_{l,0}(\nu)$, to create the $m$-averaged spectrum
$\Phi_{l}^{\mathrm{mav}}(\nu)$ for each degree $l$. If this $m$-averaging is
carried out in an unweighted manner, we will refer to the resulting set of
$m$-averaged spectra as ``unweighted, $m$-averaged spectra''. However, as we
previously described in \cite{Rho01}, it is also worth considering average
spectra computed by combining the $2l+1$ spectra $\Phi_{l,m}(\nu)$ in a
weighted manner, using as weights the inverse mean power over the 1500 to
$4500\,\,\mu$Hz frequency range. We will refer to such a set of spectra as
``weighted, $m$-averaged spectra''.

We have developed two different versions of the cross-correlation method to
generate rotational splitting coefficients which then are used in the
computation of the $m$-averaged spectra. In the first of these versions, we
cross-correlate the individual spectra over a wide range of frequencies such
that most or all of the ridges at a given degree are included, while in the
second version we carry out the cross-correlation over a narrow range of
frequencies centered about a single ridge at each degree. In the second
version, we then repeat these narrow-band cross-correlations for all of the
successive ridges at a given degree in order to build up a set of narrow-band
splitting coefficients for that degree. In the wide-band version of the
cross-correlation code we effectively are computing the averages of the
frequency splittings over all of the adjacent ridges which are located within
the frequency limits of the cross correlation (typically from 1800 to
$4800\,\,\mu$Hz). The splitting coefficients from the wide-band procedure are
called the $n$-averaged splitting coefficients, while those from the
narrow-band version of our code are called non-$n$-averaged splitting
coefficients.

\subsection{Correction for distortions introduced by latitudinal
differential rotation\label{cdsldr}}

\begin{figure}
\epsscale{1.00}
\plotone{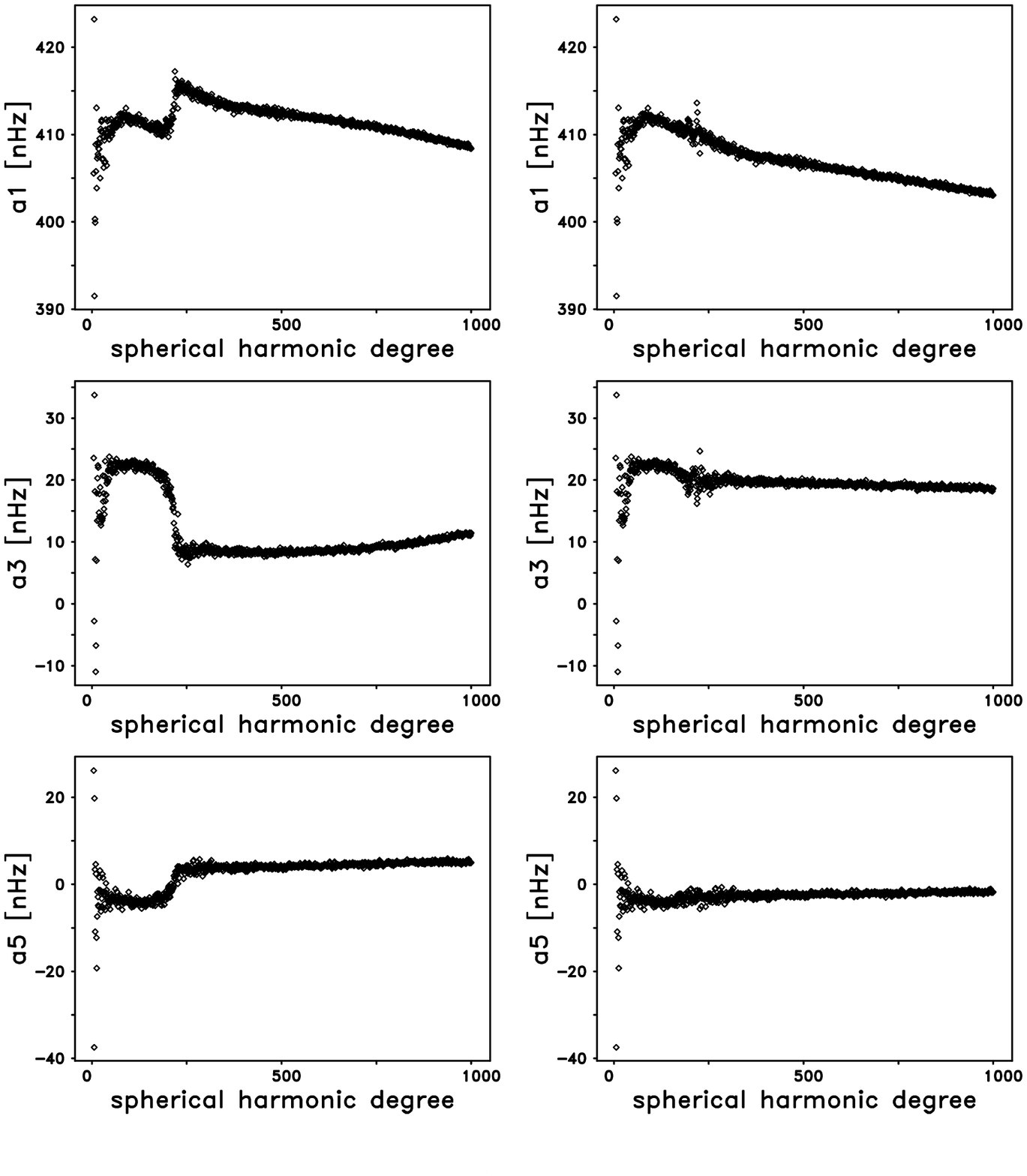}
\caption{
(left) Degree dependence of the $n$-averaged odd-order splitting coefficients
$a_1$, $a_3$, and $a_5$, in the range $5\leq l \leq 1000$ computed from the set
of un-averaged power spectra obtained from the $\cal{R}$1996\_61 observing run,
with the $a_1$ coefficient at the top. As can be clearly seen, these raw
splitting coefficients illustrate large discontinuities at about $l \approx
210$. (right) Same as shown in the left panels, but after the discontinuities
have been removed using the adjustment procedure described in Section~\ref{gmavg}.
\label{spcavg}}
\end{figure}

For the results that we will be presenting later, we generated a set of
$n$-averaged frequency-splitting coefficients by cross-correlating the
un-averaged power spectra obtained from the $\cal{R}$1996\_61 observing run
(cf. Table~\ref{tab1}). The odd-order splitting coefficients (i.e., $a_1$,
$a_3$, and $a_5$) that we obtained from this procedure are shown here in the
left three panels of Figure~\ref{spcavg}. These raw frequency-splitting
coefficients show large discontinuities in the degree range of $200\lesssim l
\lesssim 240$. Similar discontinuities were first noticed by \cite{Kor90}.
Subsequently, \cite{Rho98a} confirmed the presence of these jumps in MDI
observations. The exact location of the range of $l$ values where these jumps
occur depends primarily upon the duration of the observing run from which the
power spectra were generated, with shorter-duration observing runs showing the
jumps at lower degrees.

%Fig. 2

\cite{Woo89} pointed out that the distortion of high-degree $f$- and $p$-mode
eigenfunctions caused by a slow, antisymmetric differential rotation can be
expressed as a superposition of the unperturbed eigenfunctions of the same
radial order $n$ if the Coriolis forces are neglected. Following a discussion
of Woodard's suggestion in a preprint of \cite{Kor04}, \cite{Rei03} found that
the inclusion of this effect in the calculation of the leakage matrices had a
very dramatic impact upon the resulting frequency-splitting coefficients.
Examples of the changes introduced into the odd splitting coefficients when
corrections are made for the distortion were presented by \cite{Rei03}, who
showed that at the degrees below $l \approx 200$ the splitting coefficients
remained almost unchanged, while at the higher degrees the jumps were seen to
disappear when the mode coupling due to the differential rotation of the Sun
was taken into account. 

The results that \cite{Rei03} presented were generated using a preliminary
version of our MPTS method that we have been developing in parallel to the
WMLTP method that we are presenting in this paper. Because: 1) the MPTS method
is extremely computationally-intensive; 2) it cannot be employed upon power
spectra that come from observing runs that are as short as only three days in
duration due to the low signal-to-noise ratios that are inherent in such
low-resolution spectra; and 3) we have not yet had the opportunity of
implementing the changes that we have recently made in our WMLTP method into
the MPTS code, we have not yet employed the MPTS method to compute entire sets
of frequency-splitting coefficients. Instead, we corrected the raw splitting
coefficients that are shown in the left three panels of Figure~\ref{spcavg}
with a two-step adjustment procedure. First, for each of the five splitting
coefficients we fit a least-squares straight line to the degree range of 90 to
190 and we also fit a second least-squares straight line to the degree range
from 230 to 400. For the degrees ranging from 200 to 230 the two linear fits
for each splitting coefficient were simply connected with a third straight
line. The difference between that line and the extrapolation of the
left-hand line was subtracted from the raw coefficient values for all of the
degrees between 200 and 230. For all degrees above $l=230$ the offset employed
for $l=230$ was subtracted from each coefficient. Because these
initially-corrected splitting coefficients showed evidence of systematic
variations with increasing degree in the three odd-order coefficients (i.e.,
$a_1$, $a_3$, and $a_5$), we computed, in the second step of our correction
procedure, a low-order polynomial fit to each of the three odd-order
coefficients over the degree range of 488 to 1000. We then subtracted these
polynomial fits from the partially-corrected, odd-order splitting coefficients
and we stopped the correction process at this point. This procedure generated
the set of corrected odd-order splitting coefficients that are shown in the
right three panels of Figure~\ref{spcavg}, where it is clear that the jumps
have been removed. We refer to this set of adjusted frequency-splitting
coefficients as our set of ``corrected, $n$-averaged" coefficients. 

Using a similar adjustment procedure we also have corrected the set of raw
non-$n$-averaged frequency-splitting coefficients that we previously computed
using the narrow-band version of our cross-correlation method on the
un-averaged power spectra obtained from the $\cal{R}$1996\_61 observing run.
We refer to this set of adjusted frequency-splitting coefficients as our set of
``corrected, non-$n$-averaged'' coefficients.

\subsection{Sets of \textit{\textbf{m}}-averaged power spectra generated from
the observing runs used in this work}

From the un-averaged power spectra obtained from the observing runs 
specified in Table~\ref{tab1} we have generated a total of seven different sets
of $m$-averaged power spectra, which are listed in the second column of
Table~\ref{tab2} using a naming convention quite similar to that introduced in
Table~\ref{tab1} to refer to the individual observing runs.
These seven sets of $m$-averaged spectra differed in four key ways: 1)~origin
of the un-averaged power spectra, 2)~whether or not the raw, un-averaged
spectra were weighted prior to being averaged, 3)~whether the set of
frequency-splitting coefficients was used as computed or as corrected for the
effects of latitudinal differential rotation and other features that appeared
not to be solar in origin, and 4)~whether or not the set of frequency-splitting
coefficients was computed using a narrow or a wide frequency range at each
degree (i.e., whether those coefficients were computed for the individual
ridges or were computed in an $n$-averaged manner). We note that we will not
present any fits to the $m$-averaged spectral set $\cal{S}$1996\_61 in this
work. Rather, this set of $m$-averaged spectra is listed in Table~\ref{tab2}
only for the sake of completeness because it was a by-product of the
cross-correlation process that generated the raw, uncorrected, $n$-averaged
splitting coefficients from observing run $\cal{R}$1996\_61.

\section{Problems requiring the use of multiple peaks in the fitting profile
\label{pahds}}

\subsection{Basic considerations \label{bc}}

For the following reasons, high-degree modes cannot be observed as isolated,
sharp peaks but only as ridges of power. First, the power spectrum computed for
a specific target mode with degree $l$ and azimuthal order $m$ contains
contributions of power from modes with neighboring $l$ and $m$ because the
spherical harmonic functions used in the spatial decomposition of the observed
Dopplergrams are not orthogonal on that part of the Sun we observe (see
equation~(\ref{shc})). These unwanted contributions, or spatial leaks,
are quantified by the so-called leakage matrix (see Sect.
\ref{seclkm}). Second, with increasing degree the frequency separation of the
spatial leaks decreases, while the mode linewidth increases with both frequency
and degree. As a consequence, individual modal peaks blend together to form
ridges of power. Typically, modes begin to blend into ridges for degrees
ranging anywhere from $l\approx 20$ ($n=29$) to $l\approx 300$ ($n=0$)
depending on the radial order $n$. Since the amplitudes of the spatial leaks
are asymmetric with regard to the target mode the central frequency of a ridge
is significantly offset from the target mode frequency. Therefore, the
distribution of power in a ridge cannot be simply represented by using just a
single symmetrical or asymmetrical function of frequency. Rather, a sum of
individual overlapping profiles must be employed the relative amplitudes of
which are governed by the leakage matrix appropriate to the targeted mode.
Thus, the correct estimation of the leakage matrix is crucial in the accurate
measurement of high-degree mode parameters. Moreover, the use of a model
profile consisting of the sum of individual profiles allows the fitting of
low-, medium-, and high-degree modes in like manner. In this way systematic
errors are avoided which otherwise would be inevitably introduced if subsets of
mode parameters are to be combined each of which has been generated by using a
different fitting methodology.

\subsection{Leakage matrix\label{seclkm}}

While the determination of the leakage matrix is straightforward for low- and
medium-degrees, at high degrees the leakage matrix calculations are greatly
complicated by the necessity to take into account (1) the horizontal component
velocity, (2) the distortion of the eigenfunctions by the solar differential
rotation, and (3) instrumental effects that cause image distortion and
smearing. We will address these issues in the following paragraphs.\\

\subsubsection{Radial and horizontal component\label{rhc}}

The Fourier transform $\tilde{O}_{l,m}$ of the time series of spherical
harmonic amplitudes of a target mode with degree $l$ and azimuthal order $m$
can be written as a sum over the Fourier transform $\tilde{o}_{n',l',m'}$ of
the time series of the solar oscillation modes given by $(n',l',m')$, viz.
\begin{equation}
\tilde{O}_{l,m} = \sum_{n',l',m'} C_{n';l,m;l',m'}\,\tilde{o}_{n',l',m'},
\end{equation}
where $C_{n';l,m;l',m'}$ is the leakage matrix. As \citet{Kor04} have shown,
the leakage matrix can be written as
\begin{equation}
C_{n';l,m;l',m'} = C_{l,m;l',m'}^{\,\mbox{\scriptsize (r)}} + 
  c_t^{(n',l')} C_{l,m;l',m'}^{\,\mbox{\scriptsize (h)}},
\label{lkm}
\end{equation}
where $C_{l,m;l',m'}^{\,\mbox{\scriptsize (r)}}$ is the part coming from the
radial displacement, $C_{l,m;l',m'}^{\,\mbox{\scriptsize (h)}}$ is the part
coming from the horizontal displacement, and $c_t^{(n',l')}$ is the ratio of
the horizontal displacement to the radial displacement. It should be noted that
both $C_{l,m;l',m'}^{\,\mbox{\scriptsize (r)}}$ and
$C_{l,m;l',m'}^{\,\mbox{\scriptsize (h)}}$ are independent of the radial order
$n'$ of the mode. Using the normalization of the displacement eigenfunction
components given by \cite{Gou93}, the displacement component ratio is given by
\begin{equation}
c_t^{(n,l)} = \frac{|\xi_{n,l}^{\,\mbox{\scriptsize (h)}}(r_{\ast})|}
    {|\xi_{n,l}^{\,\mbox{\scriptsize (r)}}(r_{\ast})|}.
\label{dcr}
\end{equation}
Here, $r_{\ast}$ is the radial location of observation, and
$\xi_{n,l}^{\,\mbox{\scriptsize (h)}}$ and $\xi_{n,l}^{\,\mbox{\scriptsize
(r)}}$ are, respectively, the horizontal and radial components of the
displacement eigenfunction of the mode $(n,l)$, given by
\begin{equation}
\mbox{\boldmath $\xi$}(r,\theta,\phi) = 
    \left[\xi_{n,l}^{\,\mbox{\scriptsize (r)}}(r)\,\mbox{\boldmath $e_r$}
    + \frac{1}{L}\,\xi_{n,l}^{\,\mbox{\scriptsize (h)}}(r)
    \left(\mbox{\boldmath $e_{\theta}$}\,\frac{\partial}{\partial\theta}
    + \mbox{\boldmath $e_{\phi}$}\,
    \frac{1}{\sin\theta}\frac{\partial}{\partial\phi}\right)\right]
    Y_{l,m}(\theta,\phi),
\label{displcm}
\end{equation}
where $(r,\theta,\phi)$ are spherical polar coordinates with $r$ being the
distance to the center, $\theta$ being the co-latitude, $\phi$ being the
longitude, $\mbox{\boldmath $e_r$}$, $\mbox{\boldmath $e_{\theta}$}$,
$\mbox{\boldmath $e_{\phi}$}$ are, respectively, the unit vectors in the $r$,
$\theta$, $\phi$ directions, $Y_{l,m}(\theta,\phi)$ is a spherical harmonic of
degree $l$ and azimuthal order $m$, and $L^2=l(l+1)$. As \cite{jcd03} has
shown, the displacement component ratio can be written as
\begin{equation}
c_t^{(n,l)} = \frac{GM_{\odot}L}{4\pi^2 R_{\odot}^3\nu_{n,l}^2}=
              \left(\frac{\nu_{0,l}}{\nu_{n,l}}\right)^2,
\label{ctth}
\end{equation}
where
\begin{equation}
\nu_{0,l}=\left(\frac{GM_{\odot}L}{4\pi^2 R_{\odot}^3}\right)^{1/2}
\label{nufm}
\end{equation}
is the frequency of the $f$-mode of degree $l$ in the asymptotic high-degree
limit \citep{Gou80}, $\nu_{n,l}$ is the average frequency for the multiplet
$(n,l)$, and $M_{\odot}$ and $R_{\odot}$ are, respectively, the mass and the
radius of the Sun. For $p$-modes equation~(\ref{ctth}) implies that
$0<c_t^{(n,l)}\leq 1$ because, for fixed $l$, the $f$-mode frequency is smaller
than any $p$-mode frequency. It should be noted that $c_t^{(n,l)}$ as defined
in equation~(\ref{dcr}) is not only equal to the ratio of the displacement
eigenfunction components but is also equal to the ratio of the horizontal and
vertical components of the velocity eigenfunctions for the mode. For low- and
medium-degree modes \cite{Rho01} have shown that the observed values of
$c_t^{(n,l)}$ closely match the theoretical prediction given in
equation~(\ref{ctth}). \cite{Rab01} arrived at a similar conclusion. For
high-degree modes the agreement between the measured and theoretical
horizontal-to-vertical displacement ratio has been demonstrated by
\cite{Schou98a}.

When power spectra are to be fitted rather than Fourier spectra we have to
compute the leakage matrix, $C_{n;l,m;l',m'}^{\,\mbox{\scriptsize (ps)}}$,
relevant to power spectra, which is given by
\begin{equation}
C_{n;l,m;l',m'}^{\,\mbox{\scriptsize (ps)}} = C_{n;l,m;l',m'}^2.
\label{lkmps}
\end{equation}
Moreover, for the fitting of $m$-averaged power spectra we have to compute the
leakage matrix, $C_{n;l,l'}^{\,\mbox{\scriptsize (mavg)}}$, which measures, for
a given ridge of radial order $n$, the contribution of a mode of given $l'$ in
the power spectrum calculated for a mode of given $l$. To do so, we need to
take the sum of the squares of all the $m$-leaks in equation~(\ref{lkm}).
Using equation~(\ref{lkmps}) we get
\begin{equation}
C_{n;l,l'}^{\,\mbox{\scriptsize (mavg)}} = \sum_{m=-l}^{l}
  \sum_{m'=-l'}^{l'} C_{n;l,m,l',m'}^{\,\mbox{\scriptsize (ps)}}.
\label{lkmmavg}
\end{equation}
In practice the leaks $C_{n;l,m,l',m'}^{\,\mbox{\scriptsize (ps)}}$ fall off
rather rapidly with increasing $|m-m'|$ and increasing $|l-l'|$. Hence, the
sums in equation~(\ref{lkmmavg}) only have to be evaluated for a limited range
of both $|l-l'|$ and $|m-m'|$.\\

\subsubsection{Distortion by the solar differential rotation\label{dsdr}}

One of the conspicuous effects of solar rotation is the well-known splitting of
the oscillation frequencies, that is the dependency of the oscillation
frequencies on the azimuthal order $m$ (cf. equation~(\ref{splitc})).
Similarly, the modal eigenfunctions of the solar oscillations depart from their
customarily assumed spherical harmonic form (cf. equation~(\ref{displcm})) as a
result of solar rotation. As \citet{Woo89} has shown, the distortion of
high-degree mode eigenfunctions by a slow, axisymmetric differential rotation
can be expressed as a superposition of the unperturbed eigenfunctions of the
same radial order $n$, if Coriolis forces are neglected. He also has shown,
that the perturbed leakage matrix can be expanded in terms of the unperturbed
leakage matrix as
\begin{equation}
C_{n;l,m;l',m'}^{\,\mbox{\scriptsize (sdr)}} = 
        \sum_{l''} \gamma_{l',l''} C_{n;l,m;l'',m'},
\label{lkmexp}
\end{equation}
where
\begin{eqnarray}
\gamma_{l',l''} &=& {\left\{ \begin{array}{ll}
                    \displaystyle
                    \frac{(-1)^{p}}{2\pi}\int_{-\pi}^{\pi}
                    \cos\, [p\varkappa + Z(\varkappa)]\,{\rm d}\varkappa
                    & \hspace{2.0mm}\mbox{for}~l'-l''~\mbox{even} \\[2.0mm]
                    0 & \hspace{2.0mm}\mbox{for}~l'-l''~\mbox{odd}
                    \end{array} \right.} \label{Gamma}\\[2.0mm]
p &=& (l'-l'')/2 \\[2.0mm]
                Z(\varkappa) &=&
                \frac{x_{l'}^2}{4}\frac{m}{\partial\nu / \partial l |_{l'}} \bigg[
                (B_2 + B_4 x_{l'}^2) \sin\varkappa
                -\frac{B_4 x_{l'}^2}{8}\,\sin\,(2\varkappa) \bigg]
\label{deltaPhi}\\[2.0mm]
x_{l'} &=& 1 - \frac{m^2}{l'(l'+1)}.
\label{xlp}
\end{eqnarray}
In equation~(\ref{deltaPhi}) $\partial\nu/\partial l|_{l'}$ denotes the
derivative of frequency $\nu$ with respect to degree $l$ evaluated at degree
$l'$. It is assumed that $l$ can be treated as a continuous variable.
Actually, $\partial\nu/\partial l$ is calculated by taking the derivative of a
smooth function fitted to the march of $\nu$ versus $l$ along a ridge of given
radial order $n$ (cf. Section~\ref{srm2}). The $B$-coefficients in
equation~(\ref{deltaPhi}) result from a parametrization of the angular velocity
$\Omega(\theta)$ of the surface differential rotation as a function of
co-latitude $\theta$. Following \citet{Sno90} we have
\begin{equation}
\Omega(\theta)=2\pi (B_0 + B_2\cos^2\theta + B_4\cos^4\theta),
\label{mssr}
\end{equation}
where
\begin{equation}
B_0=473.0~\mbox{nHz},\,B_2=-77.0~\mbox{nHz},\,B_4=-57.5~\mbox{nHz}
\label{bc1}
\end{equation}
for rotation of Doppler features on the solar surface.

The dependence of the perturbed leakage matrix (\ref{lkmexp}) on the
$B$-coefficients involved in the rotational model (\ref{mssr}) implies the
following problem. On the one hand, the radial variation in the latitudinal
differential rotation profile and, hence, the surface differential rotation can
be measured through a rotational inversion of the frequency-splitting
coefficients $a^{(n,l)}_{k}$ as given in equation~(\ref{splitc}). For the
measurement of the $a^{(n,l)}_{k}$ a perturbed leakage matrix (\ref{lkmexp})
must be specified and input into the peak-bagging code. On the other hand,
equation~(\ref{deltaPhi}) demonstrates that the perturbed leakage matrix itself
and, hence, also the frequency-splitting coefficients, depend upon the
$B$-coefficients which were used to parametrize the solar differential
rotational profile in the surface layers. This mutual dependence can only be
resolved with some kind of fixed-point iteration. Nevertheless, as we discussed
earlier in Section~\ref{cdsldr}, the importance of including these effects into
the calculation of the leakage matrices was demonstrated by \cite{Rei03}, who
showed that their inclusion removed the discontinuities that were otherwise
present in the non-$n$-averaged splitting coefficients for the $n=2$ ridge.

\subsubsection{Instrumental effects\label{instreff}}

For the determination of high-degree mode parameters not only the leakage
matrix must be known but also the instrumental characteristics must be very
well understood and very precisely measured \citep[cf.][]{Rab01}. The
instrumental effects to be considered in the analysis include plate scale
error, image distortion, width and spatial non-uniformity of the instrumental
point spread function (PSF), image orientation ($P$-angle), and the finite
pixel size of the detector. Similarly, one has to consider errors in the
$P$-angle and $B_0$-angle caused by errors in the assumed orientation of the
solar rotation axis. Using data obtained with the MDI instrument, \citet{Rei03}
have shown that the inclusion of both the plate scale error and the image
distortion has a rather strong impact upon the measured splitting coefficients
for degrees $l\gtrsim 200$. 

We have found it convenient to calculate the leakage matrix by constructing
simulated images corresponding to the line-of-sight contribution of each
component of a single spherical harmonic mode and then decomposing that image
into spherical harmonic coefficients using exactly the same numerical
decomposition pipeline employed to process the observations. This approach has
the advantage that some of the above-described instrumental effects can easily
be included in the leakage matrix calculation. It should be noted, however,
that the inclusion of instrumental effects in the effective leakage matrix is,
in general, not equivalent to taking into account those effects in the pipeline
used to process the observations. 

For a thorough discussion of the instrumental effects of the MDI instrument we
refer the reader to \cite{Kor04,Kor08}.

\section{The windowed, multiple-peak, averaged-spectrum method \label{secm2}}

The Windowed, MuLTiple-Peak, averaged-spec\-trum method, which we will refer to
in the following as either Method~2 or the WMLTP method is an advancement of
our Method~1 that is briefly described in Appendix~\ref{secm1}. Several steps
were involved in coming to the design of the new method. As compared to
Method~1 the improvements incorporated in Method~2 include (1) the replacement
of a single symmetric profile with a sum of asymmetric profiles representing
the target peak as well as the peaks of the neighboring $l$-leaks, (2) a sum of
asymmetric profiles representing the peaks of the $n$-leaks neighboring the
target peak, (3) the temporal side-lobe peaks, and (4) an approximation to the
$m$-averaged leakage matrix. 
Generally speaking, an $l$-leak is a spectral peak of the same radial order as that of
the target mode but whose degree is different from that of the target mode,
while a $n$-leak is a mode of arbitrary degree whose radial order is different
from that of the target mode.
We note that all of the results that we will
present in this paper were obtained with the latest version of the Method~2
code, which we refer to as the rev6 version of this code.

By the time that we began to develop our WMLTP fitting method, the observations
of \cite{Duv93} had indicated that the peaks in the solar oscillation power
spectra showed varying amounts of asymmetry. In particular, Duvall and his
collaborators noted that the velocity and intensity power spectra revealed an
opposite sense of asymmetry. After these observations appeared, we
\citep{Rho97} modified our Method~1 fitting method in an ad hoc manner by
employing an asymmetric profile that consisted of two Lorentizan half-profiles
that had differing widths. \cite{Rho97} used that ad hoc profile to demonstrate
that the observed asymmetries in the peaks in MDI velocity power spectra
shifted the fitted frequencies by a substantial amount in the frequency range
where structural inversions are most sensitive to the observed frequencies. The
\cite{Duv93} observations were later confirmed by \cite{Nigetal98}, and then
\cite{Nig98} introduced a theoretical profile that accounted for the observed
asymmetry in a self-consistent manner. While, in principle, we could have
replaced the split-Lorentzian profile in Method~1 with the \cite{Nig98} profile
and continued to use that modified method, the problems that we described above
in Section~\ref{pahds} caused us to abandon that method as well as our use of
the ad hoc split Lorentzian profile. Instead, we continued to develop our WMLTP
method using the \cite{Nig98} profile. As we will note below, the \cite{Nig98}
profile turns into a symmetric Lorentzian profile when its asymmetry parameter
is set equal to zero; hence, its adoption allows us to fit power spectra using
both asymmetric and symmetric profiles.

\subsection{Implementation of \textit{\textbf{m}}-averaged leakage
matrices\label{imavglkm}}

As we have already mentioned in Section~\ref{instreff}, we are calculating the
leakage matrix (cf. Section~\ref{seclkm}) by using the very same numerical
decomposition pipeline that is employed to process the observations. This
approach provides initially the leakage matrix elements $C_{n;l,m;l',m'}$ from
which we then compute the $m$-averaged leakage matrix
$C_{n;l,l'}^{\,\mbox{\scriptsize (mavg)}}$ by means of equations~(\ref{lkmps})
and (\ref{lkmmavg}).

\begin{figure}
\epsscale{0.45}
\plotone{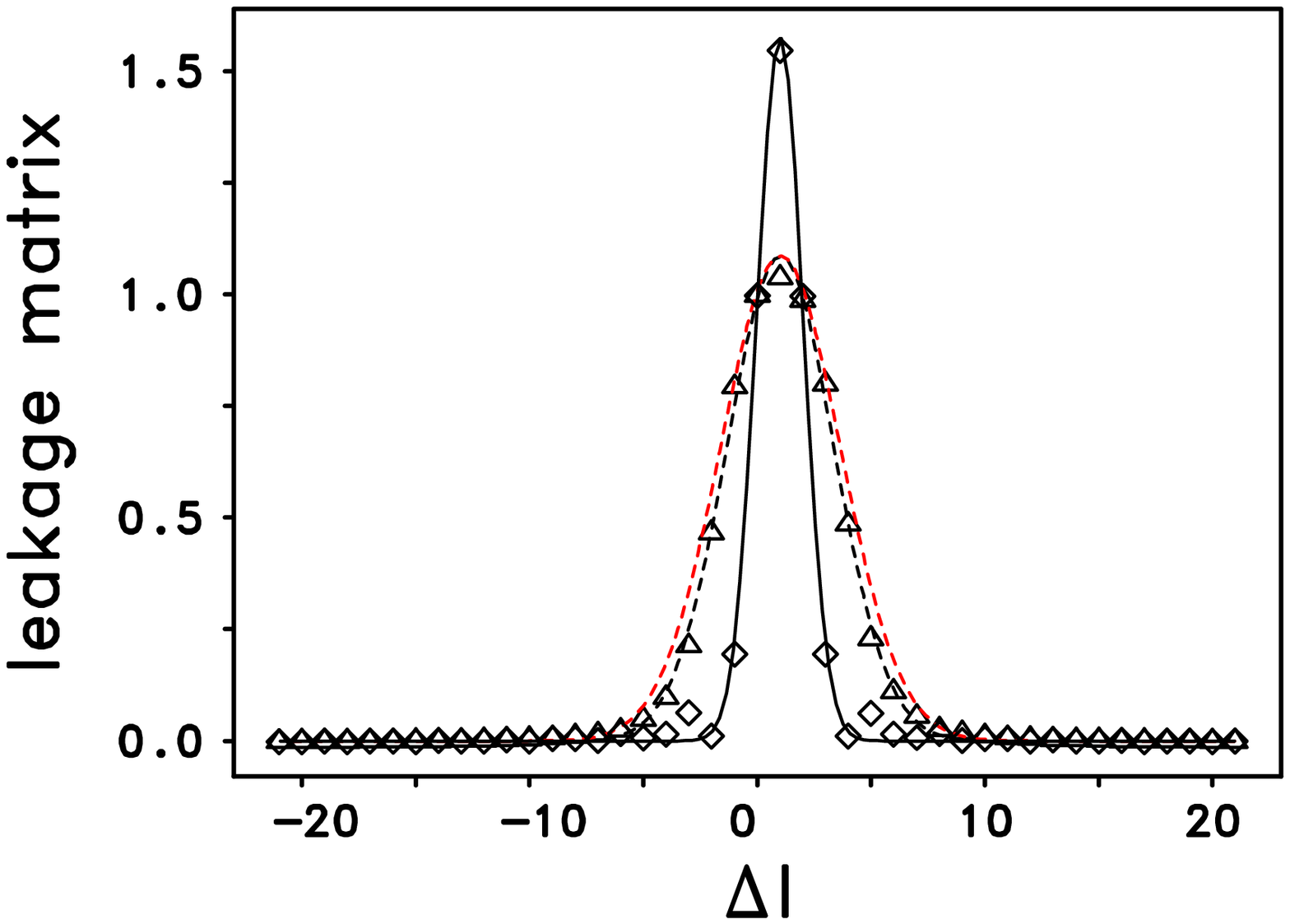}
\plotone{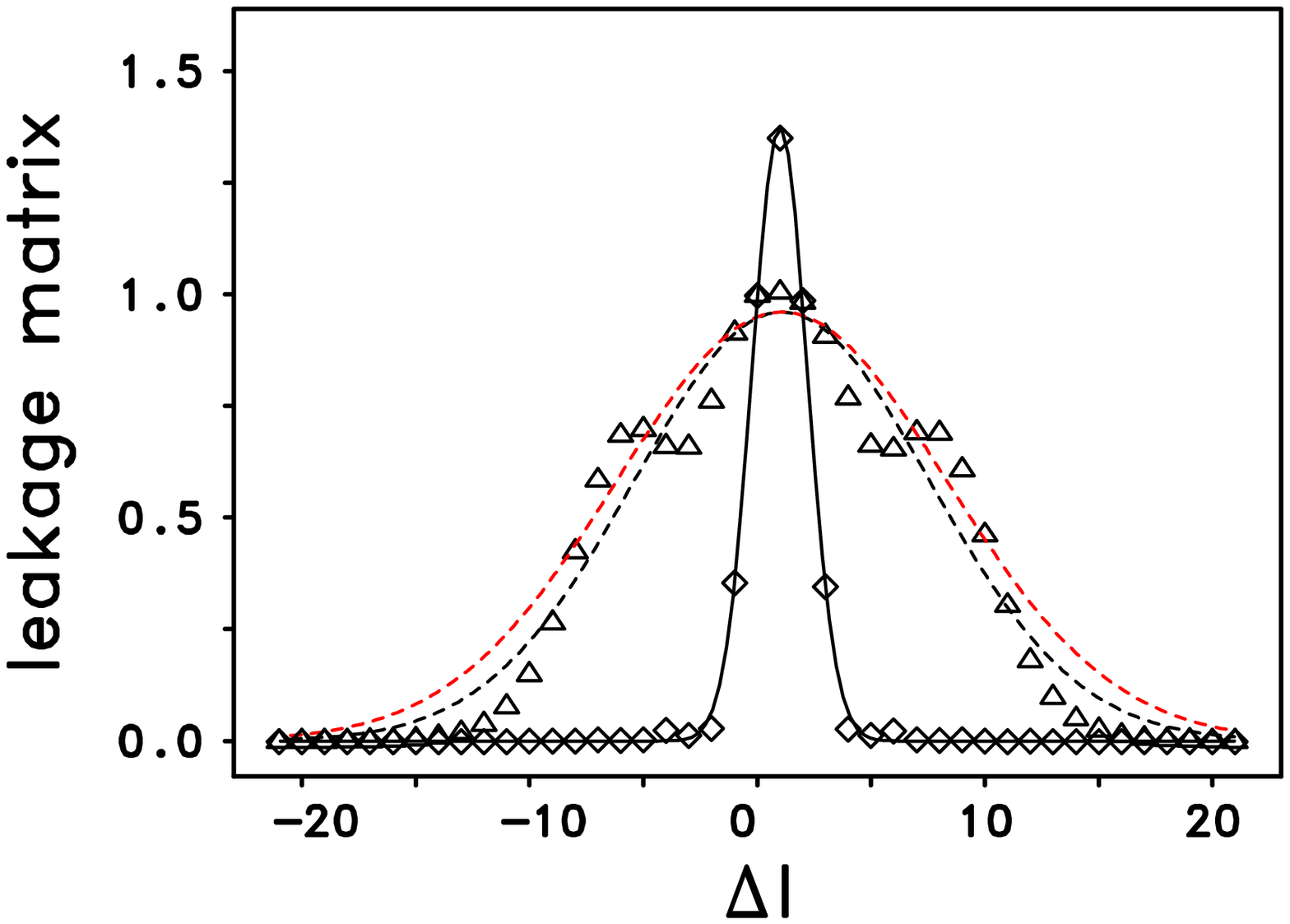}
\caption{
Dependence of $m$-averaged leakage matrix for a displacement component ratio of
$c_t=1.0$ on the difference, $\Delta l=l'-l$, between the degree $l'$ of the
spatial side-lobes and the degree $l$ of the target mode. The left panel is
for $(n,l)=(0,400)$, while the right panel is for $(2,1000)$. The diamonds
represent the original leakage matrix, while the triangles are for the leakage
matrix which is corrected for the effect of the solar differential rotation.
The full black lines are Gaussian fits to the uncorrected leakage matrices, and
the dashed black lines are Gaussian fits to the corrected leakage matrices.
The width of the corrected matrices increases considerably with increasing
degree $l$. In each panel the dashed red line represents a Gaussian the
full-width-at-half maximum (FWHM) of which is larger by a factor of 1.18 than
the FWHM of the Gaussian shown by the black dashed line. These such distorted
Gaussians are used in Section~\ref{slkm} to study the effect of changes in the
FWHM of the leakage matrix upon the fitted mode parameters.
\label{glkm}}
\end{figure}

As is demonstrated here in Figure~\ref{glkm} the $m$-averaged leakage matrix,
$C_{n;l,l'}^{\,\mbox{\scriptsize (mavg)}}$, can be approximated, for given
radial order $n$ and degree $l$, by a Gaussian profile, viz.
\begin{equation}
C_{n;l,l'}^{\,\mbox{\scriptsize (mavg)}}\approx
\Lambda_{n,l} (\Delta l,\alpha^{(n,l)},x_m^{(n,l)}) = 
    \exp\,\left[-\alpha^{(n,l)} \,(\Delta l - x_m^{(n,l)})^2\right], \quad
    \alpha^{(n,l)} > 0,
\label{lkmg}
\end{equation}
where $\alpha^{(n,l)}$ is a parameter related to the width of the leakage
matrix, $x_m^{(n,l)}$ is a parameter that accounts for the offset of the
leakage matrix due to the horizontal component of the modal velocity
eigenfunction of the Sun (cf. Section~\ref{rhc}), and $\Delta l = l'-l$ is the
distance of the spatial leak located at degree $l'$ from the target mode. In
both panels of Figure~\ref{glkm} the $m$-averaged leakage matrix is marked by
diamonds, while the fitted Gaussian profile, as given in equation~(\ref{lkmg}),
is represented by the full line. Overall, in both panels the march of the
leakage matrix is well described by the Gaussian profile. The largest
deviations are just in the ten percent range, and moreover do occur at
locations at which the amplitude of the leakage matrix is small. Hence, we
believe that those deviations can safely be neglected. In Figure~\ref{glkm} we
also show in both panels the leakage matrix that results if the distortion
introduced by the solar latitudinal differential rotation is taken into account
(cf. Section~\ref{dsdr}). The corrected leakage matrix is marked in both panels
by the triangles, and the corresponding fitted Gaussian profile, as given by
equation~(\ref{lkmg}), is represented by the dashed line. We note that for the
$(n,l)=(2,1000)$ mode the width of the corrected leakage matrix greatly exceeds
the width of the uncorrected leakage matrix, while for the $(0,400)$ mode the
corrected leakage matrix is only slightly wider than the uncorrected leakage
matrix. While the Gaussian profile represents the march of the corrected
leakage matrix quite good for the $(0,400)$ mode, in the corrected leakage
matrix for the $(2,1000)$ mode strange asymmetric ``shoulders'' do appear at
about $\Delta l=-5$ and $\Delta l=+4$, respectively, that are not well
represented by a simple Gaussian profile. We hope to re-visit the issue of the
suitability of the Gaussian profile in a later version of the WMLTP method.

%Fig. 3

\begin{figure}
\epsscale{0.45}
\plotone{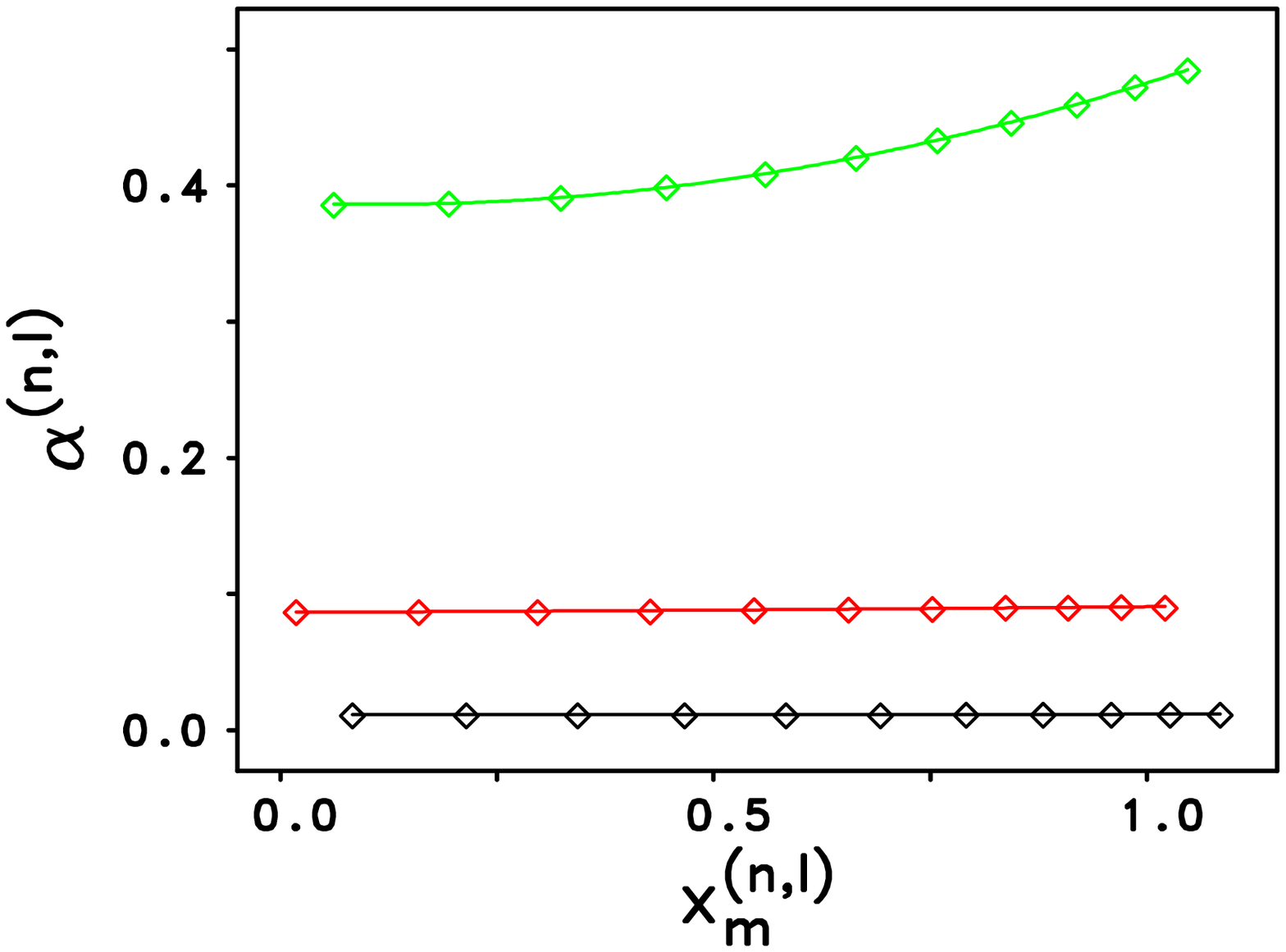}
\plotone{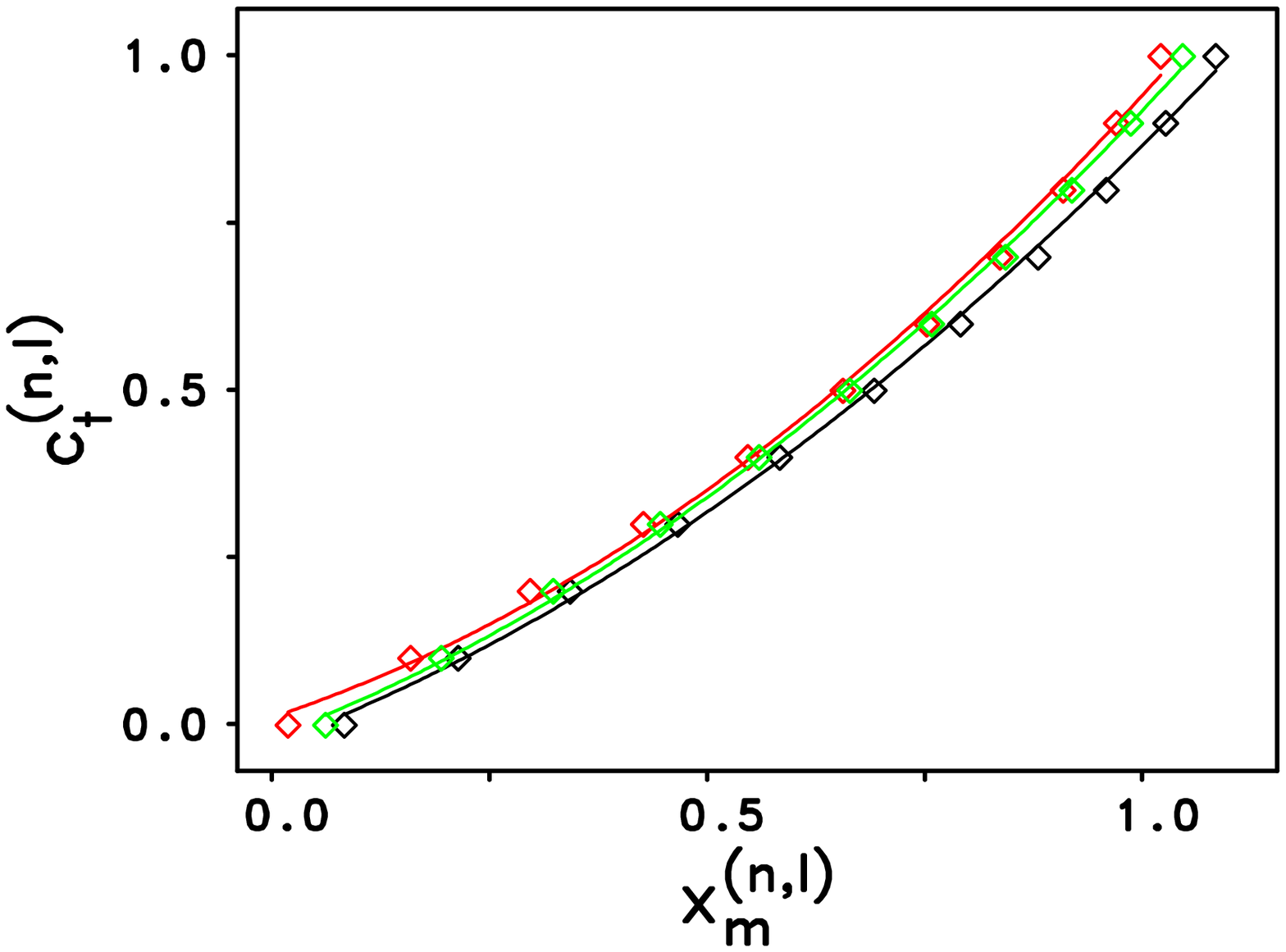}
\caption{
(left) Dependence of the parameter $\alpha^{(n,l)}$, which according to
equation~(\ref{lkmg}) is related to the width of the leakage matrix, on the
offset, $x_m^{(n,l)}$, of the leakage matrix due to the horizontal component of
the modal velocity eigenfunction for $(n,l)=(11,10)$ (green), $(0,400)$ (red),
and $(2,1000)$ (black). The diamonds are for fits to a numerical
representation of these leakage matrices that were corrected for the solar
latitudinal differential rotation. The solid lines are for the parameterization
as given in equation~(\ref{quad1}). We particularly note that for high degree
values the parameter $\alpha^{(n,l)}$ is nearly independent of the value of
$x_m^{(n,l)}$. (right) Same as left panel, but for the solar velocity
eigenfunction component ratio, $c_t^{(n,l)}$. In this panel the solid lines
are for the parameterization given in equation~(\ref{quad2}).
\label{F7}}
\end{figure}

The Gaussian approximation, as defined in equation~(\ref{lkmg}), to the
$m$-averaged leakage matrix is implemented into the WMLTP method by means of
the following approach. For given degree $l$ and radial order $n$, we first
compute for a set of discrete values of the displacement component ratio (cf.
Section~\ref{rhc}), $c_t^{(n,l)}$,
\begin{equation}
c_{t,j}^{(n,l)} = j/10,\quad j=0,\ldots,10,
\label{egrd}
\end{equation}
the $m$-averaged leakage matrix $C_{n;l,l'}^{\,\mbox{\scriptsize (mavg)}}$ that
has been corrected for the latitudinal differential rotation as described in
Section~\ref{dsdr}. Next, we fit the Gaussian profile, as given in
equation~(\ref{lkmg}), to the resulting leakage matrices to get, for each of
the values $c_{t,j}^{(n,l)}$, $j=0,\ldots,10$, the values $\alpha_{j}^{(n,l)}$
and $x_{m,j}^{(n,l)}$ of the fit parameters $\alpha^{(n,l)}$ and $x_m^{(n,l)}$,
respectively, that determine the Gaussian profile, as defined in
equation~(\ref{lkmg}). Because the three parameters $\alpha^{(n,l)}$,
$x_m^{(n,l)}$, and $c_t^{(n,l)}$ are not independent from one another, we
follow \cite{Rho01} and expand both $\alpha^{(n,l)}$ and $c_t^{(n,l)}$ in terms
of $x_m^{(n,l)}$, viz.
\begin{eqnarray}
\alpha^{(n,l)} &=& \bar\alpha_0^{(n,l)} + x_m^{(n,l)} \left(\bar\alpha_1^{(n,l)} + 
                   \bar\alpha_2^{(n,l)} x_m^{(n,l)}\right),\label{quad1}\\
c_t^{(n,l)}    &=& \bar\beta_0^{(n,l)} + x_m^{(n,l)} \left(\bar\beta_1^{(n,l)} + 
                   \bar\beta_2^{(n,l)} x_m^{(n,l)}\right),\label{quad2}
\end{eqnarray}
where $\bar\alpha_0^{(n,l)}$, $\bar\alpha_1^{(n,l)}$, $\bar\alpha_2^{(n,l)}$,
and $\bar\beta_0^{(n,l)}$, $\bar\beta_1^{(n,l)}$, $\bar\beta_2^{(n,l)}$ denote
the respective expansion coefficients. We note that it would be mathematically
equivalent to expand both $\alpha^{(n,l)}$ and $x_m^{(n,l)}$ in terms of
$c_t^{(n,l)}$. In the final step of our approach we fit the
expansion~(\ref{quad1}) to the knots
$\left(\alpha_{j}^{(n,l)},x_{m,j}^{(n,l)}\right)$, $j=0,\ldots,10$, and the
expansion~(\ref{quad2}) to the knots
$\left(c_{t,j}^{(n,l)},x_{m,j}^{(n,l)}\right)$, $j=0,\ldots,10$, to get the set
of expansion coefficients $\bar\alpha_0^{(n,l)}$, $\bar\alpha_1^{(n,l)}$,
$\bar\alpha_2^{(n,l)}$, and $\bar\beta_0^{(n,l)}$, $\bar\beta_1^{(n,l)}$,
$\bar\beta_2^{(n,l)}$. Typical fits of both $\alpha^{(n,l)}$ and $c_t^{(n,l)}$
versus $x_m^{(n,l)}$ are shown here in Figure~\ref{F7} for the modes
$(n,l)=(11,10)$, $(0,400)$, and $(2,1000)$, respectively. The fits clearly
demonstrate that the expansions given in equations (\ref{quad1}) and
(\ref{quad2}), respectively, are reasonable approximations to the variation of
both $\alpha^{(n,l)}$ and $c_t^{(n,l)}$ with respect to $x_m^{(n,l)}$.
Moreover, from the left panel in Figure~\ref{F7} it becomes evident that
$\alpha^{(n,l)}$ decreases with increasing degree $l$. Hence, according to
equation~(\ref{lkmg}) the width of the $m$-averaged leakage matrix increases
with increasing degree $l$.

Once the set of expansion coefficients $\bar\alpha_0^{(n,l)}$,
$\bar\alpha_1^{(n,l)}$, $\bar\alpha_2^{(n,l)}$, and $\bar\beta_0^{(n,l)}$,
$\bar\beta_1^{(n,l)}$, $\bar\beta_2^{(n,l)}$ has been computed for the mode
$(n,l)$, the corresponding $m$-averaged leakage matrix $\Lambda_{n,l}$, as
given in equation~(\ref{lkmg}), can easily be determined by performing the
following steps. First, we calculate the value of $c_t^{(n,l)}$ from
equation~(\ref{ctth}). Second, using this value of $c_t^{(n,l)}$ we solve
equation~(\ref{quad2}) for $x_m^{(n,l)}$. Finally, we insert this value of
$x_m^{(n,l)}$ into equation~(\ref{quad1}) to get the value of $\alpha^{(n,l)}$.

%Fig. 4

\subsection{Theoretical fitting profile\label{thfp}}

In the WMLTP method the following fitting profile is used to represent an
oscillation peak of given degree $l$ and radial order $n$:
\begin{equation}
M_{n,l}(\nu,\mbox{\boldmath $p$}) = \mathcal{M}_{n,l}(\nu,\mbox{\boldmath $p$})
    \otimes \mathcal{W}(\nu).
\label{profm2a}
\end{equation}
Here, $\nu$ is frequency, $\mathcal{W}(\nu)$ is the Fourier spectrum of the
temporal window function of the observational time series, ``$\otimes$''
denotes the convolution operator, and $\mathcal{M}_{n,l}(\nu,\mbox{\boldmath
$p$})$ is defined by 
\begin{eqnarray}
\mathcal{M}_{n,l}(\nu,\mbox{\boldmath $p$}) &=
   \underbrace{
   \sum_{\Delta l=-M}^{M}
   \Psi(A_{n,l},B_{n,l},x_{\Delta l}) \,\Theta(\Delta l)
   \,\Lambda_{n,l} (\Delta l,\alpha^{(n,l)},x_m^{(n,l)})
   }_{\mbox{targeted peak and $l$-leaks}}
   +
   \underbrace{
   \sum_{i=1}^{N} \Psi(A_{i},B_{i},x_{i}),
   }_{\mbox{$n$-leaks}}
   \nonumber\\
   &+
   \underbrace{
   a+b\,\nu +c\,\nu^2
   }_{\mbox{background}}
   , \label{profm2b}
\end{eqnarray}
where
\begin{equation}
\Psi(A,B,x) =
   A \,\frac{(1+B x)^2+B^2}{1+x^2},
   \label{profm2d}
\end{equation}
\begin{equation}
\Theta(\Delta l) =
   1 + \sum_{\mu=1}^{\mu_{\rm max}} \frac{1}{\mu!}\frac{1}{A}
   \frac{\partial^{\mu}A}{\partial l^{\mu}}\left(\Delta l\right)^{\mu},
\label{profm2c}
\end{equation}
\begin{equation}
x_{\Delta l} =
   \frac{2\left[\nu-\left(\nu_{n,l}
   +\displaystyle\sum_{\mu=1}^{\mu_{\rm max}} \displaystyle\frac{1}{\mu!}
   \frac{\partial^{\mu}\nu}{\partial l^{\mu}}
   \left(\Delta l\right)^{\mu}\right)\right]}
        {w_{n,l}\left(
   1+\displaystyle\sum_{\mu=1}^{\mu_{\rm max}} \displaystyle\frac{1}{\mu!}\frac{1}{w}
   \frac{\partial^{\mu}w}{\partial l^{\mu}}\left(\Delta l\right)^{\mu}
   \right)},
\label{profm2e}
\end{equation}
\begin{equation}
x_{i} = \frac{2\,(\nu - \nu_{i})}{w_{i}}.
\label{profm2f}
\end{equation}
\label{profm2}

\noindent
Here, $\Psi$ is the asymmetric profile of \cite{Nig98}, $\Theta$ is an
empirical adjustment to the amplitudes of the $l$-leaks, $\Lambda_{n,l}$ is an
approximation to the $m$-averaged leakage matrix that is described in
Section~\ref{imavglkm}, $2M$ is the number of the $l$-leaks, $N$ is the number
of the $n$-leaks, $A_i$, $B_i$, $\nu_{i}$, and $w_i$ are, respectively, the
amplitude, the line asymmetry parameter, the frequency, and width of the $i$-th
$n$-leak, $i=1,\ldots,N$, and $a$, $b$, $c$ are parameters describing the
background noise. We note that in equation~(\ref{profm2b}) it has been
presumed that each of the $2M+1$ modal peaks can be characterized by the same
line asymmetry parameter $B_{n,l}$. 

The total number of $l$-leaks, $2M$, included in the model profile, as given in
equation~(\ref{profm2b}), depends on the width of the leakage matrix,
$\Lambda_{n,l}$, and is typically in the range from 10 to 60. As we have noted
in Section~\ref{imavglkm}, particularly for higher degrees, $l$, the leakage
matrices are rather wide which is mainly due to the distortion of the
oscillation eigenfunctions caused by the latitudinal differential rotation.
Therefore, in equations~(\ref{profm2c}) and (\ref{profm2e}) Taylor series
expansions of fairly high order have to be employed to adequately describe the
variation of amplitude, frequency, and linewidth with degree $l$. In the
current version of the WMLTP code $\mu_{\rm max}=8$ is used. For the
determination of the total number of $n$-leaks, $N$, to be included in the
model profile given by equation~(\ref{profm2b}), we refer the reader to
Section~\ref{dnnlkfb}. In our applications we found that $N$ can vary from 0 up
to about 23, depending on both degree $l$ and radial order $n$.

In the fitting profile defined in equations (\ref{profm2a}) through
(\ref{profm2f}) a total of $4N+8$ fitting parameters are involved, viz.~the
mode amplitude $A_{n,l}$, the mode frequency $\nu_{n,l}$, the mode linewidth
$w_{n,l}$, the background noise parameters $a$, $b$, $c$, the line asymmetry
parameter $B_{n,l}$, the offset, $x_m^{(n,l)}$, of the $m$-averaged leakage
matrix due to the horizontal component of the modal velocity eigenfunction of
the Sun (cf. Sect.~\ref{imavglkm}), and the parameters $A_i$, $\nu_i$, $w_i$,
$B_i$, $i=1,\ldots, N$, representing, respectively, amplitude, frequency,
linewidth, and line asymmetry of the $N$ $n$-leaks. These fitting parameters
can be lumped together in the fitting vector 
\begin{equation}
\mbox{\boldmath $p$} =
    \left(A_{n,l},\nu_{n,l}, w_{n,l},B_{n,l}, a,b,c,x_m^{(n,l)},
    A_1,\nu_1,w_1,B_1,\ldots, A_N,\nu_N,w_N,B_N\right)^{\mbox{\footnotesize T}}.
\label{pm2b}
\end{equation}
We did not include in this fitting vector $\mbox{\boldmath $p$}$ the parameter
$\alpha^{(n,l)}$ because it depends on $x_m^{(n,l)}$ by virtue of
equation~(\ref{quad1}). We also did not include the Taylor series expansion
coefficients
\begin{equation}
\frac{1}{A}\frac{\partial^{\mu}A}{\partial l^{\mu}},\,\,\,
\frac{\partial^{\mu}\nu}{\partial l^{\mu}},\,\,\,
\frac{1}{w}\frac{\partial^{\mu}w}{\partial l^{\mu}}, \quad
\mu=1,\ldots,\mu_{\rm max},
\label{taylcoef}
\end{equation}
for the amplitude, frequency, and linewidth, respectively. Rather, these
coefficients are taken, for given degree $l$ and radial order $n$, from tables
containing initial estimates or so-called seeds that are derived from
previously computed modal parameters, and are improved upon in a fixed-point
iteration similar to that described in \cite{Rho01}. As a result, the
compilation of such seed tables is an important pre-processing step for the use
of the WMLTP method. 

In order to investigate the performance of our implemented fixed-point
iteration we have done the following analysis in which we have restricted
ourselves to the investigation of the frequency, $\nu$, and the derivative
thereof with respect to degree, $\Delta\nu/\Delta l$. For convenience, we did
not consider either the linewidth or the amplitude and the derivatives thereof
with respect to degree. In the first step of this analysis we used our seed
table for the epoch of 2010 which provided both the seed frequencies,
$\nu_{n,l}^{\rm (seed,1)}$, and the derivative thereof with respect to degree,
$\left(\Delta\nu/\Delta l\right)_{n,l}^{\rm (seed,1)}$, to fit the $m$-averaged
spectral set $\cal{S}$2001\_90. From this fitting run we obtained the
frequencies, $\nu_{n,l}^{\rm (fit,1)}$. We chose the $m$-averaged spectral set
$\cal{S}$2001\_90 for this test because the $\cal{R}$2001\_90 observing run
corresponded to the maximum phase of Solar Cycle~23. Therefore, due to the
well-known shifts in the $f$- and $p$-mode frequencies that accompany changes
in solar activity, the fitted frequencies, and the frequency derivatives, that
would result from the fitting of these power spectra, would be expected to
exhibit the largest possible differences from the frequencies and derivatives
in our 2010 seed table since the latter were generated from the
$\cal{R}$2010\_66 observing run when the mean level of solar activity was much
lower.

\begin{figure}
\epsscale{0.60}
\plotone{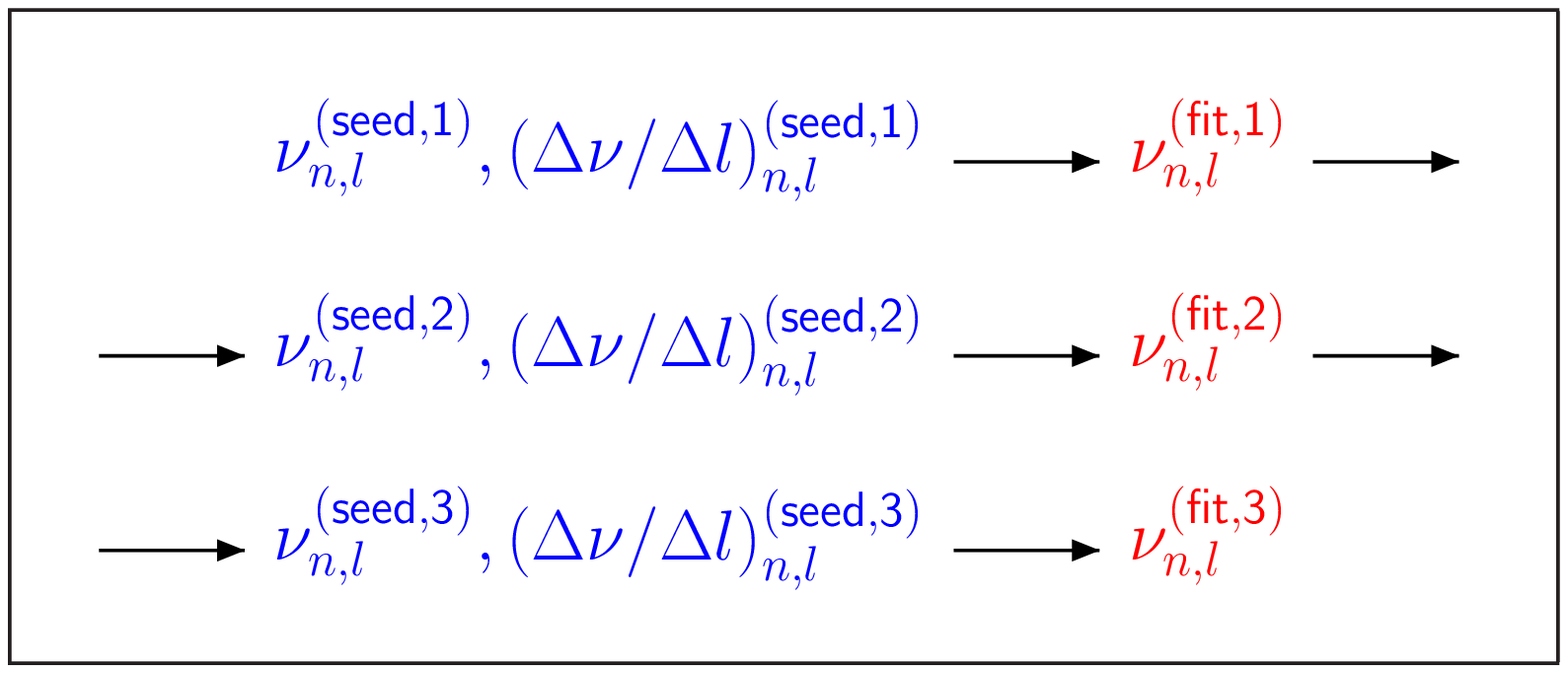}
\caption{
Flow chart of the individual steps performed in the fixed-point iteration for
the construction of accurate seed tables using the example of the frequency,
$\nu$. Similar flow charts apply to amplitude and linewidth, respectively.
\label{F32}}
\end{figure}

Next, we fitted to the frequencies, $\nu_{n,l}^{\rm (fit,1)}$, on a
ridge-by-ridge basis, a smooth function of degree, $P^{(n)}(l)$, to get new
seeds $\nu_{n,l}^{\rm (seed,2)}$ and $\left(\Delta\nu/\Delta
l\right)_{n,l}^{\rm (seed,2)}$, respectively, by setting 
\begin{align} 
\nu_{n,l}^{\rm (seed,2)} &= P^{(n)}(l), \\[1.0mm]
\left(\frac{\Delta\nu}{\Delta l}\right)_{n,l}^{\rm (seed,2)} &= 
  \left.\frac{\partial P^{(n)}(l)}{\partial l}\right|_l,  
\end{align} 
where the superscript ``$(n)$'' denotes the ridge of radial order $n$. For the
details of the construction of the function, $P^{(n)}(l)$, we refer the reader
to the end of Section~\ref{srm2}. In the second step we used this new set of
seeds to fit the $m$-averaged spectral set $\cal{S}$2001\_90 once again to get
the frequencies $\nu_{n,l}^{\rm (fit,2)}$, from which we generated yet another
set of seeds $\nu_{n,l}^{\rm (seed,3)}$ and $\left(\Delta\nu/\Delta
l\right)_{n,l}^{\rm (seed,3)}$, respectively, in the same manner as we have
done in the previous step. These new seeds were then used to fit, in the third
and final step, the $m$-averaged spectral set $\cal{S}$2001\_90 a third time to
get the fitted frequencies, $\nu_{n,l}^{\rm (fit,3)}$. For further explanation
we have illustrated the sequence of individual steps performed in our analysis
in Figure~\ref{F32}.

%Fig. 5

The results of our analysis are summarized in Table~\ref{tab3}. As can be seen
from columns~4 and 5, the two-fold update of the seed table resulted in a
significant reduction in both the average and the standard deviation of the
differences of the seed values of $\Delta\nu/\Delta l$ as well as of the scaled
differences of the fitted frequencies. As expected, the rather large changes
in the scaled differences of the fitted frequencies that accompanied the first
update of our 2010 seed table were reflective of the very large difference in
the mean levels of activity between 2001 and 2010. The fact that we could
obtain a self-consistent seed table for 2001 with only two iterations in spite
of such a large difference in the level of activity in the two epochs justifies
our approach of replacing the Taylor series expansions, as given in
equation~(\ref{taylcoef}), with such self-consistent seed values.

\subsection{Selection of fitting box widths\label{sfbw}}

The width of the selected fitting box is crucial for the successful
determination of the fitting vector $\mbox{\boldmath $p$}$ defined in
equation~(\ref{pm2b}). We have found it useful to construct the fitting box for
the mode $(n,l)$ as follows:
\begin{equation}
\nu_{n,l}^{\rm low} = \nu_{n,l}^{\rm seed} - \tau^{(n,l)}\,\frac{\Delta\nu_{n,l}}{\Delta n} \bigg |_{\rm left},
\quad
\nu_{n,l}^{\rm up} = \nu_{n,l}^{\rm seed} + \tau^{(n,l)}\,\frac{\Delta\nu_{n,l}}{\Delta n} \bigg |_{\rm right}.
\label{A1}
\end{equation}
Here, $\nu_{n,l}^{\rm seed}$ is the seed frequency of the targeted mode
$(n,l)$, $\nu_{n,l}^{\rm low}$ and $\nu_{n,l}^{\rm up}$ are the lower and upper
boundary, respectively, of the fitting box, $\Delta\nu_{n,l}/\Delta n |_{\rm
left}$ and $\Delta\nu_{n,l}/\Delta n |_{\rm right}$ denote the variation of the
mode frequency with respect to the radial order $n$ at the left (i.e., lower
frequency) and right (i.e., higher frequency) side, respectively, and
$\tau^{(n,l)}$ is given by
\begin{equation}
\tau^{(n,l)} = \left\{ \begin{array}{l@{\quad}l}
\tau_1^{(n)}               & \mbox{for~~$l\leq l_1^{(n)}$,} \\[1.0mm]
\displaystyle\frac{\tau_2^{(n)} - \tau_1^{(n)}}{l_2^{(n)} - l_1^{(n)}}\,\left(l - l_1^{(n)}\right) + 
              \tau_1^{(n)} & \mbox{for~~$l_1^{(n)} < l < l_2^{(n)},$} \\[4.7mm]
\tau_2^{(n)}               & \mbox{for~~$l\geq l_2^{(n)}$,}
               \end{array} \right.
\label{A2}
\end{equation}
where $\tau_1^{(n)}$, $\tau_2^{(n)}$, $l_1^{(n)}$, and $l_2^{(n)}$ are
predetermined parameters depending on the radial order $n$. The quantities
$\Delta\nu_{n,l}/\Delta n |_{\rm left}$ and $\Delta\nu_{n,l}/\Delta n |_{\rm
right}$ in equation (\ref{A1}) are approximated as follows:
\begin{equation}
\begin{array}{l@{\quad}l}
\displaystyle\frac{\Delta\nu_{n,l}}{\Delta n}\bigg |_{\rm left} \approx \nu_{n,l}^{\rm seed}-\nu_{n-1,l}^{\rm seed},\quad
\frac{\Delta\nu_{n,l}}{\Delta n}\bigg |_{\rm right} \approx \nu_{n+1,l}^{\rm seed}-\nu_{n,l}^{\rm seed} & \mbox{for~~$n > 0$,} \\[5.0mm]
\displaystyle\frac{\Delta\nu_{n,l}}{\Delta n}\bigg |_{\rm left} = \frac{\Delta\nu_{n,l}}{\Delta n}\bigg |_{\rm right} 
\approx \nu_{n+1,l}^{\rm seed}-\nu_{n,l}^{\rm seed} & \mbox{for~~$n = 0$,}
\end{array}
\label{A2a}
\end{equation}
where $\nu_{n,l}^{\rm seed}$, $\nu_{n-1,l}^{\rm seed}$, and $\nu_{n+1,l}^{\rm
seed}$ are taken from a seed table. As to the choice of the parameters
$\tau_1^{(n)}$, $\tau_2^{(n)}$, $l_1^{(n)}$, and $l_2^{(n)}$ in equation
(\ref{A2}) it must be kept in mind that any selected fitting box must fulfill
at least two requirements. First, the fitting box must be sufficiently wide so
that both the mode profile and the background power are well sampled. This
requirement becomes an issue if spectra are to be fitted that are derived from
an observing run the duration of which is only a few days. In this case it can
happen that the number of fit parameters included in the fitting vector
$\mbox{\boldmath $p$}$ exceeds the number of frequency bins constituting the
fitting box. Second, the fitting box must not be unduly wide in order to save
computing time which non-linearly increases with the number of frequency bins
comprising the fitting box. In practice, we determine the parameters
$\tau_1^{(n)}$, $\tau_2^{(n)}$, $l_1^{(n)}$, and $l_2^{(n)}$ for a given ridge
of radial order, $n$, by a trial-and-error method and select those values of
them which give rise to the least scatter of frequency along the ridge.

As can be seen from equations~(\ref{A1}) and (\ref{A2}), aside from variations
of both $\Delta\nu_{n,l}/\Delta n |_{\rm left}$ and $\Delta\nu_{n,l}/\Delta n
|_{\rm right}$ with degree $l$, the width of the fitting box is constant for
$l\leq l_1^{(n)}$ and $l\geq l_2^{(n)}$, and varies linearly with degree $l$
for $l_1^{(n)} < l < l_2^{(n)}$. We note here that in general
$\Delta\nu_{n,l}/\Delta n |_{\rm left} \neq \Delta\nu_{n,l}/\Delta n |_{\rm
right}$. Therefore, according to equation~(\ref{A1}) the fitting box is
generally not symmetric with respect to the seed frequency $\nu_{n,l}^{\rm
seed}$.

\begin{figure}
\epsscale{0.60}
\plotone{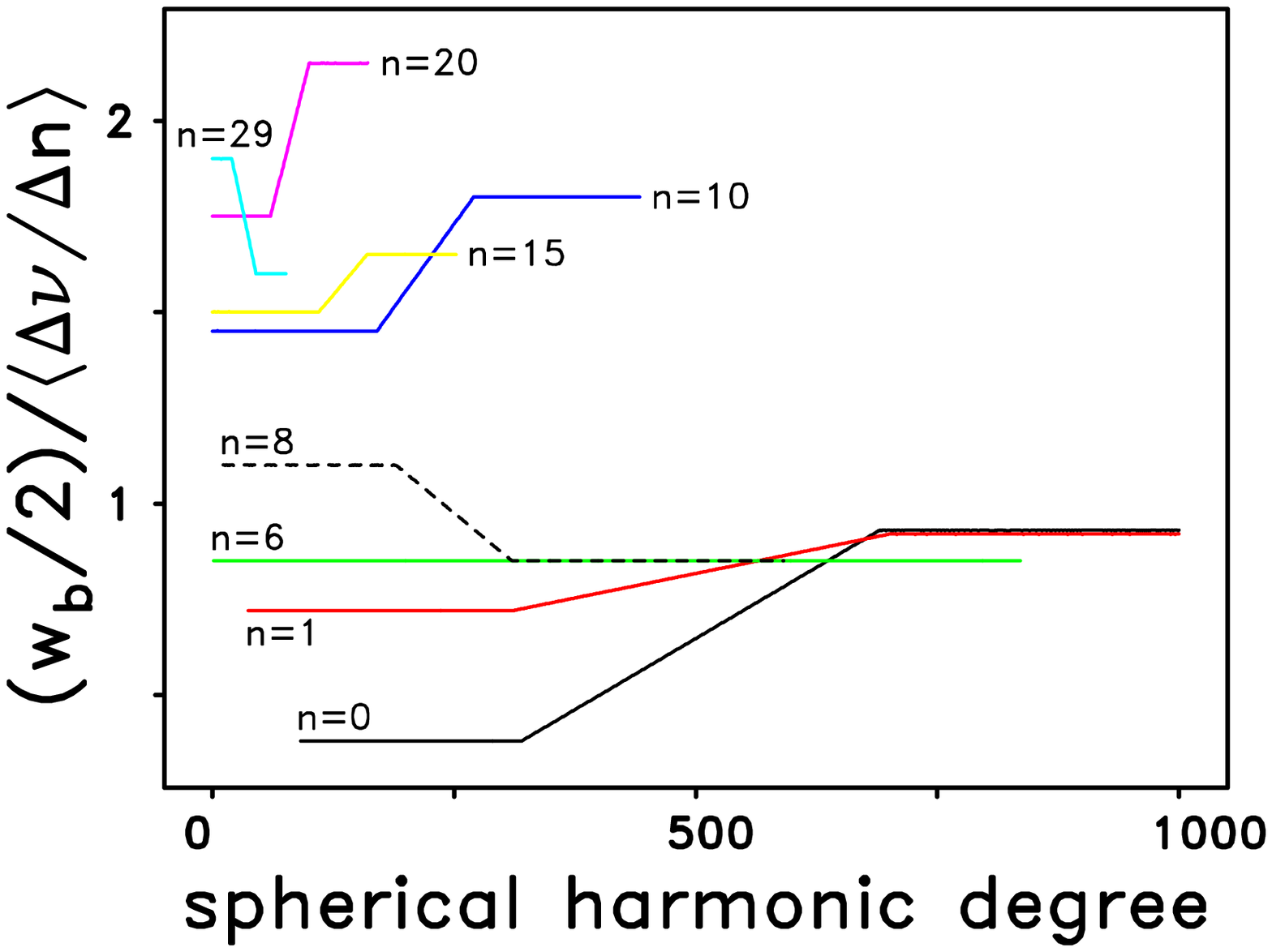}
\caption{
Half of the width, $w_b$, of the fitting box measured in terms of
$\langle\Delta\nu/\Delta n\rangle$ versus degree for a selected set of radial
orders, $n$. Here, $\langle\Delta\nu/\Delta n\rangle$ denotes the mean value of
$\Delta\nu_{n,l}/\Delta n |_{\rm left}$ and
$\Delta\nu_{n,l}/\Delta n |_{\rm right}$ with
$\Delta\nu_{n,l}/\Delta n |_{\rm left}$ and
$\Delta\nu_{n,l}/\Delta n |_{\rm right}$ being
the variation of the target mode frequency with respect to the radial order,
$n$, at the lower and the higher frequency side of the fitting box,
respectively. For most cases shown the fitting box widens with increasing
degree by only about 10 to 30\,\%, while it widens by about 150\,\% for the
$n=0$ ridge. However, for the $n=8$ and $n=29$ ridge the width decreases with
degree by about 23\,\% and 15\,\%, respectively. For the $n=6$ ridge the width
is practically constant. For the $n=20$ ridge the $n$-leaks $(n\pm 1,l)$ and
$(n\pm 2,l)$ are located within the fitting box for higher degrees.
\label{F31}}
\end{figure}

For a selected set of radial orders, $n$, we show in Figure~\ref{F31} the half
of the width, $w_b$, of the respective fitting boxes measured in terms of the
average, $\langle\Delta\nu/\Delta n\rangle$, of $\Delta\nu_{n,l}/\Delta n
|_{\rm left}$ and $\Delta\nu_{n,l}/\Delta n |_{\rm right}$. For most of the
cases shown the width of the fitting box increases with increasing degree,
while for $n=6$ the width is practically constant, and for $n=8$ and $n=29$ the
width decreases with degree. We also note that for the higher-order ridges the
$n$-leaks $(n\pm 1,l)$ are included in the fitting box. For the $n=20$ ridge
the fitting box is getting so wide that even the $n$-leaks $(n\pm 2,l)$ are
encompassed. The choice of such wide fitting boxes is expressive of the fact
that, in general, the scatter of the fitted mode parameters along a given ridge
decreases with increasing width of the fitting boxes. Therefore, it would be
desirable to fit all $n$-values simultaneously that are present in an
$m$-averaged spectrum of given degree. However, such approach is not feasible
on practical grounds.

%Fig. 6

\subsection{Determination of the number of \textit{\textbf{n}}-leaks to be
included in the fitting model profile\label{dnnlkfb}}

The number of $n$-leaks, $N$, to be included in the fitting model profile
defined by equations~(\ref{profm2a}) through (\ref{profm2f}), cannot be
determined from first principles. Therefore, a range $|l - l_0| \leq l_{\rm
range}$, $|n - n_0| \leq n_{\rm range}$, of $l$ and $n$ values is chosen around
the targeted mode $(n_0,l_0)$, and every mode $(n,l)$ within this range that
(1) has a frequency within the frequency range of the fitting box (cf.
Section~\ref{sfbw}), (2) differs from $(n_0,l_0)$, and (3) is not identical to
any of the $l$-leaks, is included as an $n$-leak in equation~(\ref{profm2b}).
In the current version of the WMLTP code we are using $l_{\rm range}=4$,
$n_{\rm range}=3$ for isolated modal peaks, $(n_0,l_0)$, and $l_{\rm range}=4$,
$n_{\rm range}=2$ otherwise. Usually, many of the $n$-leaks determined in this
manner turn out to be statistically insignificant for getting a reasonably good
fit. Because generally the $n$-leaks included in the fitting profile strongly
affect the fitted parameters of the targeted mode $(n_0,l_0)$, it is crucial to
discard the statistically insignificant $n$-leaks while keeping the
statistically significant ones. Therefore, the statistical significance of an
$n$-leak is tested by virtue of the so-called R-test \citep{Fri83}. This test
is skipped, however, for ``true'' $n$-leaks $(n_0\pm 1,l)$ and $(n_0\pm 2,l)$,
respectively, with $(n_0,l_0)$ being the targeted mode. Because the R-test does
not always work satisfactorily, we have additionally implemented into the WMLTP
code heuristic criteria for discarding an $n$-leak. For example, an $n$-leak is
discarded if its amplitude and/or width is outside a given range or if it
overlaps too much with another $n$-leak or with the targeted peak itself.
Overall, the determination of the number of $n$-leaks, $N$, is a rather time
consuming undertaking which significantly increases the compute time of the
WMLTP code.

\subsection{Implementation of numerical scaling, adjustments, and options
            \label{insao}}

The vector $\mbox{\boldmath $p$}$ defined in equation~(\ref{pm2b}) is
determined by fitting, in the least-squares sense, the model profile given in
equations~(\ref{profm2a}) through (\ref{profm2f}) to the $m$-averaged spectrum
of given degree $l$ in a fitting box (cf. Section~\ref{sfbw}) centered about
the target peak of radial order $n$. The confidence interval on each fit
parameter $p_i\in\mbox{\boldmath $p$}$ is calculated as described in
Appendix~\ref{secm1}. In order to make the solution of this least-squares
problem as robust as possible diverse provisions are made. First, all fit
parameters are scaled such as to make their values on the order of unity. This
scaling greatly improves the numerical stability of the solution of the
least-squares problem. Second, the fit parameters representing the target peak
(viz.~$A_{n,l}$, $\nu_{n,l}$, $w_{n,l}$, $B_{n,l}$) as well as the fit
parameters representing the $n$-leaks (viz.~$A_i$, $\nu_i$, $w_i$, $B_i$,
$i=1,\ldots,N$) are subject to bounds, i.e., they are constrained to lie within
prescribed intervals. This is important in order to avoid unphysical values of
any of these fit parameters, for example, negative amplitudes, frequencies too
much off from their seed values, linewidths much smaller than the spectral
resolution, or linewidths on the order of the width of the fitting box. Also,
the magnitude of the line asymmetry parameter of both the target peak and the
$n$-leak peaks is constrained to values less than or equal to 0.3. Third, the
fitting vector $\mbox{\boldmath $p$}$ is determined in several passes. In
Table~{\ref{tab4}} it is shown which parameters are fitted in the individual
passes. Any parameter not listed in a given pass as a fitted parameter in
Table~{\ref{tab4}}, but which is included in the fitting vector
$\mbox{\boldmath $p$}$, is kept fixed either to its seed value or else to its
value obtained in a previous pass. We note that the permutations given in
Table~{\ref{tab4}} were chosen on purely heuristic grounds and are, therefore,
somewhat arbitrary. Fourth, a flag can be set in the WMLTP code which enforces
the use of the symmetric Lorentzian profile. This is accomplished by setting
$B=0$ in equation~(\ref{profm2d}). Otherwise, the full asymmetric profile is
used, in which case the asymmetry parameter, $B$, in equation~(\ref{profm2d})
is employed as a fitting parameter in the least-squares problem.

For the solution of the nonlinear constrained least-squares problem which
results for the determination of the fitting vector $\mbox{\boldmath $p$}$ we
use the FORTRAN code NLPQL which is a special implementation of a so-called
sequential quadratic programming method, and which is generally designed for
solving constrained non-linear optimization problems \citep{Schitt85}.
Sequential quadratic programming is one of the most robust algorithms for the
solution of non-linear continuous optimization problems. The method is based on
solving a series of sub-problems designed to minimize a quadratic approximation
to the Lagrangian function subject to a linearization of the constraints.

In the case of either high frequencies and/or high degrees when the individual
modal peaks can no longer be resolved but rather blend together to form ridges
of power in the $m$-averaged spectra, the determination of the fitting vector
$\mbox{\boldmath $p$}$ given in equation~(\mbox{\ref{pm2b}}) becomes an
ill-defined least-squares problem. This is because in this case any change of
the parameter $x_m^{(n,l)}$ does not result in a significant change of the
fitting profile defined by equations~(\ref{profm2a}) through (\ref{profm2f})
because $\alpha^{(n,l)}$, which is related to the width of the leakage matrix,
turns out to be practically independent of $x_m^{(n,l)}$, as is demonstrated
here in the left panel of Figure~\ref{F7}. As a result, $x_m^{(n,l)}$ can no
longer be used as a fit parameter but rather must be kept fixed to its seed
value.

To make the WMLTP method more flexible in practical applications we have
implemented the following features into the code. First, if frequency splitting
coefficients are available for a mode $(n,l)$, the $m$-averaged spectrum can be
calculated from the respective zonal, sectoral, and tesseral spectra in a
weighted or unweighted fashion within the fitting box of that mode making
superfluous the calculation of the $m$-averaged spectrum in the pre-processing
step as described in Section~\ref{gmavg}. In other words, this option makes it
possible to either use splitting coefficients or else an $m$-averaged spectrum
as input to the WMLTP code. This feature is particularly useful if
non-$n$-averaged splitting coefficients are available. Second, for a mode
$(n,l)$ the expansion coefficients which are required to calculate the
corresponding $m$-averaged leakage matrix parameters $\alpha^{(n,l)}$ and
$x_m^{(n,l)}$ via equations~(\ref{quad1}) and (\ref{quad2}), can either be read
from a file generated in a previous calculation or else can be calculated as
part of the fitting procedure itself. This way a considerable amount of compute
time can be saved. Third, if the spectral peak of a mode $(n,l)$ is well
separated from the $l$- and $n$-leak peaks, the offset parameter of the leakage
matrix, $x_m^{(n,l)}$, can be invoked as a fit parameter. By then converting
the fitted values of $x_m^{(n,l)}$ into values of $c_t^{(n,l)}$ by means of
equation (\ref{quad2}), \cite{Rho01} were able to demonstrate, using a previous
version of the WMLTP code, that the such measured values of $c_t^{(n,l)}$ match
theoretical predictions. Broadly speaking, such approach is possible for modes
with $w_{n,l}\lesssim 0.8\,(\Delta\nu/\Delta l)_{n,l}$, where $w_{n,l}$ is the
mode linewidth and $(\Delta\nu/\Delta l)_{n,l}$ is an approximation to the
derivative of the mode frequency with respect to degree.

\subsection{Enforcement of symmetrical line profile for the fitting of
pseudo-modes\label{sflp}}

\begin{figure}
\epsscale{0.40}
\hspace{12.4mm}
\plotone{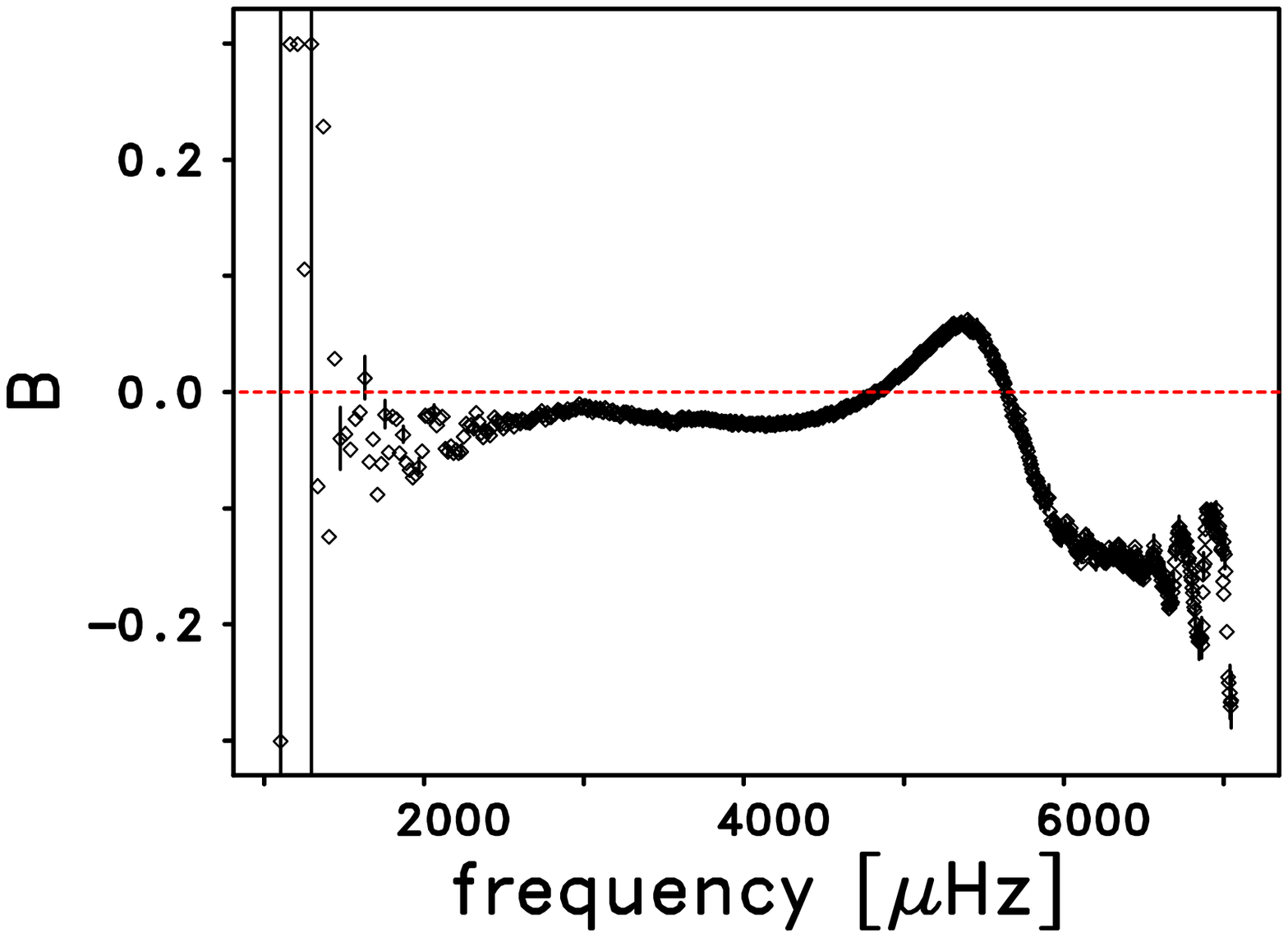}\hspace{14.0mm}
\plotone{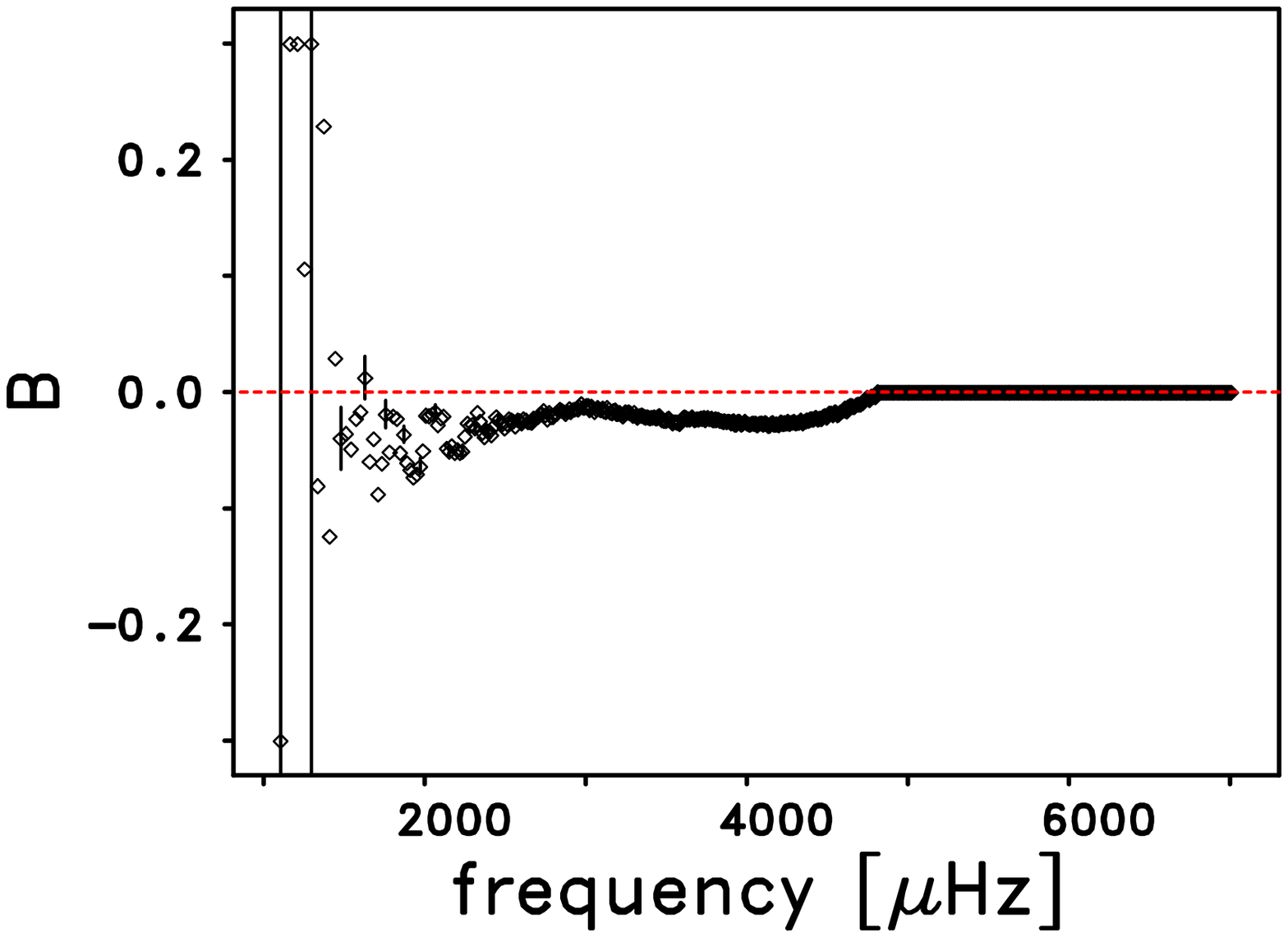}
\vspace{10.0mm}
\caption{
(left) Line asymmetry parameter $B$ versus frequency for the $n=6$ ridge
computed from the $m$-averaged spectral set $\cal{S}$2010\_66a. The error bars
shown for selected values of frequency are for $1\,\sigma$ errors. Because the
excursions of $B$ to both positive and negative values for $\nu >
5000\,\,\mu$Hz are hard to explain on physical grounds, we switched over to a
symmetric Lorentzian profile for the high-frequency portion of each ridge.
(right) Same as left panel, except that the symmetrical profile was used for
frequencies $\nu > 4815\,\,\mu$Hz for this ridge. In both panels the red dashed
line is for $B=0$.
\label{F3}}
\end{figure}

We initially employed the asymmetrical profile for all of the cases along each
ridge. However, for frequencies $\nu\gtrsim 5000\,\,\mu$Hz we found unexpected
excursions in the line asymmetry parameter, $B$, an example of which is shown
here for the $n=6$ ridge in the left panel of Figure~\ref{F3}. Because such
behavior of the line asymmetry parameter, $B$, is hard to explain on physical
grounds, we rather presume that the asymmetrical profile becomes invalid for
modes with frequencies close to or greater than the acoustic cut-off frequency.
This presumption is substantiated by the fact that for those frequencies the
assumptions made by \citet{Nig98} in the derivation of their asymmetrical
profile are not fulfilled. This is because the high-frequency peaks are not
normal modes but rather so-called pseudo-modes caused by interference between
the waves coming directly from the excitation source and waves refracted in the
interior. Hence, the pseudo-modes are not oscillations that are trapped in the
solar interior \citep{Nigetal98,Nig98}. Moreover, the pseudo-mode spectral
peaks are essentially symmetric. Asymptotically, at high frequencies, their
profile is described by a $\sin^2$ function. Unfortunately, there is no simple
fitting formula describing both normal modes and pseudo-modes. Because we need
to leave the rather complicated issue of the proper profile to be employed as a
subject of further investigation, we rather decided to resort to the workaround
to simply force the fitting code to switch from the asymmetrical to the
symmetrical profile at a prescribed frequency, $\nu_s$, along a given ridge.
This switchover frequency is chosen to coincide with the first zero-crossing of
the line asymmetry parameter, $B$, that occurs for frequencies $\nu\gtrsim
4600\,\,\mu$Hz along a given ridge. For example, for the $n=6$ ridge the
switchover frequency is $\nu_s = 4815\,\,\mu$Hz, as is shown here in the right
panel of Figure~\ref{F3}. Expectedly, a sharp kink is introduced in the run of
$B$ at the switchover frequency. Of course, a more physically motivated fitting
profile would give rise to a gradual transition between the asymmetrical and
the symmetrical profile. We point out that the switchover frequency $\nu_s$
depends not only on the radial order $n$ but also on the epoch of the
observation, because the mode parameters are affected by the mean level of
solar activity during each observing run.

%Fig. 7

At a glance, the tiny error bars as shown in the left panel of Figure~\ref{F3}
seem to indicate that the excursions of the line asymmetry parameter, $B$, to
both positive and negative values for frequencies $\nu\gtrsim 5000\,\,\mu$Hz are a
statistically significant effect. This is not the case, however. Rather this
``significant'' asymmetry in the fitted line profiles is a further argument
that the asymmetric profile of \citet{Nig98} is invalid for the fitting of
pseudo-modes. Namely, if it were valid, the resulting line asymmetry should
be statistically compatible with $B=0$ for frequencies $\nu\gtrsim 5000\,\,\mu$Hz because
it is known that the pseudo-mode peaks are essentially symmetric.

\begin{figure}
\epsscale{1.00}
\plotone{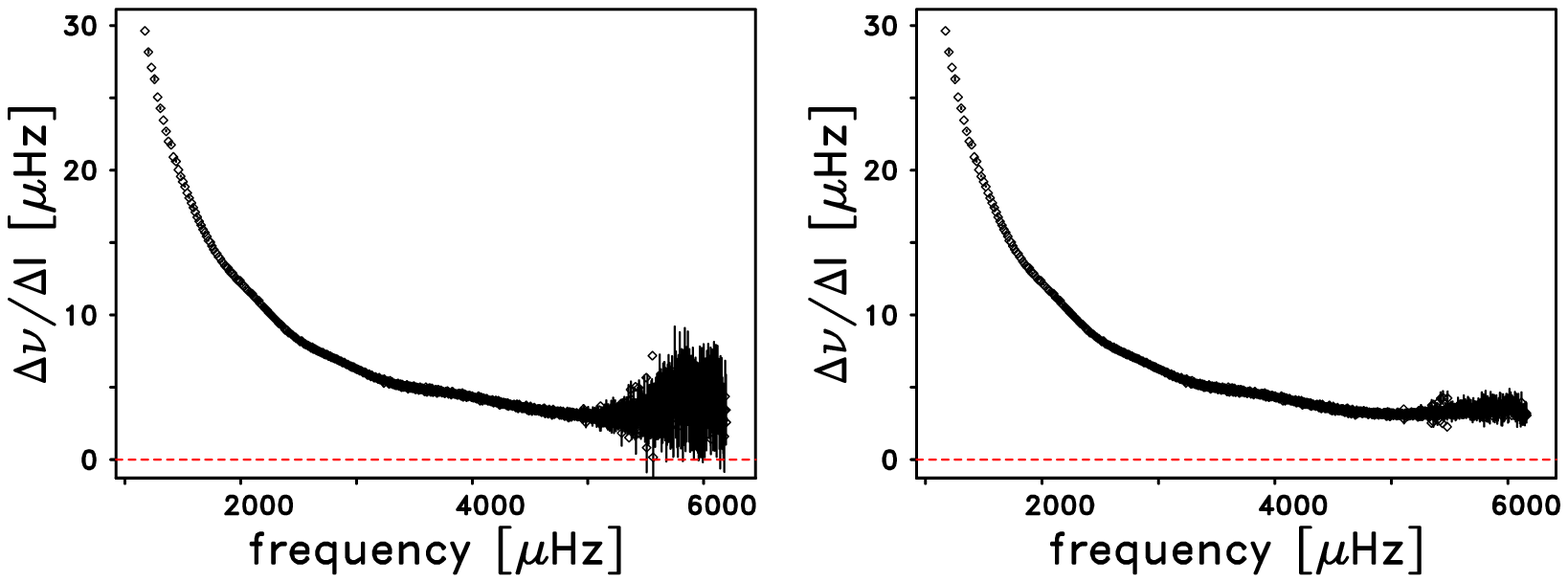}
\caption{Frequency dependence of the numerical derivative of the frequency with
respect to degree, $\Delta\nu/\Delta l$ (in $\mu$Hz), for the $n=4$ ridge. In
the left panel the asymmetrical profile has been used for fitting all of the
cases for this ridge, while in the right panel the asymmetrical profile has
been replaced with the symmetrical profile for frequencies $\nu >
4700\,\,\mu$Hz. Both sets of fits were computed from the $m$-averaged spectral
set $\cal{S}$2010\_66a. The error bars that are shown for selected cases are
$1\,\sigma$ errors. The dashed red line is for $\Delta\nu/\Delta l = 0$.
\label{sctfn04}}
\end{figure}

The use of the symmetrical profile for frequencies above the switchover
frequency, $\nu_s$, along a given ridge not only resolves the issue of the
unexpected excursions in the line asymmetry parameter, $B$, but also results in
a substantial reduction in the frequency scatter observed in the high-frequency
portion of each ridge. For the study of the variation in the frequency scatter
along a given ridge we found it useful to evaluate the point-to-point scatter,
$\varSigma$, as defined by equation~(\mbox{\ref{A3}}), not for the frequency
$\nu$ itself but rather for the numerical derivative of the frequency with
respect to degree, $\Delta\nu/\Delta l$. An example of the reduction in the
scatter of $\Delta\nu/\Delta l$ for frequencies above the switchover frequency
is shown here for the $n=4$ ridge in Figure~\ref{sctfn04}. For this ridge the
switchover frequency is $\nu_s=4700\,\,\mu$Hz. In the left panel of
Figure~\ref{sctfn04} the asymmetrical profile has been used for all of the
cases along this ridge, while in the right panel the asymmetrical profile has
been replaced with the symmetrical profile for frequencies $\nu > \nu_s$, i.e.,
for degrees $l\ge 562$. In order to evaluate the obvious reduction in the
frequency scatter quantitatively, we computed $\varSigma$ for the
high-frequency portion of the $n=4$ ridge by setting $l_{\rm min}=562$, $l_{\rm
max}=1000$, and $\upsilon\equiv \Delta\nu/\Delta l$ in
equation~(\mbox{\ref{A3}}). When we did so, we found after eliminating some
obvious gross outliers that $\varSigma=0.639$ for the asymmetrical profile, but
only $0.214$ for the symmetrical profile. Hence, by simply changing over from
the asymmetrical to the symmetrical profile at $\nu=\nu_s$, we were able to
reduce the scatter in $\Delta\nu/\Delta l$ by a factor of about 3. We also saw
similar reductions in the scatter in the high-frequency cases of
$\Delta\nu/\Delta l$ when we employed the symmetrical profile in place of the
asymmetrical profile for the high-frequency portions of the $f$-mode and the
other $p$-mode ridges.

%Fig. 8

\subsection{Determination of the background portion of the theoretical 
model profile\label{dbpmp}}

\begin{figure}
\epsscale{0.40}
\hspace{9.0mm}\plotone{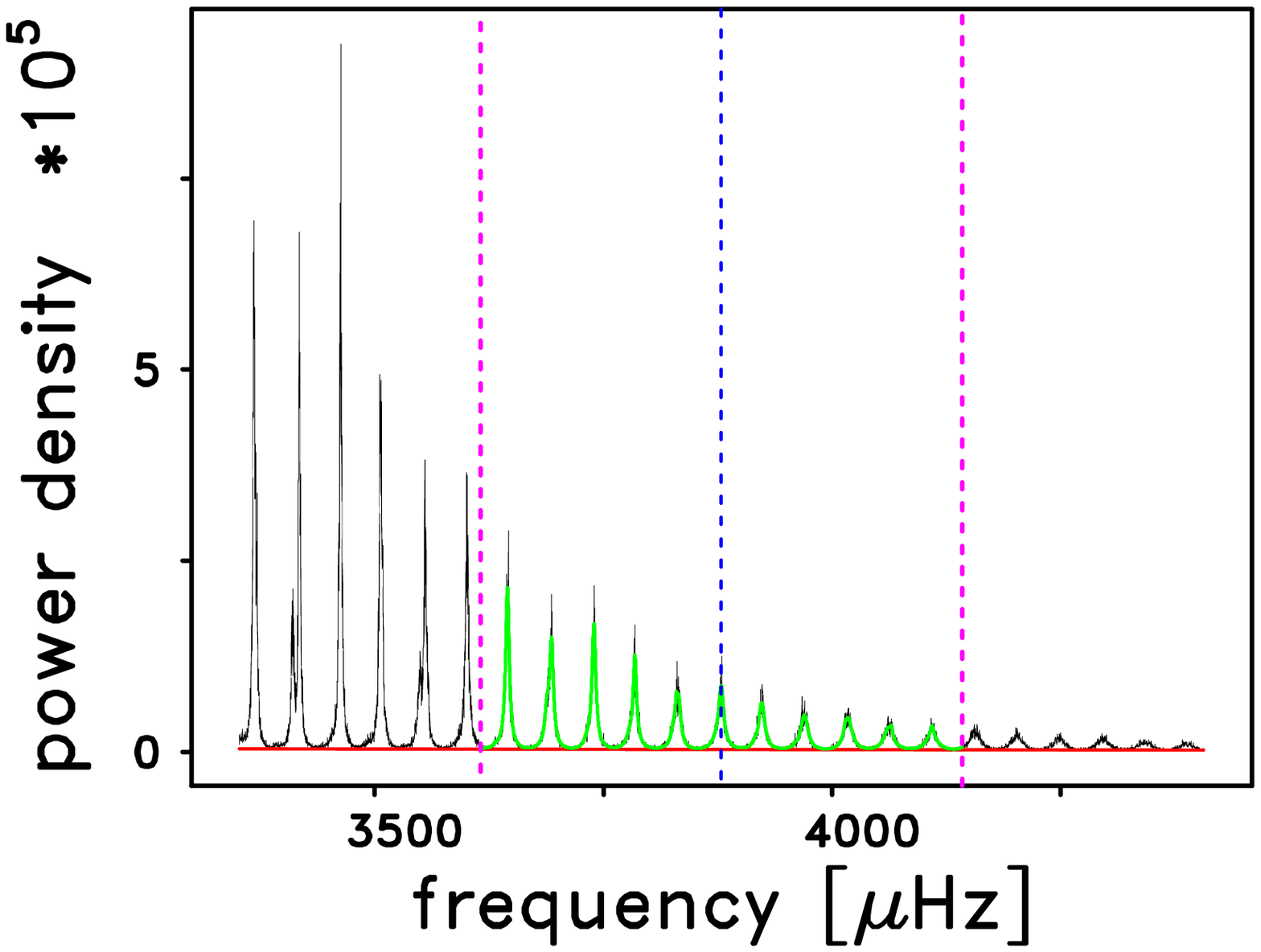}\hspace{16.0mm}
              \plotone{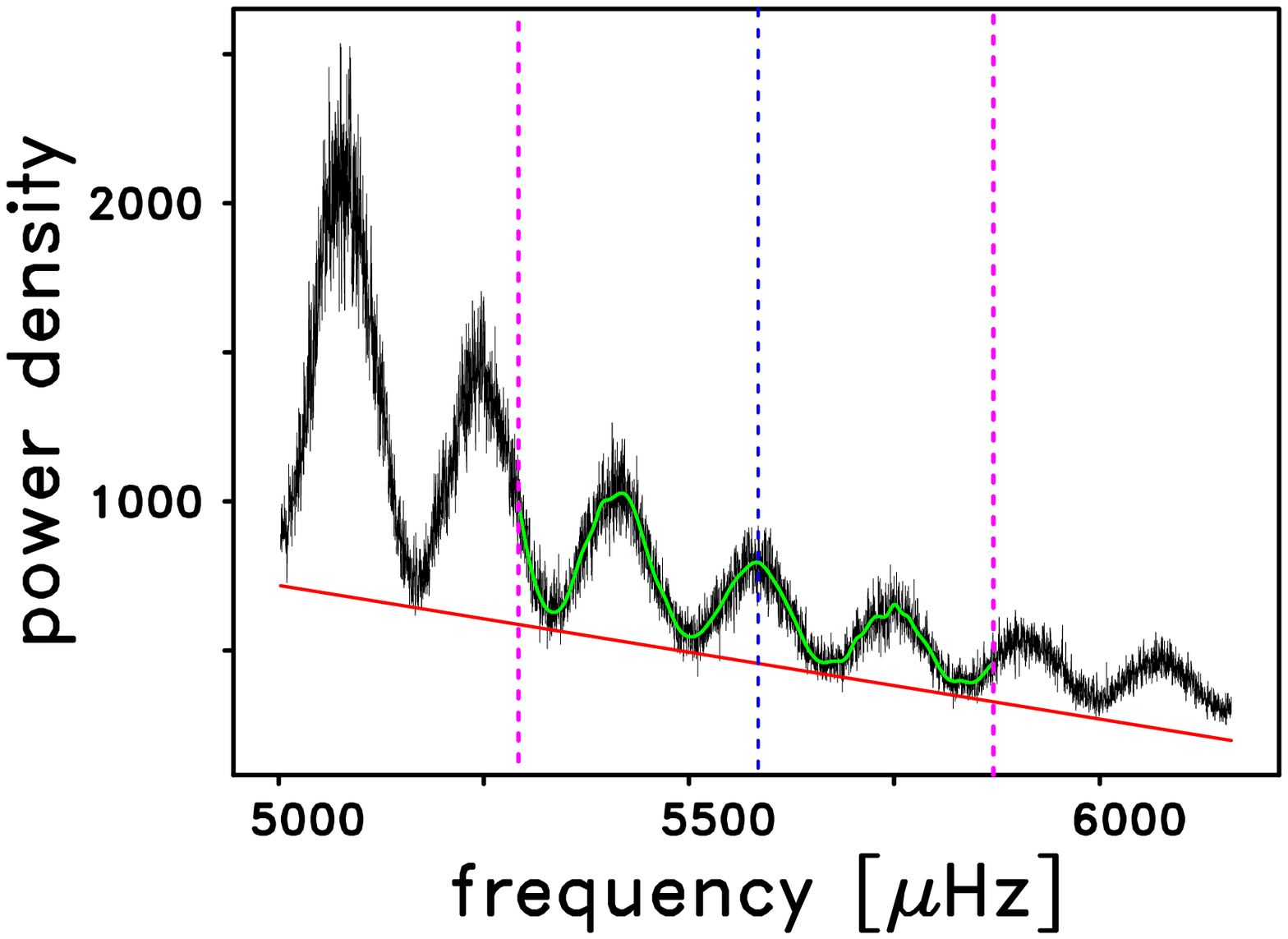}
\vspace{11.0mm}
\caption{
(left) Straight line fit to the spectral power density in the troughs of the spectrum
of the $(n,l)=(24,8)$ mode, using the example of the $m$-averaged spectral set
$\cal{S}$2010\_66a. The spectral power density is shown in black, while the
fitted line is shown in red. The slope of this fitted line is used as an
estimate of the slope, $b$, of the background portion of the theoretical model
profile given in equations~(\ref{profm2a}) through (\ref{profm2f}), when
this profile is fitted to the spectrum in the fitting box indicated by the
vertical dashed lines in magenta. The fitted profile is shown as the green
line, and the vertical blue dashed line is for the resulting fitted frequency.
(right) Same as left panel, except that it is for the $(24,56)$ mode.
\label{F4}} 
\end{figure}

A power spectrum computed from a time series of Dopplergrams of length $T$ has
a spectral resolution of $\Delta\nu = 1/T$. If the Dopplergrams are observed at
a cadence of $\Delta t$ the spectrum covers a frequency range from zero up to
the Nyquist frequency $\nu_{\rm Ny} = 1/(2\Delta t)$, which is $8333\,\,\mu$Hz
for the MDI instrument, as we have already pointed out in
Section~\ref{gunavgspc}. On the other hand, the maximum width, $W_{\rm max}$,
of the fitting boxes used for fitting the modal peaks in the set of spectra
obtained from an observing run is on the order of a few hundred $\mu$Hz at
most. Hence, for $T$ being greater than a day or so, we have
\begin{equation}
\Delta\nu \ll W_{\rm max} \ll \nu_{\rm Ny}.
\label{wfb}
\end{equation}
The inequality $W_{\rm max} \ll \nu_{\rm Ny}$ in the above equation~(\ref{wfb})
implies that we can safely approximate the background noise present in the
measured power spectra as a quadratic function of frequency within a selected
fitting box. We note, however, that generally a linear background model, i.e.,
$c=0$ in equation~(\ref{profm2b}), is an adequate choice. We also note that
the background portion of the theoretical model profile, as given in
equations~(\ref{profm2a}) through (\ref{profm2f}), is not a reliable estimate
of the actual frequency variation of the background noise within the selected
fitting box around the targeted peak. This is evident from the fact that the
fit parameters $a$, $b$, and $c$ in equation~(\ref{profm2b}), which describe
the background portion of the model profile, can change significantly from one
degree to the next along a given ridge. This scatter in the background portion
of the model profile is most likely an artificial effect, and unfortunately
translates into a similar scatter in other fitting parameters included in the
model profile, such as frequency and linewidth, for example. To mitigate this
scatter, we devised the following heuristic approach. First, we fit a straight
line to the spectral power in the troughs of the spectrum in a frequency range
that is twice as large than the fitting box centered around the targeted peak
to estimate the slope of the background noise. This is demonstrated here in
Figure~\ref{F4} where we show, for the modes $(n,l)=(24,8)$, and $(24,56)$,
respectively, the measured power spectrum in black with the such fitted
straight line overlaid in red. While the troughs in a spectrum are not
indicative of the background noise, but rather correspond to the intersection
of the wings of the peaks, we believe that they are indicative of the slope of
the background. Second, if $b_{\rm e}$ denotes the slope of the such determined
straight line and $\Delta b_{\rm e}$ the uncertainty thereof, we constrain the
slope $b$ of the background term in equation~(\ref{profm2b}) by 
\begin{equation} 
b_{\rm e}-3\,\Delta b_{\rm e} \leq b \leq b_{\rm e}+3\,\Delta b_{\rm e},
\label{bgca}
\end{equation}
where the term $3\,\Delta b_{\rm e}$ allows for uncertainties in the value of
$b_{\rm e}$. Third, we constrain the background term in
equation~(\ref{profm2b}) by
\begin{equation}
a+b\,\nu +c\,\nu^2 \geq 0 \quad \mbox{for all frequencies $\nu$ within the fitting box},
\label{bgcb}
\end{equation}
just to ensure that the estimated background noise is positive throughout the
entire fitting box.

%Fig. 9

\section{Sensitivity of Method~2 in terms of the
\textit{\textbf{m}}-averaging procedure and the effective leakage
matrix\label{sfmp}}

\subsection{Sensitivity of Method~2 to details of the
\textit{\textbf{m}}-averaging procedure\label{smavg}}

In order to study the sensitivity of Method~2 to the details of the
$m$-averaging procedure, as described in Section~\ref{gmavg}, we first used the
Method~2 code to fit the four different sets of $m$-averaged power spectra
$\cal{S}$2010\_66a through $\cal{S}$2010\_66d (cf. Table~\ref{tab2}). From
these fits we obtained four tables of fitted $f$- and $p$-mode parameters:
frequencies, linewidths, amplitudes, line asymmetries, and their associated
uncertainties. From these four tables of fitted parameters, we collected the
frequencies and their uncertainties into four frequency tables,
$\cal{F}$2010\_66a through $\cal{F}$2010\_66d, which we then inter-compared on
a mode-by-mode basis. As one example of these four frequency tables, our table
$\cal{F}$2010\_66a covered the degree range of 0 to 1000, the radial order
range of 0 to 29, and the frequency range of 965 to $7000\,\,\mu$Hz. Within
these ranges of $l$, $n$, and $\nu$ we were able to obtain a total of 12,359
sets of converged frequencies and frequency uncertainties.

We note that we have limited our sensitivity study to the investigation of the
impact of changes in the $m$-averaged spectra upon the fitted frequencies
because the estimates of the mode frequencies are an essential part of any
structural inversion.

\subsubsection{The influence of the weighting of the un-averaged
spectra\label{s6sub1}}

\begin{figure}
\epsscale{1.00}
\plotone{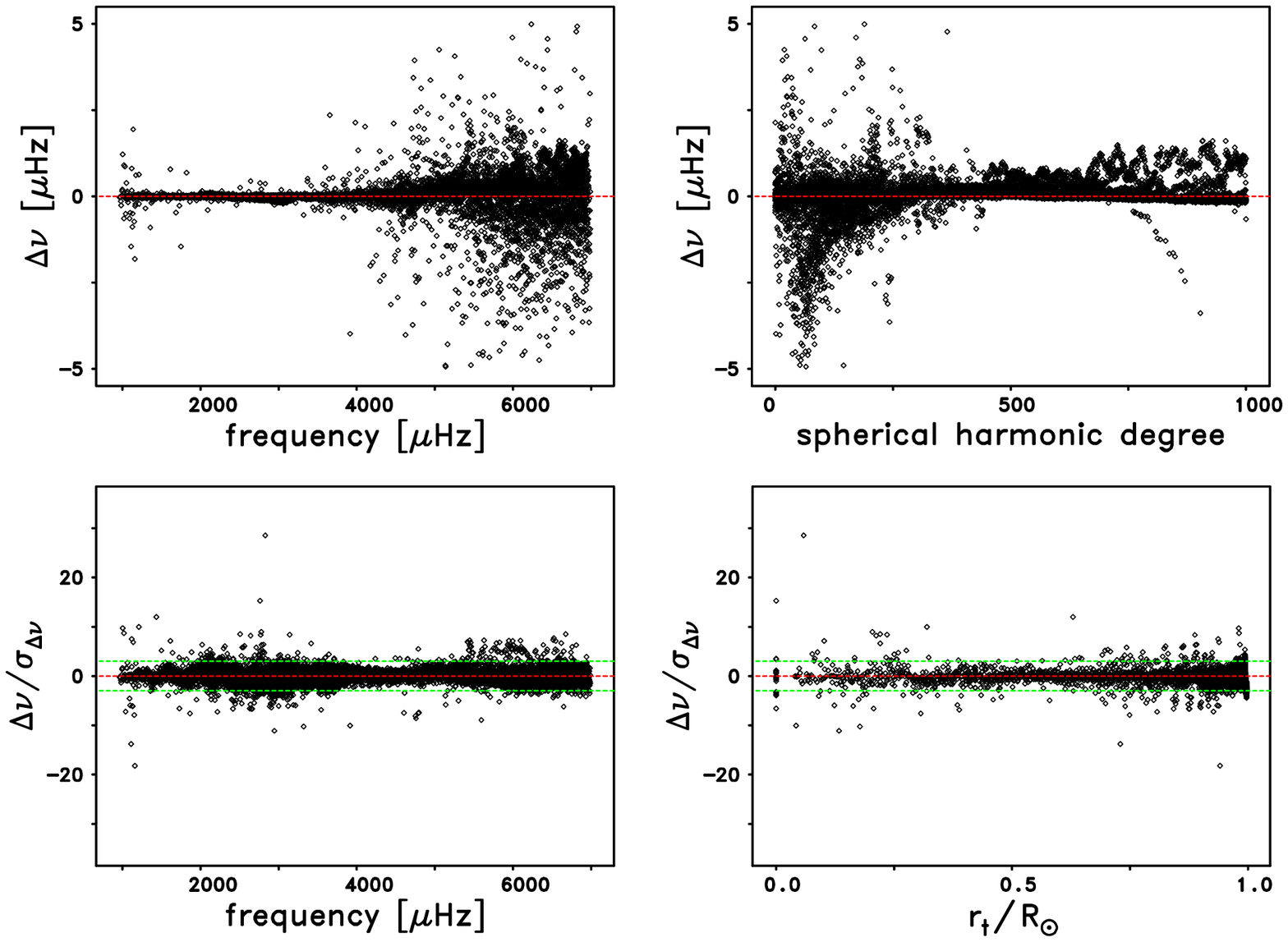}
\caption{
(upper-left) Frequency dependence of the frequency differences, $\Delta\nu$,
which resulted from frequency tables $\cal{F}$2010\_66a and $\cal{F}$2010\_66b,
in the sense of ``weighted'' minus ``unweighted''. In each case, the same set
of corrected, $n$-averaged frequency splitting coefficients has been used for
the collapsing of the spectra. (upper-right) Degree dependence of the set of
frequency differences, $\Delta\nu$, shown in the upper-left panel. (lower-left)
Frequency dependence of the normalized frequency differences,
$\Delta\nu/\sigma_{\Delta\nu}$. The normalization was carried out by dividing
the raw frequency differences, $\Delta\nu$, as shown in the upper-left panel,
by the formal error, $\sigma_{\Delta\nu}$, of each difference. (lower-right)
Dependence of the low-frequency subset of the normalized frequency differences
upon the fractional inner turning-point radii, $r_{\rm t}/R_{\odot}$, of those
modes. Here, $R_{\odot}$ is the radius of the Sun. Only mode frequencies less
than $4500\,\,\mu$Hz were included in this panel. In all four panels the red dashed
line is for a frequency difference of zero. In the two panels in the lower row
the green dashed lines show the $\pm 3\,\sigma$ values.
\label{dfsvsnuwtunwt}}
\end{figure}

In order to study the possible influence upon the fitted frequencies of the
weighting of the $2l+1$ un-averaged spectra at each degree in the computation
of the $m$-averaged power spectra, as described in Section~\ref{gmavg}, we
aligned our frequency tables $\cal{F}$2010\_66a and $\cal{F}$2010\_66b on a
mode-by-mode basis, and we then subtracted the frequency of each mode in
$\cal{F}$2010\_66a from the corresponding frequency in $\cal{F}$2010\_66b. The
raw frequency differences that resulted, are shown in the two upper panels of
Figure~\ref{dfsvsnuwtunwt}, where they are plotted, in the sense of ``weighted''
minus ``unweighted'', as functions of both frequency and degree. Inspection of
these two panels illustrates that the largest frequency differences
corresponded primarily to the high-frequency portions of the ridges for degrees
less than 350.

This concentration of the largest raw frequency differences at the higher
frequencies is illustrated numerically in section C1 of Table~\ref{tab5}, where
we show in row~1 that the average magnitude of the differences for the
frequency range $\nu<7000\,\,\mu$Hz is six times larger than the average magnitude of
the cases for which $\nu<4500\,\,\mu$Hz. On the other hand, when we normalized these
raw frequency differences by dividing each of them by its formal error, as is
shown in the lower left panel of Figure~\ref{dfsvsnuwtunwt}, we found that only
4.6\,\% of them were statistically significant. This is also demonstrated in
row~2 of Table~\ref{tab5}, where we show that the average magnitude of the
normalized frequency differences was about $1\,\sigma$ for both frequency
ranges $\nu<7000\,\,\mu$Hz and $\nu<4500\,\,\mu$Hz, respectively.

Because most structural inversions have been limited to frequencies that have
been less than $4500\,\,\mu$Hz, we also wanted to study the radial distribution
of the normalized frequency differences for the cases that were below this
limit. We found that 94.5\,\% of the cases that remained in this restricted
frequency range were within the range of $\pm 3\,\sigma$. Furthermore, when we
plotted the remaining normalized frequency differences as a function of the
fractional inner turning-point radius of the corresponding modes, as shown in
the lower-right panel of  Figure~\ref{dfsvsnuwtunwt}, we found that the
majority of the cases that were the most significant were concentrated in the
solar convection zone. In fact, very few of the normalized frequency
differences exceeded $\pm 3\,\sigma$ inward of the base of the convection zone.
Overall, Figure~\ref{dfsvsnuwtunwt} indicates that the weighting of the
un-averaged power spectra in the computation of the $m$-averaged power spectra
had a very modest effect upon the fitted frequencies.

%Fig. 10

\subsubsection{Influence of correction of frequency-splitting
coefficients\label{s6sub2}}

\begin{figure}
\epsscale{1.00}
\plotone{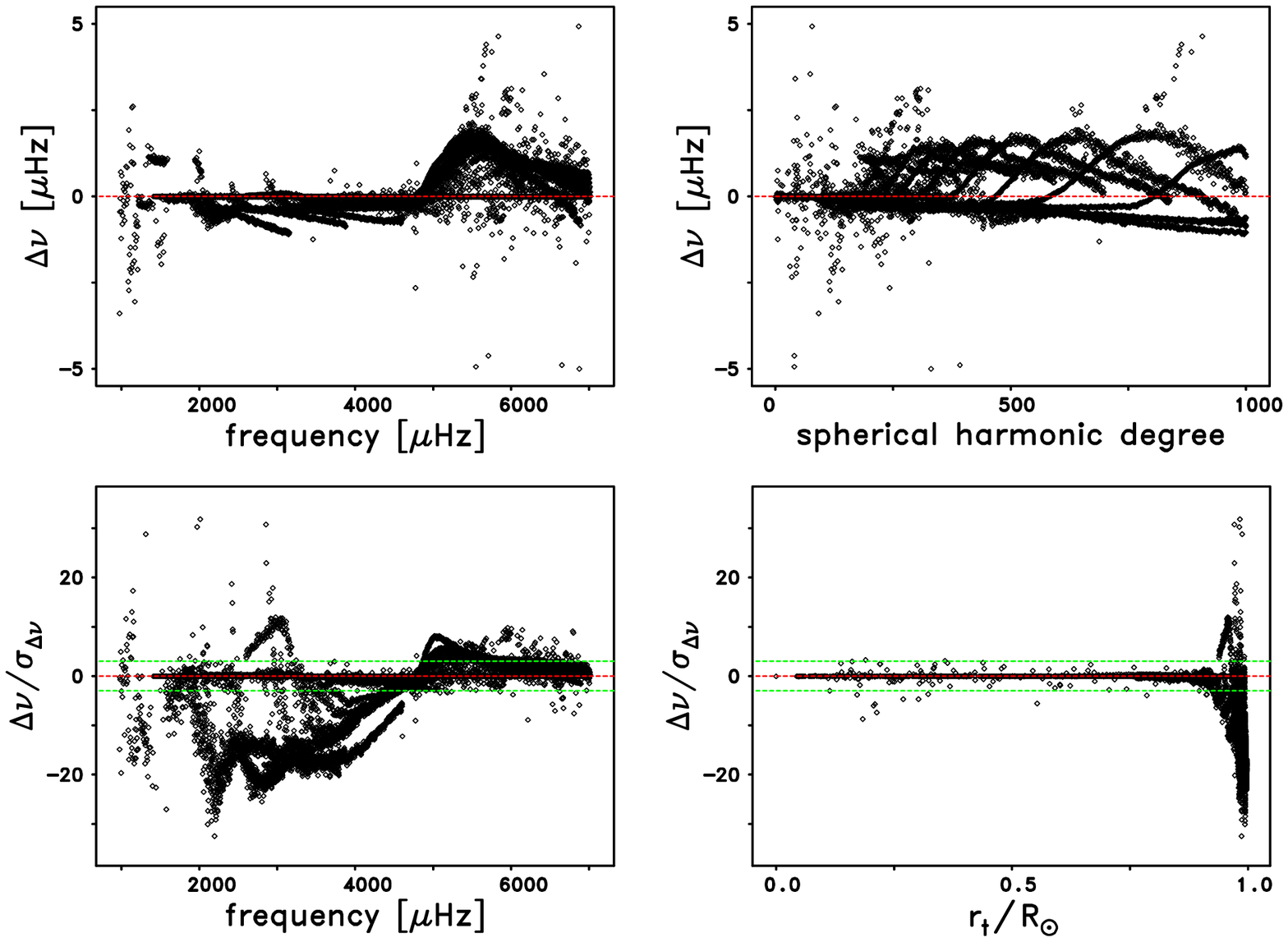}
\caption{
Same as Figure~\ref{dfsvsnuwtunwt}, except that it shows the dependence of both
the raw and normalized frequency differences, $\Delta\nu$, which resulted from
frequency tables $\cal{F}$2010\_66c and $\cal{F}$2010\_66d, in the sense of
``corrected'' minus ``raw'' with respect to frequency, degree, and fractional
inner turning-point radius.
\label{dfsvsnunonavgadj}}
\end{figure}

In contrast to the situation that we just described in Section~\ref{s6sub1} for
the construction of the $m$-averaged power spectra using both unweighted and
weighted tesseral and sectoral power spectra, the correction of the
un-corrected, non-$n$-averaged frequency splitting coefficients for the
distortions introduced by the latitudinal differential rotation resulted in
highly significant frequency differences for those cases that are used in
structural inversions, as we have demonstrated in
Figure~\ref{dfsvsnunonavgadj}. In the upper-left panel of this figure, we show
the frequency dependence of the un-normalized frequency differences in the
sense of ``corrected'' minus ``un-corrected'' that resulted when we subtracted
the frequencies in table~$\cal{F}$2010\_66c from those in
table~$\cal{F}$2010\_66d. In contrast to the upper-left panel of
Figure~\ref{dfsvsnuwtunwt}, the majority of these frequency differences for the
modes having $\nu<4500\,\,\mu$Hz were negative and the majority of the
high-frequency differences were positive. In another contrast with the
upper-left panel of Figure~\ref{dfsvsnuwtunwt}, there is a strong frequency
dependence of the high-frequency differences shown in the upper-left panel of
Figure~\ref{dfsvsnunonavgadj}. In a third contrast with the situation shown in
Figure~\ref{dfsvsnuwtunwt}, the upper-right panel of
Figure~\ref{dfsvsnunonavgadj} shows that the positive frequency differences
were not restricted to the lower-degree modes alone, but instead were present
for all modes having degrees greater than 200. The negative frequency
differences that are shown for $2500 < \nu < 4500\,\,\mu$Hz in the upper-left
panel of Figure~\ref{dfsvsnunonavgadj} corresponded to the $n=0$ through $n=3$
ridges. These negative differences indicate that the frequencies that were
computed from the $m$-averaged power spectra that were computed using the
corrected splitting coefficients were systematically smaller than were the
frequencies for the same ridges that were fit to the averaged spectra that were
generated using the un-corrected splitting coefficients. These systematic
frequency differences resulted from the fact that the corrected splitting
coefficients in this frequency range were generally smaller than their
un-corrected counterparts.

When we normalized these raw frequency differences by dividing each of them by
the formal uncertainty of the difference, we found that the most significant
normalized differences corresponded to modes which have frequencies between
1800 and $4500\,\,\mu$Hz, as is shown here in the lower-left panel of
Figure~\ref{dfsvsnunonavgadj}. In particular, the normalized frequency
differences were highly significant for the $n=0$ through $n=2$ ridges. In the
contrast to the situation for the raw frequency differences that are shown in
the upper-left panel of this figure, the normalized frequency differences
of the high-frequency modes were mainly less than $\pm 3\, \sigma$. When we
restricted these normalized frequency differences to those having frequencies
less than $4500\,\,\mu$Hz and plotted the remaining cases as a function of the
fractional inner turning-point radius of the modes, as shown in the lower-right
panel of Figure~\ref{dfsvsnunonavgadj}, we found that the cases with the
highly-significant negative ratios were all concentrated in the shallow
subphotospheric layers.

The statistics of both the raw and normalized frequency differences are listed
in section C2 (i.e., rows~3 and 4) of Table~\ref{tab5}. Here, we show in row~4
that the average magnitude of the normalized frequency differences for the
frequency range $\nu<7000\,\,\mu$Hz was equal to $4.9\,\sigma$, while the average
magnitude of the normalized frequency differences for the cases for which
$\nu<4500\,\,\mu$Hz was equal to $7.8\,\sigma$. Furthermore, this average of
$7.8\,\sigma$ is over 79 times its standard error away from zero. Clearly, the
correction of the frequency-splitting coefficients that are used in the
generation of $m$-averaged power spectra is an essential step.

A previous comparison of the sets of frequencies computed using both raw and
corrected, $n$-averaged splitting coefficients had shown small enough frequency
differences that led us to believe that the rather time-consuming correction of
more than 12,000 raw, non-$n$-averaged splitting coefficients would not be
necessary. However, in response to the suggestion of the anonymous referee that
we investigate the effects of making such corrections, we found significant
differences in the sets of frequencies computed using the raw and corrected,
non-$n$-averaged splitting coefficients, as is demonstrated here in the fourth
row of Table~\ref{tab5}, (i.e., the second row of Section C2).

%Fig. 11

\subsubsection{Influence of non-n-averaged frequency-splitting
coefficients\label{s6sub3}}

\begin{figure}
\epsscale{1.00}
\plotone{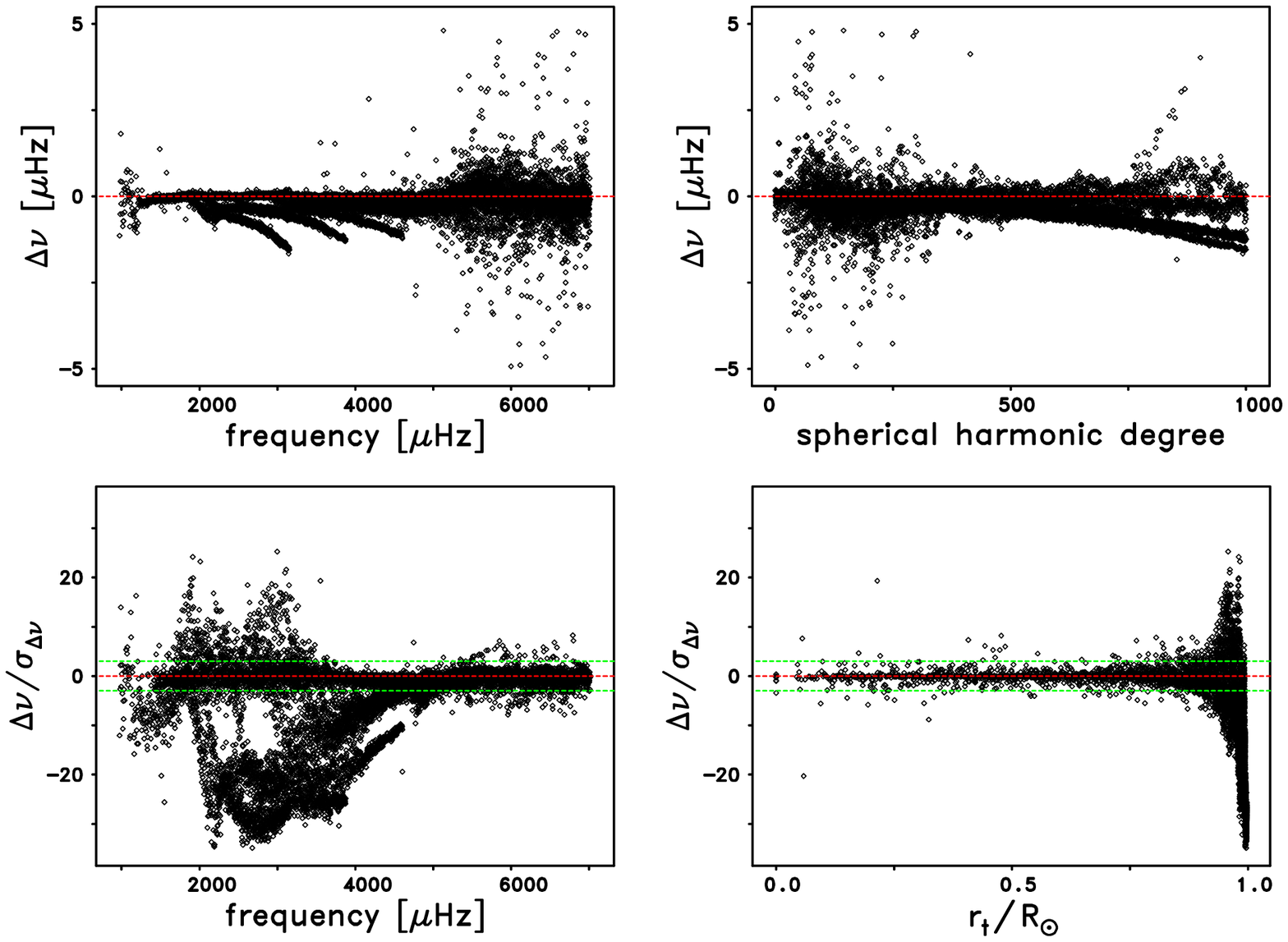}
\caption{
Same as Figure~\ref{dfsvsnuwtunwt}, except that it shows the dependence of both
the raw and normalized frequency differences, $\Delta\nu$, which resulted from
frequency tables $\cal{F}$2010\_66b and $\cal{F}$2010\_66d, in the sense of
``non-$n$-averaged'' minus ``$n$-averaged'' with respect to frequency, degree,
and fractional inner turning-point radius.
\label{dfsvsnunonavgavg}}
\end{figure}

The influence upon the fitted frequencies of the use of non-$n$-averaged
frequency-splitting coefficients in the generation of $m$-averaged power
spectra is illustrated in Figure~\ref{dfsvsnunonavgavg}. The raw frequency
differences, in the sense of ``non-$n$-averaged'' minus ``$n$-averaged'', are
shown as functions of frequency and degree in the upper-left and upper-right
panels, respectively. These frequency differences resulted when we subtracted
the frequencies in table $\cal{F}$2010\_66b from those in table
$\cal{F}$2010\_66d. The principal difference between these raw frequency
differences and those shown in the upper panels of
Figure~\ref{dfsvsnunonavgadj} is the absence of the strong, wave-like frequency
dependence of the high-frequency cases that was seen at the right side of the
upper-left panel of Figure~\ref{dfsvsnunonavgadj}. It was this wave-like
behavior of the high-frequency differences that was also visible as the series
of shifted peaks that are visible in the upper-right panel of
Figure~\ref{dfsvsnunonavgadj}. Hence, the absence of such a similar wave-like
variation in the high-frequency differences that are shown in the upper-left
panel of Figure~\ref{dfsvsnunonavgavg} resulted in the absence of a similar
series of shifted peaks in the upper-right panel of
Figure~\ref{dfsvsnunonavgavg}. In turn, the absence of this series of positive
peaks in the upper-right panel of Figure~\ref{dfsvsnunonavgavg} meant that the
majority of the high-degree frequency differences that are shown in the
upper-right panel of Figure~\ref{dfsvsnunonavgavg} were negative.

%Fig. 12

The frequency dependence of the normalized frequency differences that resulted
from the use of the non-$n$-averaged frequency-splitting coefficients is shown
in the lower-left panel of Figure~\ref{dfsvsnunonavgavg}. Close comparison of
this panel with the corresponding panel of Figure~\ref{dfsvsnunonavgadj}
illustrates that both sets of normalized frequency differences had a very
similar frequency dependence. As was the case in Figure~\ref{dfsvsnunonavgadj},
the most significant of these normalized frequency differences were negative
and corresponded to $\nu < 4500\,\,\mu$Hz.

The dependence of the subset of the normalized frequency differences for which
$\nu < 4500\,\,\mu$Hz upon the fractional inner turning-point radius is shown in the
lower-right panel of Figure~\ref{dfsvsnunonavgavg}. As was the case in
Figure~\ref{dfsvsnunonavgadj}, the most significant subset of these normalized
frequency differences was concentrated in the outer portion of the convection
zone. Very few of the normalized frequency differences that had inner
turning-point radii inward of $0.875\,R_{\odot}$ were outside the range of $\pm
3\,\sigma$.

The statistics of both the raw and normalized frequency differences are given
in section C3 (i.e., rows~5 and 6) of Table~\ref{tab5}. Here, we can see in
row~6 that the average magnitude of the normalized frequency differences for
the frequency range $\nu<7000\,\,\mu$Hz was equal to $5.8\,\sigma$, while the
average magnitude of the normalized frequency differences for the cases for
which $\nu < 4500\,\,\mu$Hz was equal to $10.2\,\sigma$. Comparison of both
averages with those for the same frequency range in row~4 of Table~\ref{tab5}
shows that the effects of using non-$n$-averaged splitting coefficients were
only slightly more significant than were the effects of the correction of the
raw frequency splitting coefficients. Other than this minor difference, the
influence of the non-$n$-averaged splitting coefficients upon the fitted
frequencies was very similar to that of the correction of the raw frequency
splitting coefficients themselves.

We have just shown in both Figure~\ref{dfsvsnunonavgavg} and in section C3
(i.e., rows~5 and 6) of Table~\ref{tab5} that the use of corrected,
non-$n$-averaged splitting rather than the corrected, $n$-averaged splitting
coefficients in the generation of the $m$-averaged power spectra produced
significant frequency differences for the low-order ridges. In order to
determine which ridges were most sensitive to the changes in the splitting
coefficients due to the use of non-$n$-averaged coefficients, we compared the
differences between the frequencies in Tables $\cal{F}$2010\_66b and
$\cal{F}$2010\_66d with the differences in the two sets of corrected splitting
coefficients themselves, and we found that the use of the $n$-averaged
splitting coefficients was equivalent to using only the splitting coefficients
of the higher-order ridges in the generation of the $m$-averaged spectra. The
use of the non-$n$-averaged splitting coefficients for the low-order ridges is
what resulted in the introduction of the significant, systematic frequency
differences.

\subsection{Sensitivity of Method~2 to an increase in the widths of leakage
matrix peaks\label{slkm}}

\begin{figure}
\epsscale{1.00}
\plotone{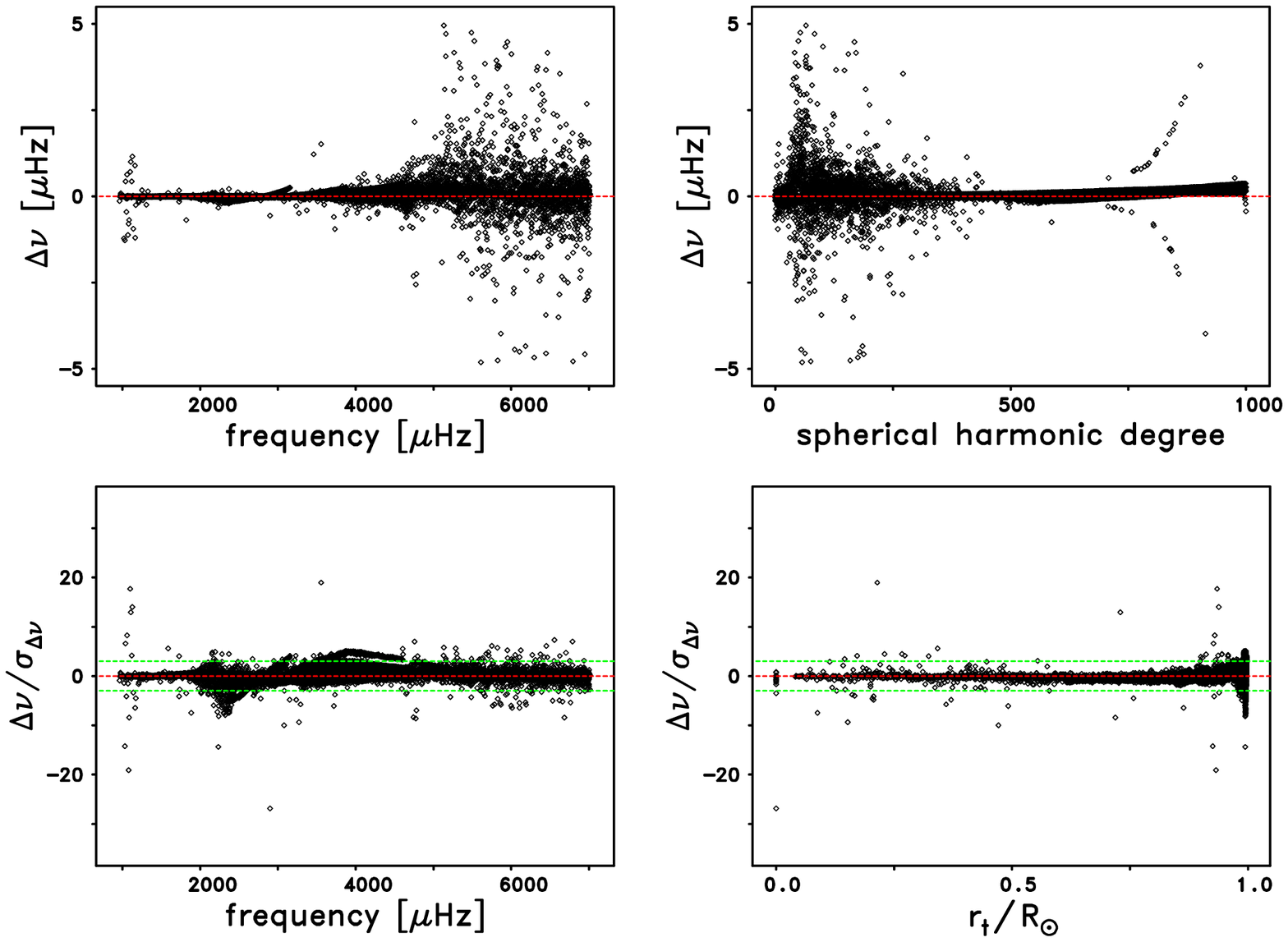}
\caption{
Same as Figure~\ref{dfsvsnuwtunwt}, except that it shows the dependence of both
the raw and normalized frequency differences, $\Delta\nu$, which resulted from
frequency tables $\cal{F}$2010\_66b and $\cal{F}$2010\_66b.lkm, in the sense of
``wider Gaussian'' minus ``original Gaussian'' with respect to frequency,
degree, and fractional inner turning-point radius.
\label{dfsvsnulkm}}
\end{figure}

In addition to studying the influence on the fitted frequencies of the details
of the computation of the $m$-averaged power spectra, we also investigated the
effects of an increase in the width of the Gaussian approximation that we
employ to represent the peaks of the leakage matrices (cf.
Section~\ref{imavglkm}). To this end, we refitted the $m$-averaged spectral
set $\cal{S}$2010\_66b with the Method~2 code using a Gaussian approximation
the width of which has been increased artificially by 18\,\%. This fitting run
resulted in the frequency table $\cal{F}$2010\_66b.lkm. By subtracting the
frequencies in table $\cal{F}$2010\_66b from those in table
$\cal{F}$2010\_66b.lkm we obtained raw frequency differences the frequency and
degree dependencies of which are shown, in the sense of ``wider
Gaussian'' minus ``original Gaussian'', in the upper-left and upper-right
panels, respectively, in Figure~\ref{dfsvsnulkm}. Comparison of these two
panels with the upper panels of Figure~\ref{dfsvsnuwtunwt} shows that these raw
frequency differences were similar to those shown in
Figure~\ref{dfsvsnuwtunwt}, with the main differences being an increase in the
number of cases at both high frequencies and moderate degrees for which
$\Delta\nu$ was positive and a corresponding decrease in the number of cases
for which $\Delta\nu$ was negative. Further comparison of the two upper panels
of Figure~\ref{dfsvsnulkm} with the upper panels of both
Figures~\ref{dfsvsnunonavgadj} and \ref{dfsvsnunonavgavg} shows that the
increase in the width of the leakage matrix peak did not result in systematic
frequency differences for the low-order ridges.

The normalized frequency differences that resulted from the use of the
different approximations to the peaks of the leakage matrices are shown in the
lower-left panel of Figure~\ref{dfsvsnulkm}. Comparison of this panel with the
lower-left panel of Figure~\ref{dfsvsnuwtunwt} confirms the absence of any
significant frequency differences in both figures. The dependence of the subset
of the normalized frequency differences for which $\nu < 4500\,\,\mu$Hz upon
the fractional inner turning-point radius is shown in the lower-right panel of
Figure~\ref{dfsvsnulkm}. This panel demonstrates that, with the exception of a
small concentration of significant differences just below the photosphere,
there is no obvious radial dependence to the remaining differences that
exceeded $\pm 3\,\sigma$.

The statistics of the frequency differences that are shown in
Figure~\ref{dfsvsnulkm} are summarized in section C4 of Table~\ref{tab5}. Here,
row~7 shows that the average magnitude of the raw frequency differences was
$0.149\,\,\mu$Hz for the cases for which $\nu<7000\,\,\mu$Hz and dropped to
$0.060 \,\,\mu$Hz for the cases for which $\nu<4500\,\,\mu$Hz, while row~8
confirms our statement above that these frequency differences were not
significant, since the average ratio was equal to 1.00 over the entire
frequency range, and was equal to only 1.37 for the low-frequency subset.

%Fig. 13

\section{Sample results from Method~2 and comparison with Method~1\label{SR}}

Before we shall present sample results from Method~2 in Section~\ref{srm2}, we
will demonstrate in Section~\ref{cmpm1m2} the substantial improvements that
Method~2 makes in the frequencies, linewidths, and amplitudes that we generated
with this method by comparing them with corresponding quantities that we
generated using our Method~1. For a description of Method~1 we refer the reader
to Appendix~\ref{secm1}.

\subsection{Comparison of results from Methods 1 and 2\label{cmpm1m2}}

One key difference between Method~1 and our Method~2 is the fact that Method~1
is restricted to use only symmetric Lorentzian profiles, while Method~2
includes the option of using either symmetric or asymmetric profiles. A second
key difference between the two methods is the use of both narrow and wide
fitting ranges in Method~1, while such wide fitting ranges are never employed
in Method~2. Instead, in Method~2 we employ the procedure described in
Section~\ref{sfbw} to determine the width of each fitting box individually. As
a result, the widths of the fitting boxes vary systematically as functions of
both the degree and the frequency in Method~2. 

\begin{figure}
\afterpage{
\epsscale{0.95}
\plotone{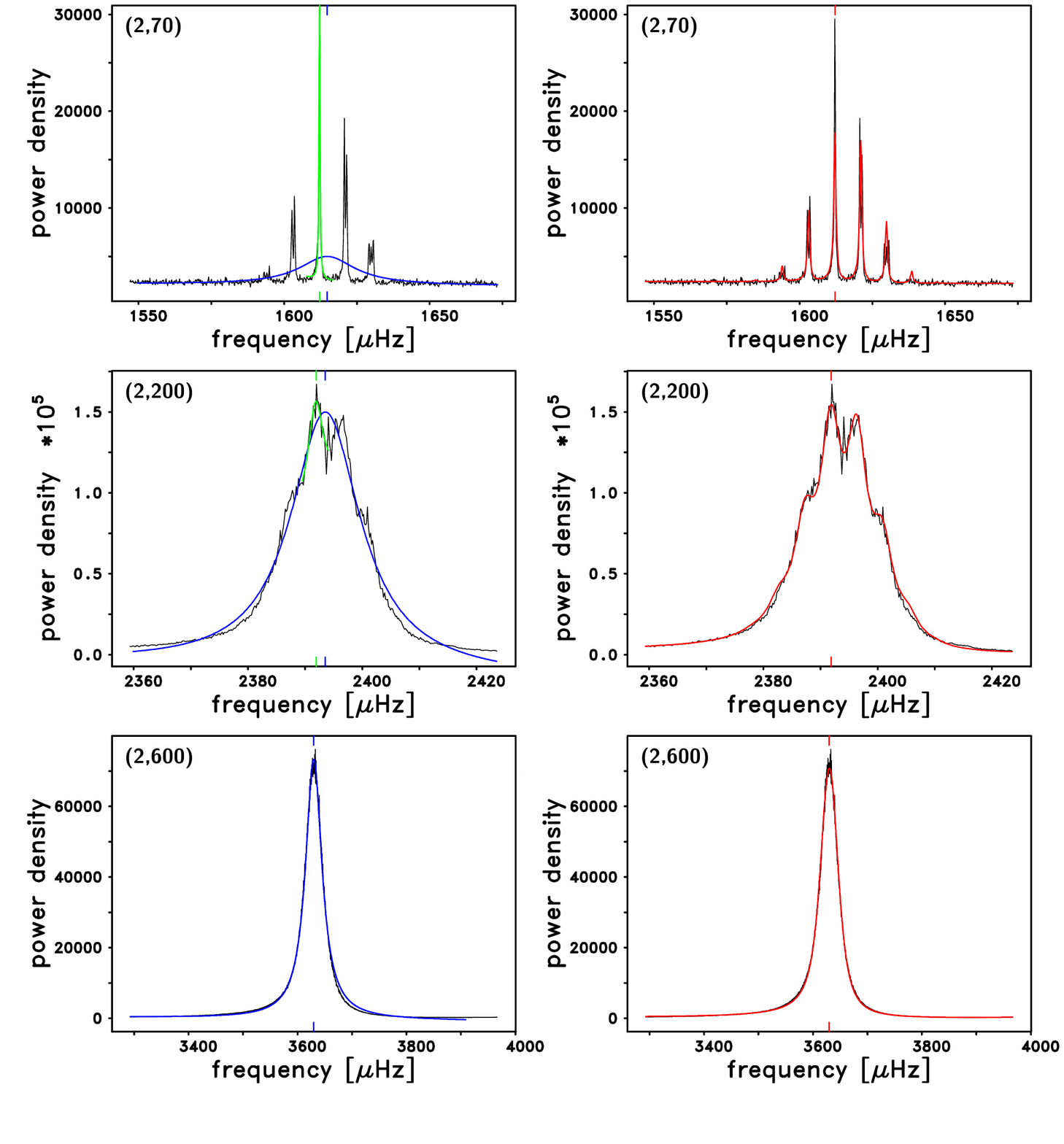}
\figcaption{
Fits to segments of three different spectra from the $m$-averaged spectral set
$\cal{S}$2010\_66a that were centered around the frequency of the
$(n,l)=(2,70)$ mode (top panels), the $(2,200)$ mode (middle panels), and the
$(2,600)$ mode (bottom panels). In each panel the black line is for the
$m$-averaged spectrum. In the top-left and middle-left panel the green line is
for the fit using the narrow fitting range in Method~1, while the blue line is
for the fit using the wide fitting range in Method~1 to simulate the effect of
fitting broad ridges of power. The segment centered on the $(2,600)$ mode has
been fit using only the wide fitting range in Method~1, and the fitted profile
is shown as the blue curve in the bottom-left panel. All of the computed
profiles shown in the left-hand panels were generated using the symmetric
profile. The red lines in the three right-hand panels were all computed using
Method~2 employing the asymmetric profile of \cite{Nig98}. For all six panels
the fitted frequencies are indicated by the colored tick marks that are located
along both the top and bottom axes of each plot. 
\label{F2}}
}
\end{figure}

In the three left-hand panels of Figure~\ref{F2}, we compare the fitted
profiles that Method~1 produced with segments of three different spectra from
our $m$-averaged spectral set $\cal{S}$2010\_66a. In the three right-hand
panels of the same figure we compare the profiles that Method~2 produced for
the same three spectral segments using the asymmetric profile of \cite{Nig98}.
The normalized comparison of Method~1 and Method~2 fit results shown in
Table~\ref{tab6} for the same three modes whose fits are shown in
Figure~\ref{F2} demonstrates that the Method~2 frequencies and linewidths do
agree more closely with the Method~1 results that were computed using the
narrow fitting range than they do with the Method~1 results that were computed
using the wide fitting range. Table~\ref{tab6} also shows that, with the
exception of the frequency that was computed for the $(n,l)=(2,70)$ mode using
the narrow fitting range in Method~1, all of the other frequencies and
linewidths are systematically different between Method~1 and Method~2.

%Fig. 14

\begin{figure}
\epsscale{1.00}
\plotone{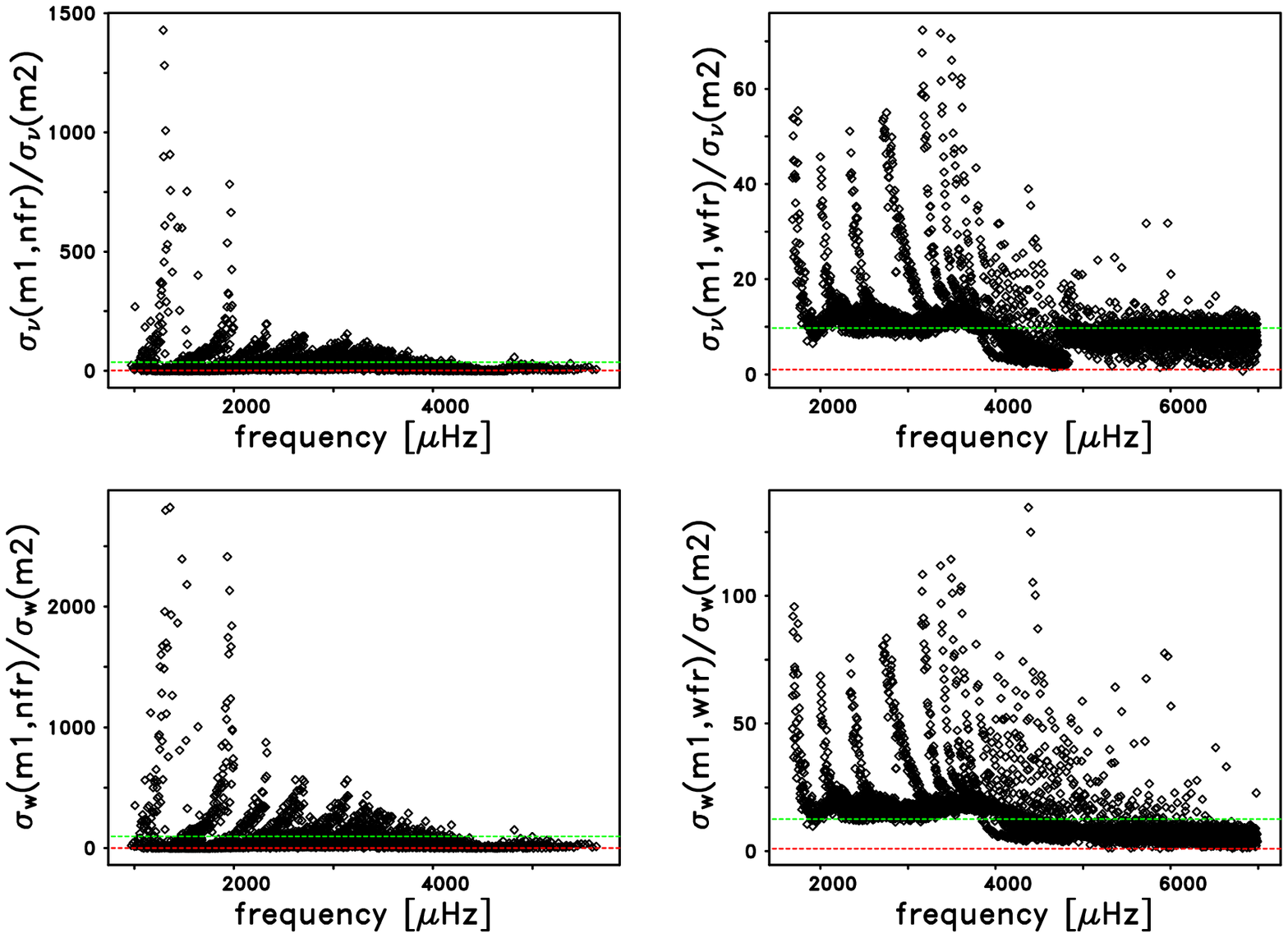}
\caption{
(upper-left) Frequency dependence of the ratios of the Method~1 narrow fitting
range frequency uncertainties divided by the Method~2 frequency uncertainties.
(upper-right) Frequency dependence of the ratios of the Method~1 wide fitting
range frequency uncertainties divided by the Method~2 frequency uncertainties.
(lower-left) Frequency dependence of the ratios of the Method~1 and Method~2
linewidth uncertainties for the narrow fitting range cases. (lower-right)
Frequency dependence of the ratios of the Method~1 and Method~2 linewidth
uncertainties for the wide fitting range cases. In all four panels the dashed
green lines show the average ratios, while the dashed red lines show the error
ratios of unity. Note that the vertical scales are different in all four
panels. In all four panels the $m$-averaged spectral set $\cal{S}$2010\_66a has
been fitted.
\label{m1m2rat}}
\end{figure}

In Figure~\mbox{\ref{m1m2rat}} we illustrate the improvements that Method~2
made in both the formal frequency and linewidth uncertainties relative to
the corresponding quantities that were computed using Method~1. In the
upper-left panel of Figure~\ref{m1m2rat} we show the frequency dependence of
the ratios of the Method~1 and Method~2 frequency uncertainties for the cases
in which the narrow fitting range was employed in Method~1, while we show the frequency
dependence of the similar ratios for the cases in which the wide fitting range
was employed in the upper-right panel. The average ratio of the two sets of
uncertainties in the upper-left panel of this figure was $36.1\pm 1.5$, while
the average ratio of the two sets of uncertainties in the upper-right panel was
$9.7\pm 0.1$.

The frequency dependence of the ratios in the Method~1 and Method~2 linewidth
uncertainties are shown in the lower-left panel of Figure~\ref{m1m2rat} for the
cases in which the narrow fitting range was used in Method~1, and we show the frequency
dependence of the ratios of the Method~1 and Method~2 linewidth uncertainties
for the cases in which the wide fitting range was employed in the lower-right
panel of the same figure. The average ratio of the two sets of linewidth
uncertainties in the lower-left panel was $96.2\pm 4.3$, while the average
ratio of the two sets of linewidth uncertainties in the lower-right panel was
$12.6\pm0.1$. Combining both the narrow and wide fitting range ratios in
overall sets of uncertainty ratios, we found that Method~2 reduced the
frequency uncertainties by an average ratio of $15.2\pm 0.3$ and it reduced the
linewidth uncertainties by an average ratio of $29.8\pm 0.9$. These sizeable
reductions in both the frequency and linewidth uncertainties testify to the
importance of including more than a single peak in the theoretical profile that
is used to fit both modes and ridges of power.

%Fig. 15

\begin{figure}
\epsscale{1.00}
\plotone{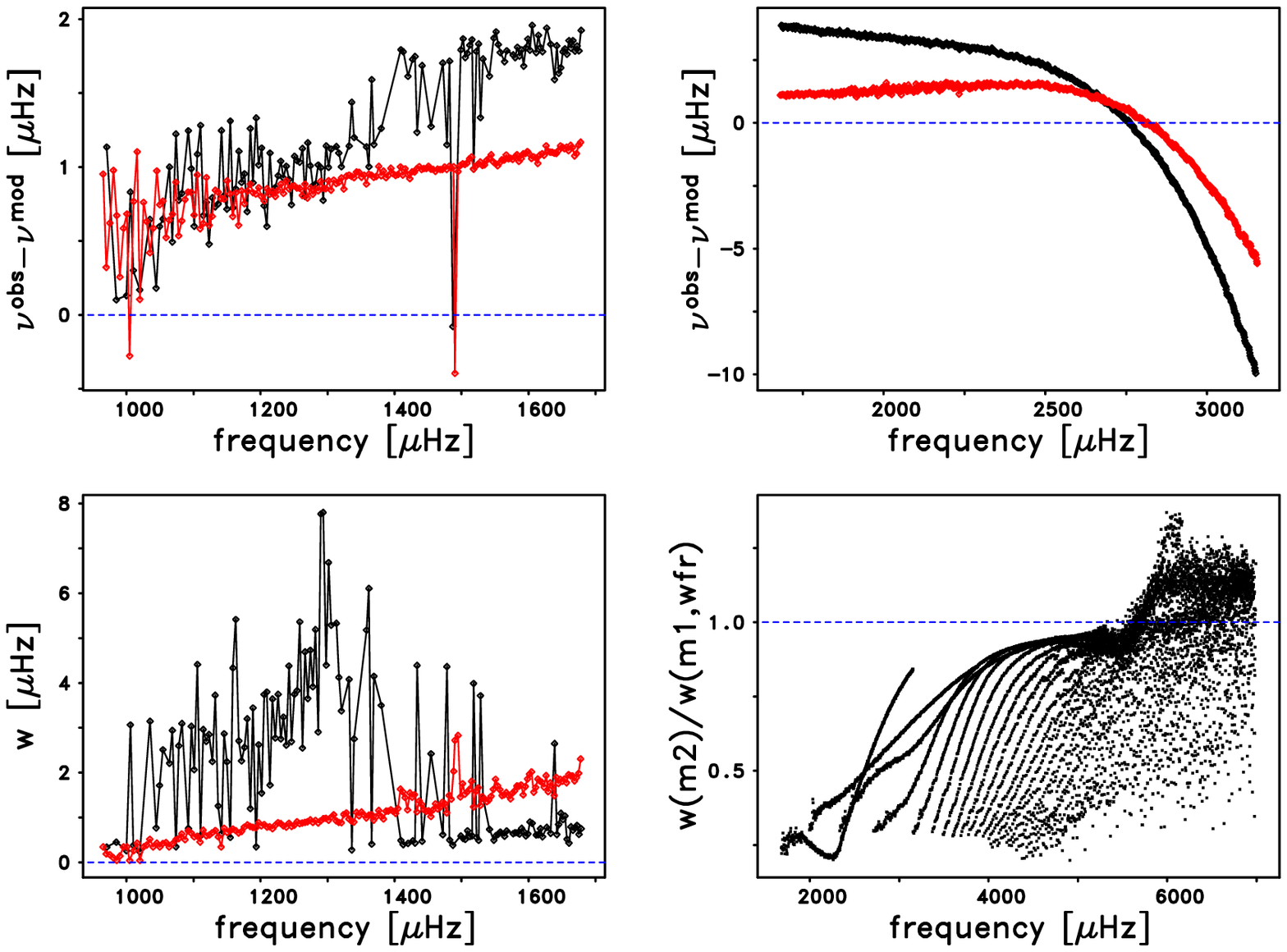}
\caption{
(upper-left) Frequency dependence of two sets of frequency differences,
$\Delta\nu=\nu^{\rm obs}-\nu^{\rm mod}$, between our fitted frequencies,
$\nu^{\rm obs}$, and the corresponding Model~S frequencies, $\nu^{\rm mod}$,
for the $n=0$ ridge. The $\Delta\nu$ values as computed by Method~1 using the
narrow fitting range are shown as the black line, while the $\Delta\nu$ values
computed using Method~2 are shown as the red line. The outlier in both the
Method~1 and Method~2 frequencies at about $1500\,\,\mu$Hz is for $l=219$, and
is caused by a glitch in the underlying spectrum. In both upper panels the
dashed blue line represents a difference of zero. (upper-right) Same as
upper-left panel, except that the wide fitting range was used for computing the
Method~1 frequencies. (lower-left) Frequency dependence of the linewidths
computed using Method~1 with the narrow fitting range (the black line) and
using Method~2 (the red line). The dashed blue line represents linewidths of
zero. (lower-right) Frequency dependence of the ratio of the linewidths as
computed using Method~2 divided by the linewidths computed using Method~1 with
the wide fitting range. The dashed line represents a linewidth ratio of unity.
In all four panels, the fits were computed using the $m$-averaged spectral set
$\cal{S}$2010\_66a. 
\label{m1m2diff}}
\end{figure}

In the upper two panels of Figure~\ref{m1m2diff} we show the improvements that
Method~2 introduced into the frequencies of the $n$=0 ridge in comparison with
the frequencies computed using Method~1. We demonstrate these improvements by
comparing both sets of fitted frequencies with theoretical frequencies that
have been computed from Model~S of \cite{jcd96} employing the approach
described by \cite{Kos99}. Specifically, in both of these panels we show the
frequency dependence of the frequency differences $\Delta\nu = \nu^{\rm
obs}-\nu^{\rm mod}$, where $\nu^{\rm obs}$ refers to either our Method~1 or
Method~2 fitted frequencies and where $\nu^{\rm mod}$ is the
corresponding theoretical frequency computed using Model~S. The
$\Delta\nu$ values for the Method~1 fits are shown as the black diamonds, while
the $\Delta\nu$ values for Method~2 are shown as the red diamonds. The
upper-left panel is for the portion of the $n=0$ ridge that was originally fit
using the narrow fitting range in Method~1, while the upper-right panel is for
the frequency range of the same ridge that was initially fit using the wide
fitting range in Method~1. The upper-left panel shows that the Method~2
frequencies are much smoother than the Method~1 frequencies and that they also
agree more closely with the theoretical frequencies. The upper-right panel
shows that, with the exception of a range of frequencies around
$2700\,\,\mu$Hz, the Method~2 frequencies agreed more closely with the
theoretical frequencies than did the Method~1 frequencies.

In the lower-left panel of Figure~\ref{m1m2diff} we show the improvements that
Method~2 introduced into the linewidths of the $n=0$ ridge. Since there are no
theoretical linewidths that we can compare the fitted linewidths with, we have
simply plotted both sets of fitted linewidths as a function of frequency in
this panel. This comparison demonstrates that the Method~2 linewidths (shown as the
red diamonds) varied in a much smoother manner with increasing frequency than
did the Method~1 linewidths (shown as the black diamonds). It also demonstrates that
for the majority of the modes shown, the Method~2 linewidths were smaller than
were the corresponding linewidths computed using Method~1.

In the lower-right panel of Figure~\ref{m1m2diff} we extend our comparison of
the Method~1 and Method~2 linewidths to the entire set of 30 ridges that were
fit using Method~2 and using both fitting ranges in Method~1. Here we have
plotted the frequency dependence of the ratios of the Method~2 linewidths
divided by the Method~1 linewidths. This panel clearly shows that for all of
the modes with frequencies below the acoustic cut-off frequency the fits
computed using Method~2 produced linewidths that were smaller than the
corresponding Method~1 linewidths by factors ranging between~2 and 5. The
smaller linewidths all correspond to longer estimated lifetimes for these
modes.

%Fig. 16

in order to show the entire frequency range spanned by both fitting ranges with a common
scale for the y-axis"

\begin{figure}
\afterpage{
\epsscale{0.90}
\plotone{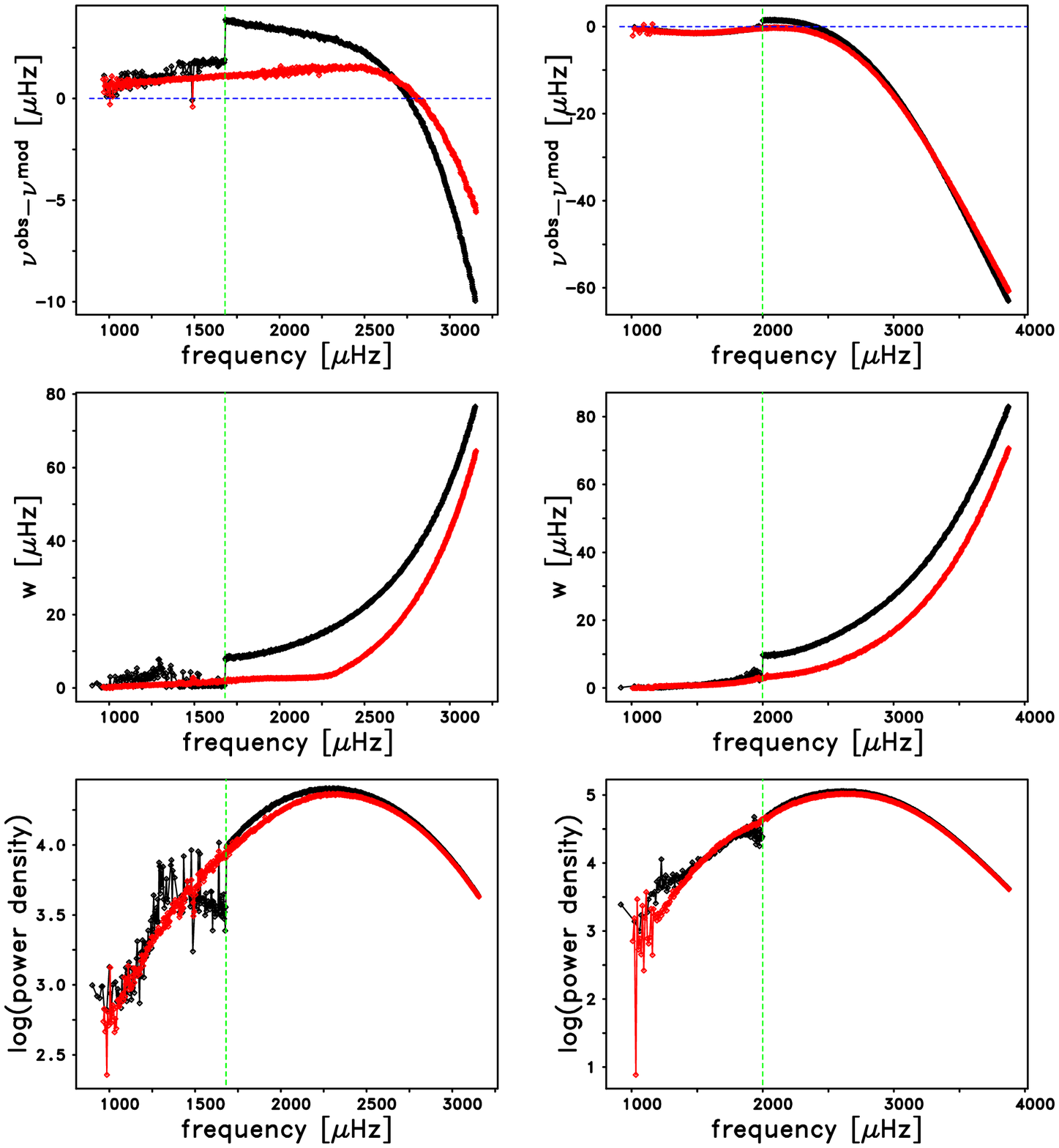}
\figcaption{
(upper-left) Frequency dependence of two sets of frequency differences,
$\Delta\nu=\nu^{\rm obs}-\nu^{\rm mod}$, between our fitted frequencies,
$\nu^{\rm obs}$, and the corresponding Model~S frequencies, $\nu^{\rm mod}$,
for the $n=0$ ridge. The $\Delta\nu$ values computed using our Method~1 fits
are shown as the black line, while the $\Delta\nu$ values computed using
Method~2 are shown as the red line. The Method~1 frequency differences exhibit
a pronounced discontinuity at the location of the switch-over in the fitting
range that was employed in Method~1. This location is marked by the vertical
green dashed line in all three of the left-hand panels. (upper-right) Same as
upper-left panel, but for the $n=1$ ridge. In this panel, and in the other two
right-hand panels, the vertical green dashed line marks the location of the
transition from the narrow fitting range to the wide fitting range in Method~1
for this ridge. In both top panels the dashed blue line indicates a frequency
difference of zero. Note that the vertical scales are different in the two top
panels. (middle-left) Frequency dependence of the linewidths computed using
Method~1 (black) and Method~2 (red) for the $n=0$ ridge. The Method~1
linewidths exhibit a sharp discontinuity precisely where the narrow fitting
range was replaced with the wide fitting range. By contrast, the Method~2
linewidths do not exhibit any such discontinuity at the same frequency.
(middle-right) Same as middle-left panel, but for the $n=1$ ridge. The
discontinuity in the linewidths occurs at the same frequency as in the
upper-right panel. (lower-left) Frequency dependence of the logarithms of the
amplitudes, or power densities, computed using Method~1 (black) and using
Method~2 (red) for the $n=0$ ridge. The Method~1 amplitudes exhibit a
discontinuity at the precise frequency where we switched from the narrow
fitting range to the wide fitting range. (lower-right) Same as lower-left
panel, but for the $n=1$ ridge. The discontinuity in these amplitudes occurs at
the same frequency as in the upper-right and middle-right panels. In all six
panels, the fits were computed using the $m$-averaged spectral set
$\cal{S}$2010\_66a.
\label{dfm1ms}}
}
\end{figure}

In the upper-left panel of Figure~\ref{dfm1ms} we have combined both portions
of our comparisons of Method~1 and Method~2 frequencies with the Model~S
frequencies for the $n=0$ ridge into a single panel in order to show the entire
frequency range spanned by both fitting ranges with a common scale for the
vertical axis to illustrate yet another important improvement of Method~2 over
Method~1 -- namely the complete absence in the Method~2 frequencies of the
substantial discontinuity that is present in the Method~1 frequencies at
precisely the frequency where the transition was made from the use of the
narrow fitting range to the use of the wide fitting range in that method. In
the upper-right panel of Figure~\ref{dfm1ms} we show a similar comparison of
Method~1 and Method~2 frequencies with the corresponding Model~S frequencies
for the $n=1$ ridge. As with the upper-left panel, there is a pronounced
discontinuity in the Method~1 frequencies at the location of the switch-over
from the narrow fitting range to the wide fitting range that is absent in the
Method~2 frequencies.

%Fig. 17

In addition to the introduction of jumps into the fitted frequencies, the
switch-over in the width of the fitting ranges that were employed in Method~1
also caused a corresponding discontinuity in the Method~1 linewidths, as we
show here in the two middle panels of Figure~\ref{dfm1ms}. The jumps in the
linewidths that are shown in black in these two panels occurred for exactly the
same modes for which the frequencies also exhibited the jumps in the two upper
panels. In contrast to the Method~1 linewidths, the Method~2 linewidths
exhibited no such discontinuities, as we show in red in the middle panels of
Figure~\ref{dfm1ms}. As we also showed in the lower panels of
Figure~\ref{m1m2diff}, the Method~2 linewidths agreed more closely with the
Method~1 linewidths that were computed using the narrow fitting range than they
did with the Method~1 linewidths that were computed using the wide fitting
range.

We also demonstrate in the two lower panels of Figure~\ref{dfm1ms} that the use
of the single-peak profile in Method~1 introduced discontinuities in the
amplitudes (as shown in black), or power densities, that disappeared when the
multiple-peak profile of Method~2 was used instead (as shown in red). Not only
did the single-peak profile introduce discontinuities into the amplitudes, it
also caused the amplitudes that were computed using the narrow fitting range to
show more scatter as a function of degree along the ridges to the left of the
jumps than was the case with our Method~2 amplitudes.

\begin{figure}
\afterpage{
\epsscale{0.90}
\plotone{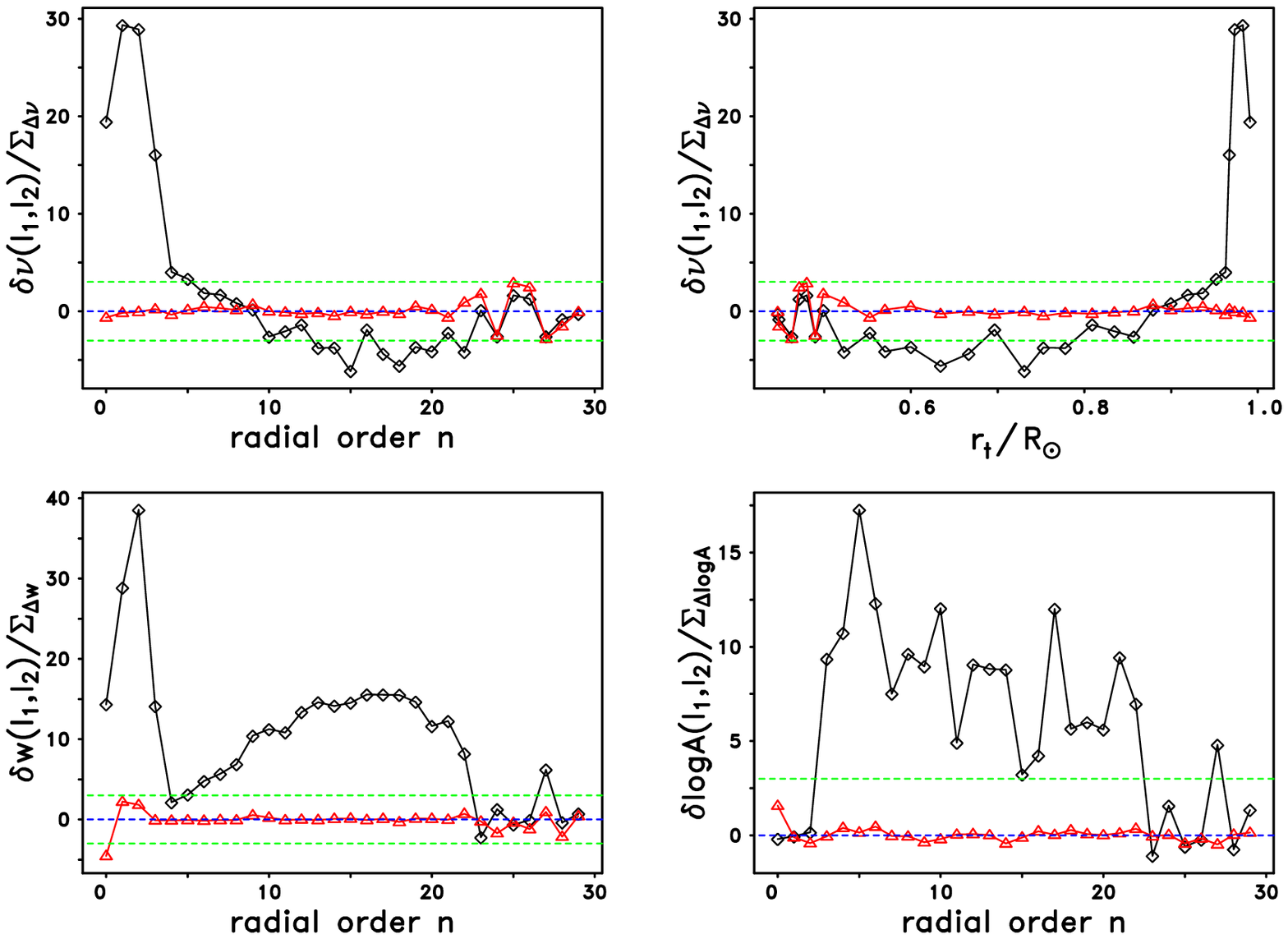}
\figcaption{
(upper-left) Radial order dependence of the normalized discontinuities in
Method~1 (black diamonds) and Method~2 (red triangles) frequencies obtained
from the $m$-averaged spectral set $\cal{S}$2010\_66a. These normalized
frequency discontinuities, $\delta\nu(l_1,l_2)/ \varSigma_{\Delta\nu}$, were
computed for the ridges $n=0$ through $n=29$ using the procedure that is
described in Section~\ref{cmpm1m2}. The degree ``$l_1$" in all four panels is
the degree of the highest-degree mode that we were able to fit for a given
ridge using the narrow fitting range in Method~1, while the degree ``$l_2$"
represents the degree of the lowest-degree mode that we were able to fit for
the same ridge using the wide fitting range in that method. In all four panels
the two dashed green lines represent normalized frequency discontinuities of
$\pm3$, while the dashed blue line is plotted for normalized frequency
discontinuities of zero. (upper-right) The same two sets of normalized
frequency discontinuities that were shown in the upper-left panel are shown
here as functions of the fractional inner turning-point radius, $r_{\rm
t}/R_{\odot}$, of the $l_1$ mode for each ridge. Here, $R_{\odot}$ is the
radius of the Sun. (lower-left) Same as upper-left panel, but for the
normalized discontinuities in the linewidth $w$. (lower-right) Same as
upper-left panel, but for the normalized discontinuities in the logarithm of
the amplitude $A$.
\label{m1ridgejumps}}
}
\end{figure}

In Figure~\ref{m1ridgejumps} we demonstrate that the replacement of the narrow
fitting range with the wide fitting range in Method~1 did not just introduce
distinct discontinuities in the fitted frequencies, linewidths, and amplitudes
for the $n=0$ and $n=1$ ridges, but rather introduced similar discontinuities
in all three quantities for nearly all of the ridges that we fit. We found it
useful to measure these discontinuities in terms of the normalized
discontinuities defined as
\begin{equation}
\delta\varUpsilon(l_1,l_2)/\varSigma_{\Delta\varUpsilon},
\label{dje1}
\end{equation}
where $\varUpsilon$ can be either frequency $\nu$, linewidth $w$, or else the
logarithm of the amplitude $A$, and where 
\begin{equation}
\delta\varUpsilon(l_1,l_2)=
\Delta\varUpsilon(l_2)-\Delta\varUpsilon(l_1)=\varUpsilon^{\rm \,fit}(l_2)-\varUpsilon^{\rm \,seed}(l_2)-
\left[\varUpsilon^{\rm \,fit}(l_1)-\varUpsilon^{\rm \,seed}(l_1)\right], 
\label{dje2}
\end{equation}
with $l_1$ being the degree of the highest-degree mode that we fit using the
narrow fitting range for a given ridge, with $l_2$ being the degree of the
lowest-degree mode that we fit using the wide fitting range for the same ridge,
and with $\varUpsilon^{\rm \,fit}$ and $\varUpsilon^{\rm \,seed}$ denoting,
respectively, the fitted and the corresponding seed value of $\varUpsilon$, and
where $\varSigma_{\Delta\varUpsilon}$ is the point-to-point scatter in the
$\Delta\varUpsilon$ values for the same ridge, as computed from
equation~(\ref{A3}) using 
\begin{equation}
\upsilon\equiv\Delta\varUpsilon/\Delta l,
\label{dje3}
\end{equation}
with $\Delta\varUpsilon=\varUpsilon^{\rm \,fit}-\varUpsilon^{\rm \,seed}$.
The use of the differences $\varUpsilon^{\rm \,fit}-\varUpsilon^{\rm \,seed}$
rather than the fitted values, $\varUpsilon^{\rm \,fit}$, themselves in the
above equations~(\ref{dje2}) and (\ref{dje3}) has the advantage that the
changes $\Delta\varUpsilon/\Delta l$ are approximately zero along the ridge
under consideration and, thus, gives rise to an unbiased value of the scatter,
$\varSigma_{\Delta\varUpsilon}$, as pointed out in Appendix~\ref{A3}. We note
that the values of $\varUpsilon^{\rm \,seed}$ that we have employed are
smoothly varying with degree, and we also note that we used a different pair of
$l_1$ and $l_2$ values for each of the analyzed ridges, depending on the degree
at which the narrow fitting range is replaced with the wide fitting range in
Method~1.

%Fig. 18

In the upper-left panel of Figure~\ref{m1ridgejumps} we show the normalized
discontinuities, $\delta\nu(l_1,l_2)/ \varSigma_{\Delta\nu}$, in both the
Method~1 and Method~2 frequencies in dependence of the radial order $n$. These
normalized discontinuities were computed from equations~(\ref{dje1}) through
(\ref{dje3}) using $\varUpsilon\equiv\nu$, with $\nu$ denoting the fitted
frequency from either Method~1 or Method~2. The hereto required values of the
seed frequency, $\nu^{\rm seed}$, were taken from a suitable seed table (cf.
Section~\ref{thfp}). In order to help indicate which of the normalized
frequency discontinuities represent statistically significant values, we have
included the two dashed green lines at ratios of $\pm 3$. Inspection of the
upper-left panel of Figure~\ref{m1ridgejumps} shows that for Method~1 (black
diamonds) 14 of the 30 normalized frequency discontinuities exceeded $\pm
3\,\sigma$. By contrast, of the 30 normalized frequency discontinuities that we
generated using Method~2, and which are shown as the set of red triangles in
the upper-left panel of Figure~\ref{m1ridgejumps}, none of them exceeded $\pm
3\,\sigma$. Clearly, the multiple-peak profile that we are using in Method~2
has allowed us to overcome the discontinuities that we were forced to introduce
into the fitted frequencies when we were using the single-peak profile of
Method~1.

The introduction of the jumps into the Method~1 frequencies that corresponded
to the change-over in the fitting range that we had to employ with Method~1
meant that those frequencies could not be employed in structural inversions
since these jumps would introduce unphysical oscillations in the radial profile
of any quantity determined in such inversions, e.g., internal sound speed. We
have re-plotted the normalized Method~1 frequency discontinuities that are
shown in the upper-left panel of Figure~\ref{m1ridgejumps} as a function of the
inner turning-point radius of the $l_1$ mode in the upper-right panel of
Figure~\ref{m1ridgejumps}. It is clear from this panel that one group of
significant Method~1 frequency discontinuities corresponds to modes that span
the lower portion of the convection zone, while the other group of
statistically significant discontinuities corresponds to modes that span the
sub-surface shear layer. On the other hand, the fact that none of the 30
normalized frequency discontinuities computed using Method~2 exceeded $\pm
3\,\sigma$ means that this method is capable of producing frequencies that can
be employed in successful structural inversions, as we will show later in
section~\ref{hinv}.

For the linewidths, as shown in the lower-left panel of
Figure~\ref{m1ridgejumps}, 23 of the 30 normalized Method~1 discontinuities,
$\delta w(l_1,l_2)/ \varSigma_{\Delta w}$, exceeded $\pm 3\,\sigma$, while only
the $n=0$ value exceeded $\pm 3\,\sigma$ for Method~2. For the logarithm of the
amplitudes, as shown in the lower-right panel of Figure~\ref{m1ridgejumps}, 21
of the 30 normalized Method~1 discontinuities, $\delta\log A(l_1,l_2)/
\varSigma_{\Delta\log A}$, exceeded $\pm 3\,\sigma$. However, none of the
normalized Method~2 discontinuities exceeded $\pm 3\,\sigma$. Clearly, Method~2
is able to produce both linewidths and amplitudes which do not exhibit the
systematic discontinuities that Method~1 introduced into both quantities.

\begin{figure}
\epsscale{0.48}
\plotone{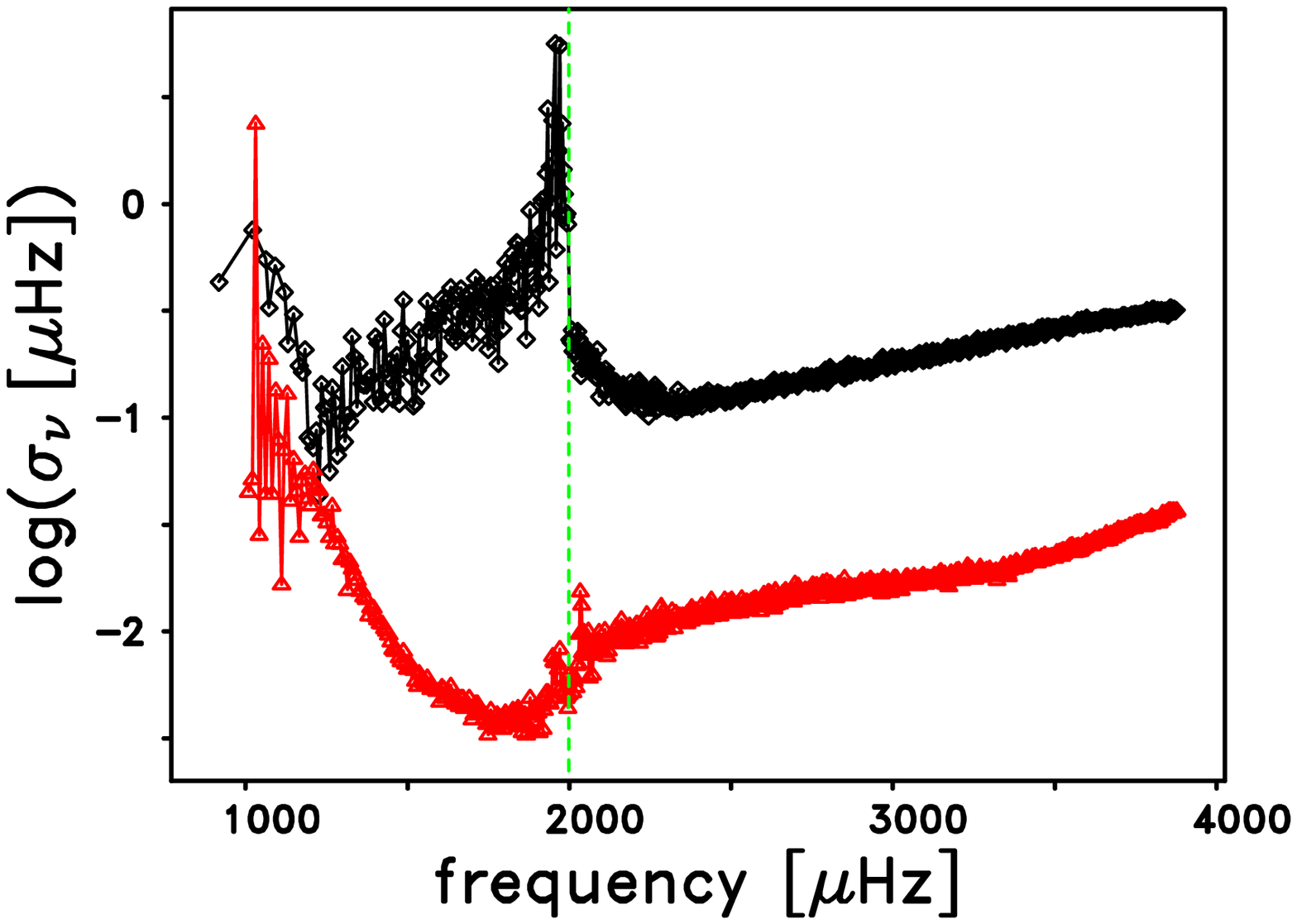}\hspace{0.0mm}
\plotone{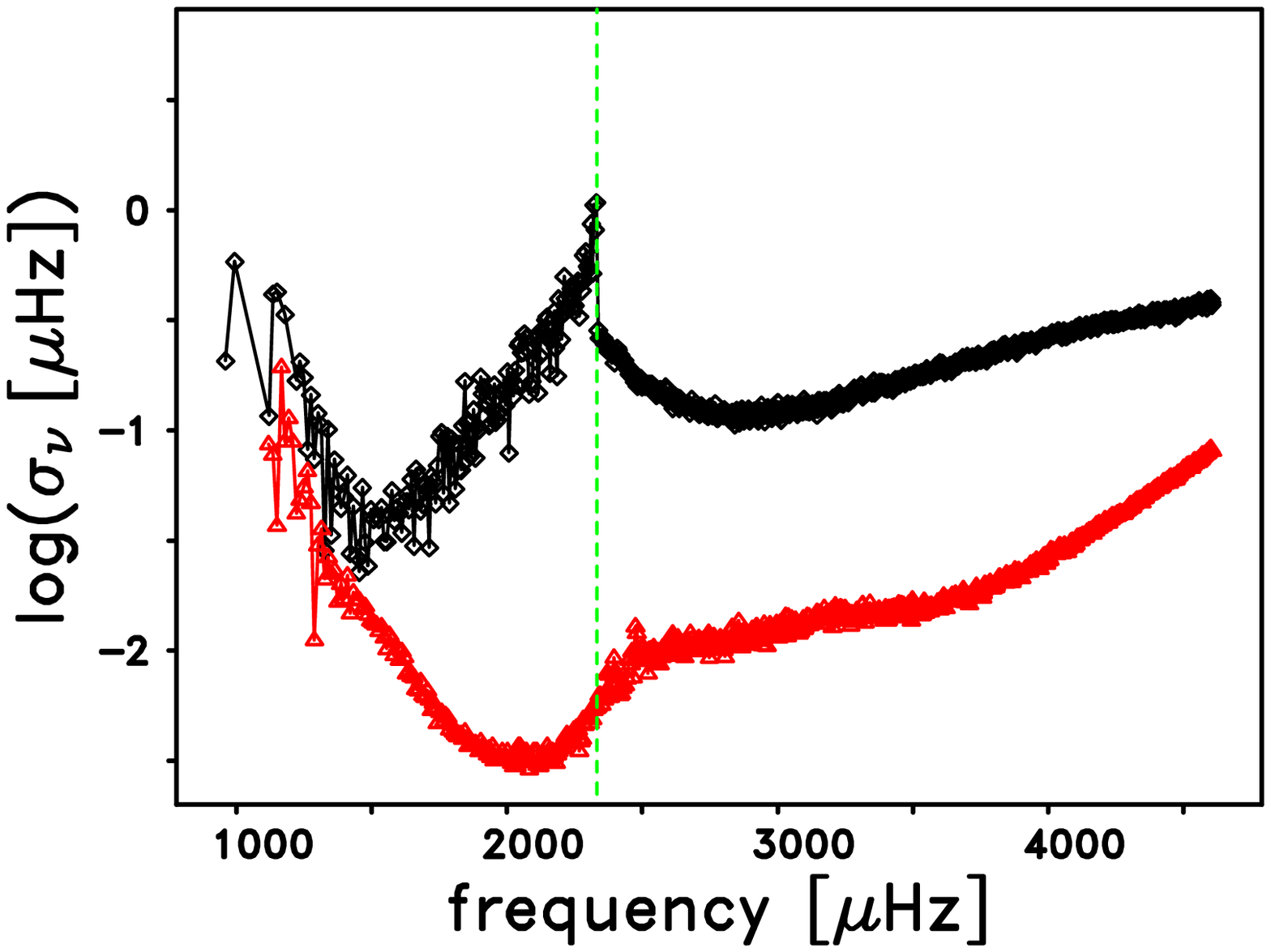}
\caption{
(left) The frequency dependence of the logarithm of the Method~1 frequency
uncertainties for the $n=1$ ridge is shown as the black curve, while the
frequency dependence of the logarithm of the corresponding Method~2
uncertainties is shown as the red curve. The vertical dashed line denotes the
frequency where the wide fitting range replaced the narrow fitting range in
Method~1. (right) Same as the left panel, except that the two sets of frequency
uncertainties were for the $n=2$ ridge. Both sets of these frequency
uncertainties resulted from fits to the $m$-averaged spectral set
$\cal{S}$2010\_66a.
\label{sigjumps}}
\end{figure}

In addition to all of the discontinuities in the frequencies, linewidths, and
amplitudes, Method~1 also introduced similar discontinuities into the frequency
uncertainties, as we show in both panels of Figure~\ref{sigjumps}. In these
panels we show the frequency dependence of the logarithm of the Method~1
frequency uncertainties that were computed from the fits to the $m$-averaged
spectral set $\cal{S}$2010\_66a as the black curves. The left-hand panel is for
the $n=1$ ridge, while the right-hand panel is for the $n=2$ ridge. There are
clearly discontinuities in both curves at the locations of the vertical dashed
lines, which are located at the frequencies of the transition from the narrow
fitting range to the wide fitting range for both ridges. By contrast, the
corresponding frequency uncertainties that we computed for the same power
spectra for both of these ridges using Method~2 are shown as the two red
curves. It is clear that neither of these curves contains a discontinuity at
the location of the vertical lines.

%Fig. 19

\subsection{Sample results from Method~2\label{srm2}}

The results that we will now describe were obtained from fitting the
$m$-averaged spectral sets $\cal{S}$2010\_66a and $\cal{S}$2010\_03 using
Method~2. These two spectral sets have been computed from the 66-day long
observing run $\cal{R}$2010\_66 and from the 3-day long observing run
$\cal{R}$2010\_03, respectively. The rationale behind the fitting of spectra
from both a 66-day long time series and a 3-day long time series is to
investigate the impact of spectral resolution upon the performance of Method~2.
In each of these two sets of $m$-averaged spectra we considered modes in the
range $0\le l\le 1000$, $0\leq n\leq 29$, $900\leq\nu\leq 7000\,\,\mu$Hz. In
this range of frequencies, radial orders, and degrees the seed table that we
had available to us contained a total of 12,533 entries.

In the low-degree, low-frequency portion of the dispersion plane, the mode
amplitude becomes comparable to the background noise, and at the same time the
mode linewidth becomes smaller than the spectral resolution. This can easily
lead to the confusion of a noise spike with a modal peak. By a careful
inspection of the quality of the fits we therefore set, for each radial order,
$n$, a lower degree limit, $l_{\rm min}^{(n)}$, below which we did not try to
fit a mode for that ridge. In a similar analysis of the high-degree,
high-frequency portion of the dispersion plane we defined, for each ridge, an
upper degree limit, $l_{\rm max}^{(n)}$, above which we did not try to fit a
mode for that ridge. Based upon this approach we came up with a total of 12,359
modes to be fit for the 66-day spectra, and a slightly smaller total of 12,242
modes to be fit for the 3-day spectra because of the lower signal-to-noise
ratios present in this set of spectra. We found that for all of the modes that
we attempted to fit in the case of the 66-day spectra and the 3-day spectra,
respectively, the Method~2 code converged on a solution. As a result, we ended
up with a total of 12,359 sets of fitted mode parameters for the 66-day
spectra, and a total of 12,242 sets of fitted mode parameters for the 3-day
spectra. This impressively demonstrates that Method~2 is working in a stable
manner over wide ranges of frequency, degree, and spectral resolution.

\begin{figure}
\epsscale{0.60}
\plotone{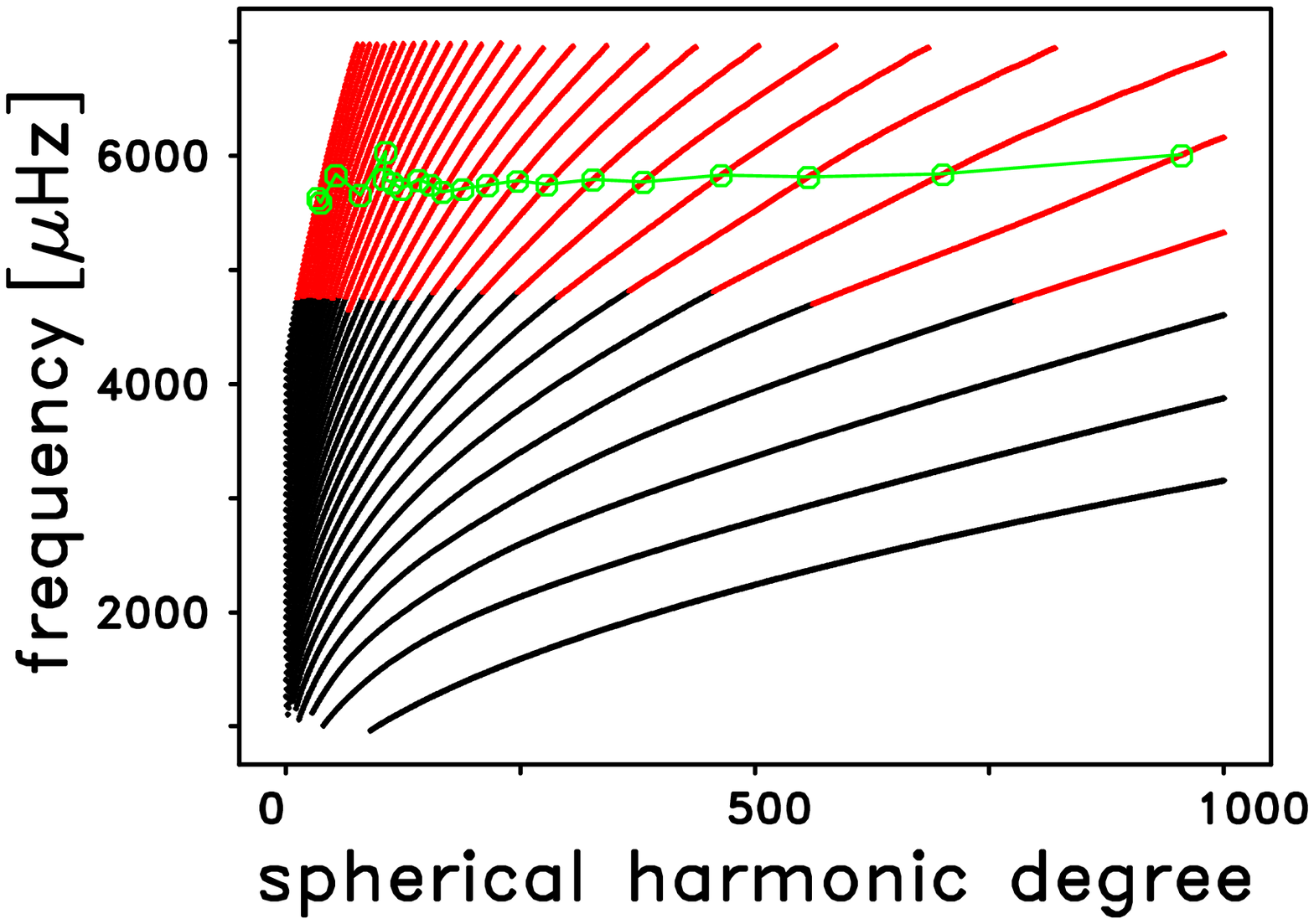}
\caption{
Dispersion plane coverage of the set of frequencies obtained with Method~2 when
applied to the $m$-averaged spectral set $\cal{S}$2010\_66a. In the range $0\le
l\le 1000$, $0\leq n\leq 29$, $965\leq\nu\le 7000\,\,\mu$Hz we were able to
successfully fit a total of 12,359 modes. The black diamonds are for fits
employing the asymmetric profile of \cite{Nig98}, while the red diamonds are
for those fits for which the symmetric Lorentzian profile was used. The
frequencies where the linewidths of most of the ridges exhibited either local
or global maxima are shown as the green open circles that are connected by the
green solid line. All of these frequencies can be seen to lie above the
corresponding frequencies where the fitting profiles were switched. The maxima
in the linewidths are shown as a function of frequency in the upper-left panel
of Figure~\ref{awpvsf}, while the maximum in the $n=14$ ridge is shown in the
upper-right panel of Figure~\ref{ridge14}.
\label{mdi66dnuvsl}}
\end{figure}

We collected the 12,359 frequencies and their corresponding uncertainties that
we generated with Method~2 from the $\cal{S}$2010\_66a $m$-averaged power
spectra into the frequency table $\cal{F}$2010\_66a (cf. Section~\ref{smavg}).
The dispersion plane coverage of this set of frequencies is illustrated in
Figure~\ref{mdi66dnuvsl}.  For the reasons explained in Section~\ref{sflp}, we
employed the symmetric Lorentzian profile for fitting the pseudo-modes. For all
of the cases for which the Lorentzian profile was used, the line asymmetry was
defined to be zero, i.e., $B=0$. Those cases are shown as the red diamonds in
Figure~\ref{mdi66dnuvsl}. For all of the other cases, which are shown as the
black diamonds, the asymmetric profile of \cite{Nig98} was employed.

%Fig. 20

\begin{figure}
\epsscale{1.000}
\plotone{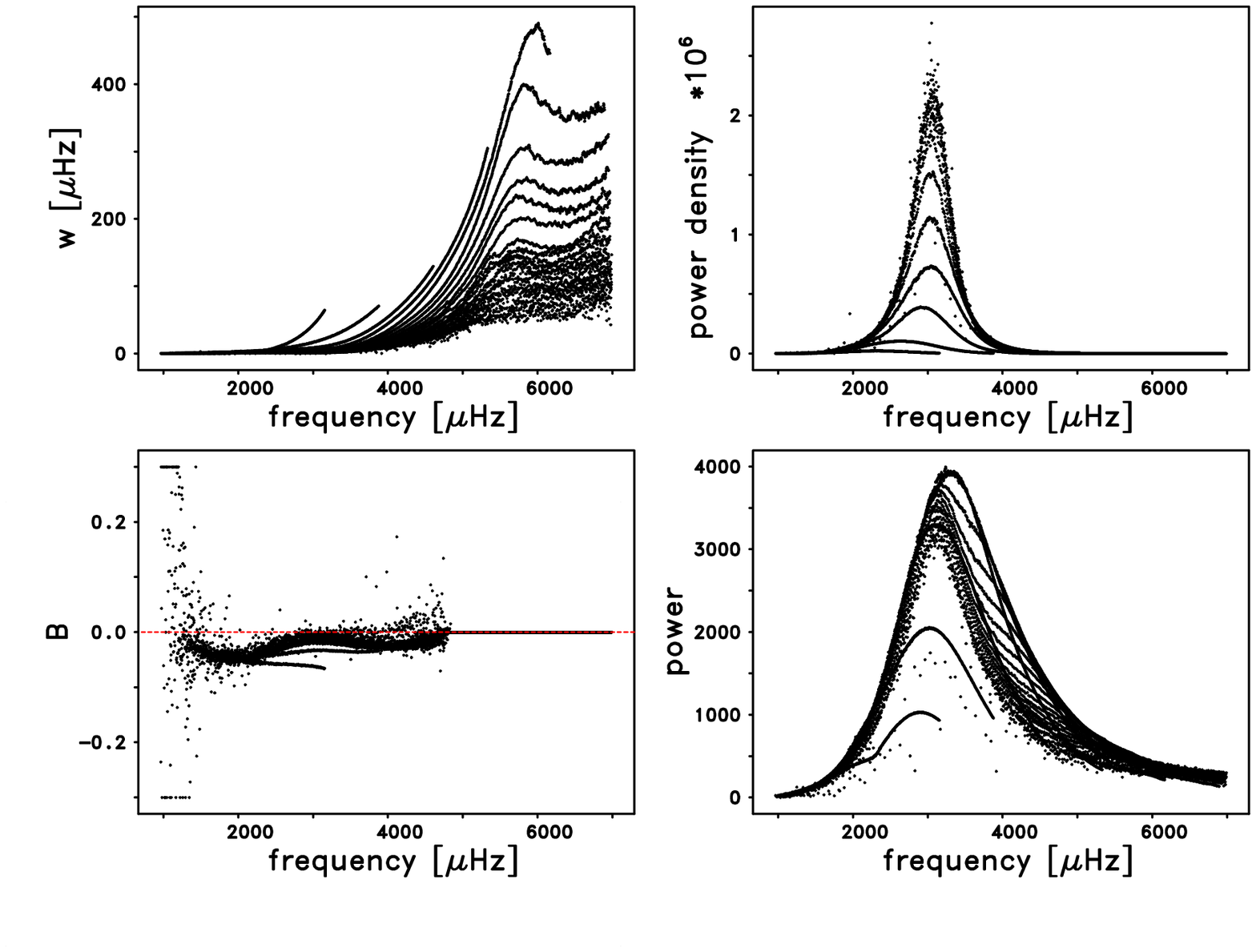}
\caption{The frequency dependencies of the linewidths (upper-left), line
asymmetries (lower-left), amplitudes, or power densities (upper-right), and
powers (lower-right) of the 12,359 fits that were made using Method~2 on the
$m$-averaged spectral set $\cal{S}$2010\_66a. The linewidths of many of the
ridges exhibit either local or global maxima near frequencies around 5800 to
$6000\,\,\mu$Hz before decreasing slightly at even higher frequencies. For $\nu
< 4638\,\,\mu$Hz, only the asymmetric profile was used and the vast majority of
the line asymmetries were less than zero. For reasons described in
Section~\ref{sflp}, for $4638<\nu<4842\,\,\mu$Hz, the asymmetric profile was
used for some of the ridges, while the symmetric Lorentzian profile was
employed for others. For $\nu>4842\,\,\mu$Hz only the symmetric profile was
used, so $B$ was set equal to zero for all of those cases. The dashed red line
in the lower-left panel is for a line asymmetry $B=0$.
\label{awpvsf}}
\end{figure}

In addition to the frequencies that we computed from the $\cal{S}$2010\_66a
$m$-averaged power spectra using Method~2, we show the corresponding linewidths
and line asymmetries in the upper-left and lower-left panels of
Figure~\ref{awpvsf}, respectively. The upper-left panel shows that the
linewidths of the higher-order ridges exhibited either local or global maxima
for frequencies between 5800 and $6000\,\,\mu$Hz before decreasing slightly at
higher frequencies. The frequencies where the linewidths of the ridges
exhibited either local or global maxima are shown as the green open circles in
Figure~\ref{mdi66dnuvsl} that are connected by the solid green line. All of
these frequencies can be seen to lie well above the range of frequencies where
the fitting profile was switched from an asymmetric to a symmetric profile,
which was $4638 < \nu < 4842\,\,\mu$Hz.

The maxima in the linewidths are very similar to peaks in the high-frequency
linewidths that were shown in Figure~4 of \cite{Duv91}. If we compare the
frequencies of the peaks in the linewidths in the degree range that they
employed (i.e., $11\le l \le 150$) with the frequencies as given by the green
line in Figure~\ref{mdi66dnuvsl}, we see that our frequencies are about
$200\,\,\mu$Hz larger than the frequencies given by \cite{Duv91}. This small
difference is most likely due to the different levels of solar activity in the
two observing epochs. The \cite{Duv91} observations were acquired in late 1987,
while the 66-day data set that we have employed, viz. $\cal{S}$2010\_66a, was
acquired in mid-2010. We find it intriguing that the solid green line that
connects the maxima in the linewidths in Figure~\ref{mdi66dnuvsl} looks very
similar to the degree dependence of the chromospheric modes as shown in
Figure~5 of \cite{Ulr77}; however, the frequencies of these linewidth maxima
are all much closer to the frequency of the second chromospheric mode, which
\cite{Ulr77} computed to be $5570\,\,\mu$Hz than they are to the frequency of
the first mode, which they computed to be $4138\,\,\mu$Hz. Furthermore, there
is no evidence in Figure~\ref{awpvsf} for the first chromospheric mode in the
linewidths, nor is there any evidence in Figure~\ref{mdi66dnuvsl} for either
chromospheric mode in the frequencies themselves. Since there is no other
evidence for either chromospheric mode in any of our frequency tables, the
origin of these local maxima in the linewidths is not clear, but is probably
related to the transition from standing waves below the acoustic cut-off
frequency to pseudo-modes above. The lower-left panel of Figure~\ref{awpvsf}
shows that most of the line asymmetries were negative, while they were zero by
construction for those cases for which the symmetric Lorentzian profile was
used in Method~2. 

The amplitudes, or power densities, as is shown in the upper-right panel of
Figure~\ref{awpvsf}, peak very close to $3000\,\,\mu$Hz, in agreement with all
previous studies of the solar oscillation amplitudes. In contrast, the power
values that are shown in the lower-right panel go through their peak at about
$3300\,\,\mu$Hz because the power values are computed from the products of the
amplitudes and the linewidths and the increase in the linewidths with
increasing frequency partially compensates for the decreasing amplitudes. As we
summarized at the beginning of this section, we also obtained similar sets of
linewidths, line asymmetries, amplitudes, and power values for the set of 3-day
power spectra in addition to the results we have shown in this section from the
66-day power spectra.

%Fig. 21

\begin{figure}
\epsscale{1.00}
\plotone{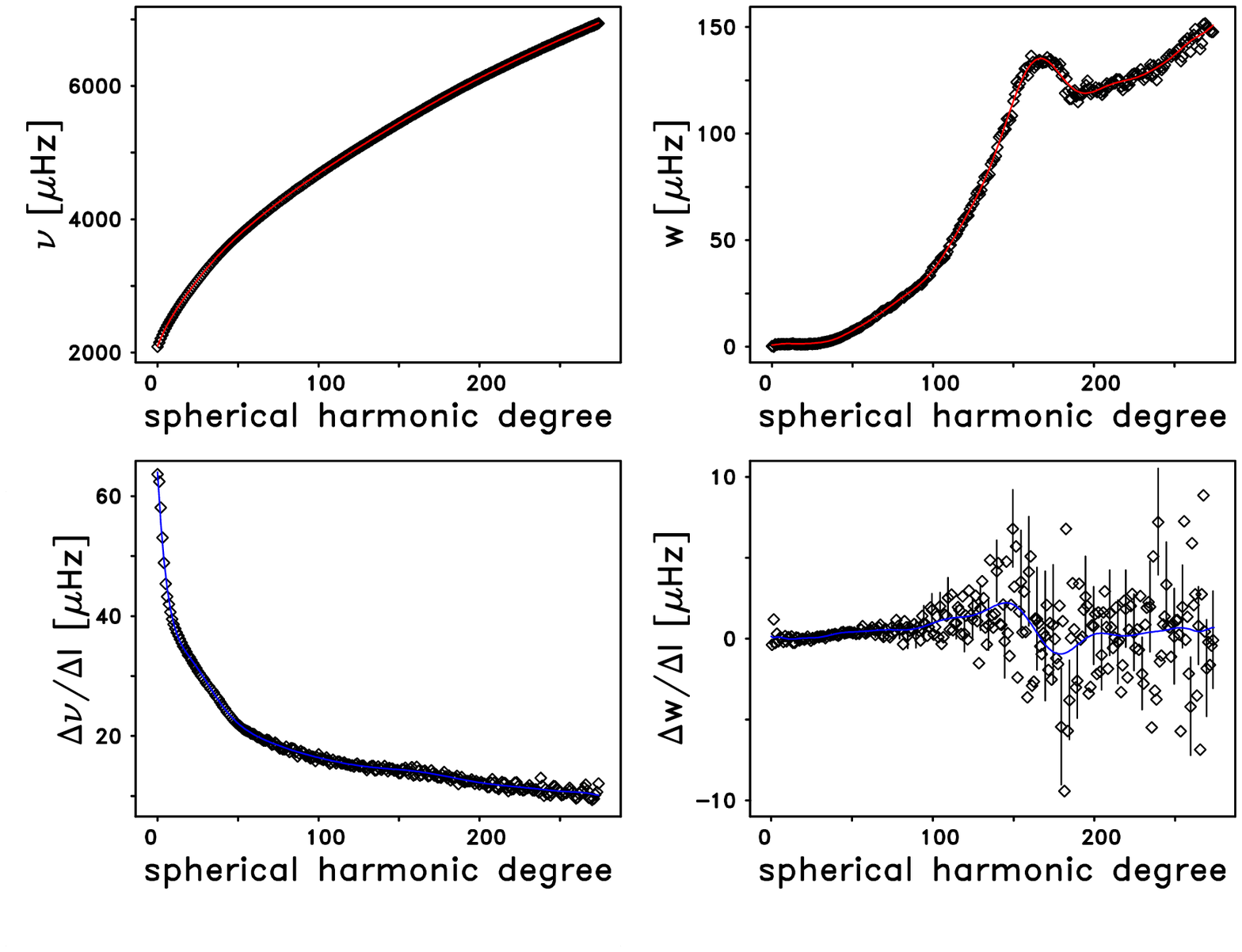}
\caption{
(upper-left) Degree dependence of frequency, $\nu$, for the $n=14$ ridge from
fitting the $m$-averaged spectral set $\cal{S}$2010\_66a using Method~2. The
black diamonds represent the fitted frequencies in the range $0\leq l\leq 274$.
The red line is for a smooth fit to these fitted frequencies. (lower-left)
Degree dependence of the frequency slopes, $\Delta\nu/\Delta l$, for the same
ridge. The values of $\Delta\nu/\Delta l$ are plotted as the black diamonds at
the abscissae $l+0.5$, and are computed by subtracting two consecutive values
of the fitted frequency shown in the upper-left panel, i.e., $\Delta\nu/\Delta
l|_{l+0.5} = \nu(l+1)-\nu(l)$. The blue line is the derivative with respect to
$l$ of the smooth curve that has been fitted to the frequencies and which is
shown as the red line in the upper-left panel. (upper-right) Degree dependence
of the modal linewidth, $w$. We note that the linewidth goes through a relative
maximum at $l\approx 167$ which corresponds to a frequency of $5686\,\,\mu$Hz.
This is an example of the linewidth maxima that were illustrated in
Figure~\ref{awpvsf}. (lower-right) Degree dependence of the linewidth slopes,
$\Delta w/\Delta l$, plotted as the black diamonds at the abscissae $l+0.5$.
They were computed by subtracting two consecutive values of the fitted
linewidth, i.e., $\Delta w/\Delta l|_{l+0.5} = w(l+1)-w(l)$. The blue line is
the derivative of the smooth curve that has been fitted to the linewidths and
which is shown as the red line in the upper-right panel.
\label{ridge14}} 
\end{figure}

The smoothness in both the fitted frequencies and linewidths that resulted from
the fits to the $\cal{S}$2010\_66a $m$-averaged power spectra using Method~2 is
illustrated in Figure~\ref{ridge14} using the example of the $n=14$ ridge. The
upper-left panel shows the degree dependence of the frequencies for that ridge,
and it also includes the smooth fit to those frequencies as the red curve. The
lower-left panel shows the degree dependence of the frequency slope,
$\Delta\nu/\Delta l$, that was computed as described in the figure caption,
and it also includes, as the blue curve, the derivative with respect to degree
of the red curve shown in the upper-left panel. The fact that the derivative
with respect to degree (blue line in the lower-left panel) of the smooth curve
(red line in the upper-left panel) fitted to the frequencies is an excellent
representation of the numerical values of $\Delta\nu/\Delta l$ (black diamonds
in lower-left panel) at all of the degrees shown, impressively demonstrates the
smoothness of the fitted frequencies (black diamonds in upper-left panel).
The upper-right panel of Figure~\ref{ridge14} shows the degree dependence of
the linewidths for the $n=14$ ridge, and it also includes the smooth fit to
those linewidths as the red curve. This panel shows that this ridge was one of
the ridges for which the linewidth exhibited a local maximum before decreasing
toward the high-degree ends of those ridges as we just described in
Figure~\ref{awpvsf}. The lower-right panel of Figure~\ref{ridge14} shows the
linewidth slopes, $\Delta w/\Delta l$, which were computed as described in the
figure caption, as a function of degree, and it also includes, as the blue
curve, the derivative of the smooth curve with respect to degree that is shown
as the red curve in the upper-right panel. As was the case for the frequencies,
the linewidths are also smooth enough that the derivative of the smooth fitting
curve provides an excellent representation to the values of $\Delta w/\Delta
l$.

%Fig. 22

As we noted in Section~\ref{thfp}, the derivatives with respect to degree of
the frequency, the linewidth, and the amplitude are an essential ingredient of
Method~2. We compute these derivatives on a ridge-by-ridge basis by fitting, as
a function of degree, higher-order Chebyshev polynomials to the frequencies,
linewidths, and amplitudes, respectively, as they have been determined with
Method~2, and by then differentiating the resulting polynomials with respect to
degree. The red curves in the upper panels of Figure~\ref{ridge14} are examples
of such fits to the frequency and linewidth, respectively, while the blue curves
shown in the lower panels are examples of the first derivative of the fitted
polynomials. Because higher-order polynomials are involved, this approach is
prone to deliver wildly oscillating fitting curves. Therefore, the degree of
the polynomials has to be selected rather carefully, and the least-squares fits
of the polynomials need to be constrained as to avoid oscillating curves.

\section{Helioseismic inversion for solar structure\label{hinv}}

In this Section we will present the results of an inversion for the radial
structure of the Sun using a subset of the frequencies that are contained in
our table $\cal{F}$2010\_66a. We have already mentioned in Section~\ref{srm2}
that this entire table contains a total of 12,359 sets of fitted modal
parameters, and we showed the dispersion plane coverage of this complete set of
frequencies in Figure~\ref{mdi66dnuvsl}. This is the largest mode-set to have
ever been fit thus far in helioseismology. To carry out our structural
inversion, we selected a subset of 6,366 frequencies which contained degrees
that ranged from 0 to 1000, radial orders from 0 to 29, and frequencies from
969 to $4500\,\,\mu$Hz.  Since this frequency upper limit of $4500\,\,\mu$Hz is
less than the frequency of the lowest symmetrical fit, which was
$4638\,\,\mu$Hz for the $n=17$ ridge, all of the frequencies in this subset
were computed using the asymmetric profile of \citet{Nig98}. By using the
Optimally Localized Averaging (OLA) technique we inverted this subset of modes
to determine the spherically symmetric structure of the Sun. We limited the
frequency range of the subset of modes by $4500\,\,\mu$Hz because this
threshold is well below the acoustic cut-off frequency which separates the
$p$-modes from the pseudo-modes and which is about $5000\,\,\mu$Hz. We note
that the choice of the upper frequency limit of the mode set to be inverted is
not critical since modes with frequency above $4000\,\,\mu$Hz do not contribute
significantly to the inversion results due to their large errors. Details of
the inversion procedure, including calculation of the sensitivity kernels and
test results, are presented by \cite{Kos99}.

While we had to select from our full mode-set a suitable subset of frequencies
for our structural inversion, we note, however, that tables of fitted mode
parameters covering a similar range of degrees and frequencies as our full
mode-set presented here, are an indispensable prerequisite for the study of
temporal changes in the sensitivities of the mode parameters to corresponding
changes in the levels of solar activity \citep[see,
e.g.,][]{Rab08a,Rho11,Rab11}.

\subsection{Analysis of systematic errors in fitted frequencies and associated
uncertainties \label{ocs}}

In an inversion for solar internal structure both the frequencies and the
associated uncertainties of the solar oscillation modes are employed as input
parameters. Therefore, systematic errors in these input parameters may affect
the resulting inferences of the Sun's internal structure. As we have explained
in Section~\ref{instreff}, such systematic errors may arise from instrumental
effects. However, they also are likely to arise from the methodology employed
for fitting the spectra. For example, the use of an inadequate fitting box
around the target peaks, an inadequate treatment of the background power within
the selected fitting boxes, and/or contributions from $n$- and $l$-leaks that
are not adequately accounted for in the fitting model profile, as given in
equations~(\ref{profm2a}) through (\ref{profm2f}), may lead to systematic
errors in both the fitted frequencies and the uncertainties thereof. On the
other hand, the presence of a group of outliers in a table of frequencies and
associated uncertainties that is designed to be used as input to a solar
structural inversion, may result in the appearance of unphysical oscillations
in the radial profiles of the resulting sound speed or other thermodynamic
quantities. Such a set of outliers may also have the unwanted side effect of
preventing a stable, regularized solution from being found outside of a very
narrow range of the regularization parameter that is being employed in the
inversion procedure. In order to avoid the appearance of such unphysical
oscillations in the inverted sound speed profile, we developed a five-step
procedure that allows us to substantially alleviate the issue of systematic
errors in both the fitted frequencies and their associated uncertainties in
tables that will be employed as input data sets for structural inversions. The
rather high complexity of this procedure mainly results from our desire to keep
the approach as objective as possible, while keeping unavoidable subjective
elements at a minimum.

The basic idea behind our correction scheme is to compare the observed
frequency for the $(n,l)$ mode, $\nu_{n,l}^{\rm obs}$, where the term
``observed frequency'' refers to our original set of fitted frequencies, with
the theoretical model frequency for the same mode, $\nu_{n,l}^{\rm mod}$, on a
ridge-by-ridge basis. For this comparison we employ two sets of frequency
differences between the observed frequencies, $\nu_{n,l}^{\rm obs}$, and the
model frequencies, $\nu_{n,l}^{\rm mod}$, viz. the set of unscaled differences,
\begin{equation}
\Delta\nu_{n,l}=\nu_{n,l}^{\rm obs}-\nu_{n,l}^{\rm mod}, 
\label{fdiffu}
\end{equation}
and the set of scaled differences
\begin{equation}
\delta\nu_{n,l}=(\nu_{n,l}^{\rm obs} - \nu_{n,l}^{\rm mod}) \,q_{n,l}
\left(\frac{\nu_{n,l}^{\rm mod}}{1000}\right)^{-3.5} \equiv
\Delta\nu_{n,l} \,q_{n,l} \left(\frac{\nu_{n,l}^{\rm mod}}{1000}\right)^{-3.5}.
\label{fdiffs}
\end{equation}
In equations~(\ref{fdiffu}) and (\ref{fdiffs}) $\nu_{n,l}^{\rm mod}$ is the
theoretical frequency of mode $(n,l)$ that has been computed from Model~S of
\cite{jcd96} employing the approach described by \cite{Kos99}, and $q_{n,l}$ is
the inertia of mode $(n,l)$, normalized to the inertia of mode $(n,0)$. We
note that both $\nu_{n,l}^{\rm mod}$ and $q_{n,l}$ are computed using the same
solar model that is also used to compute the kernel functions employed in the
structural inversion. The rationale behind the use of two different sets of
frequency differences, $\Delta\nu_{n,l}$ and $\delta\nu_{n,l}$, respectively,
is to exploit the fact that, for a given ridge, the run of $\Delta\nu_{n,l}$
with respect to degree differs markedly from that of $\delta\nu_{n,l}$. This
allows the identification of outliers that would otherwise remain undetected if
only one set of frequency differences is used. The rationale behind the definition of the
scaled frequency differences, $\delta\nu_{n,l}$, is to remove the frequency
differences between the solar model and the Sun that come from the dominant
near-surface effects \citep[e.g.,][]{jcd84} by approximating these differences
using a frequency power-law function scaled with the mode inertia.

Before we delve into the details of our correction scheme, we briefly summarize
its major steps. First, for each ridge we fit, as a function of degree $l$, a
Chebyshev polynomial, $\varXi_{\varpi}^{(n)}(l)$, of degree $\varpi$ to the
unscaled frequency differences, $\Delta\nu_{n,l}$, and a Chebyshev polynomial,
$\varPi_{\pi}^{(n)}(l)$, of degree $\pi$ to the scaled frequency differences,
$\delta\nu_{n,l}$. Although both $\varXi_{\varpi}$ and $\varPi_{\pi}$ are
Chebyshev polynomials of the same kind but generally of different degree, we
have introduced different symbols for them in order to highlight that they are
used for the fitting of different sets of frequency differences.

Next, from the two fit polynomials, $\varXi_{\varpi}^{(n)}(l)$ and
$\varPi_{\pi}^{(n)}(l)$, respectively, we generate, for each ridge, two sets of
emulated (imitated) frequencies, $\nu_{n,l}^{\rm em1}$ and $\nu_{n,l}^{\rm
em2}$, respectively. Here, $\nu_{n,l}^{\rm em1}$ corresponds to an observed
frequency that would fall exactly onto the fitted curve
$\varXi_{\varpi}^{(n)}(l)$, while $\nu_{n,l}^{\rm em2}$ corresponds to an
observed frequency that would fall exactly onto the fitted curve
$\varPi_{\pi}^{(n)}(l)$. At this point we have five sets of frequency
differences at our disposal, namely,
\begin{align}
\Delta_{n,l}^{(1)} &= \Delta\nu_{n,l}-\varXi_{\varpi}^{(n)}(l),\\
\Delta_{n,l}^{(2)} &= \delta\nu_{n,l}-\varPi_{\pi}^{(n)}(l),\\
\Delta_{n,l}^{(3)} &= \nu_{n,l}^{\rm obs}-\nu_{n,l}^{\rm em1},\\
\Delta_{n,l}^{(4)} &= \nu_{n,l}^{\rm obs}-\nu_{n,l}^{\rm em2},\\
\Delta_{n,l}^{(5)} &= \nu_{n,l}^{\rm em1}-\nu_{n,l}^{\rm em2}, 
\end{align}
from which we identify outlying modes by applying suitable statistical criteria.

Next, we refit, by means of the WMLTP method employing updated seed tables,
all of the modes that have been flagged as outliers.

Finally, we use the refitted frequencies and associated uncertainties to
update all of the outlying modes. In some cases we only update the frequencies
of those outliers and in other cases we update only the associated
uncertainties, and in still other cases we update both the frequencies and
their uncertainties at the same time. We also encountered a few cases, however,
for which neither the frequency nor the uncertainty needed to be altered.

\subsubsection{Step 1\label{ocsstp1}}

\begin{figure} 
\epsscale{0.450} 
\plotone{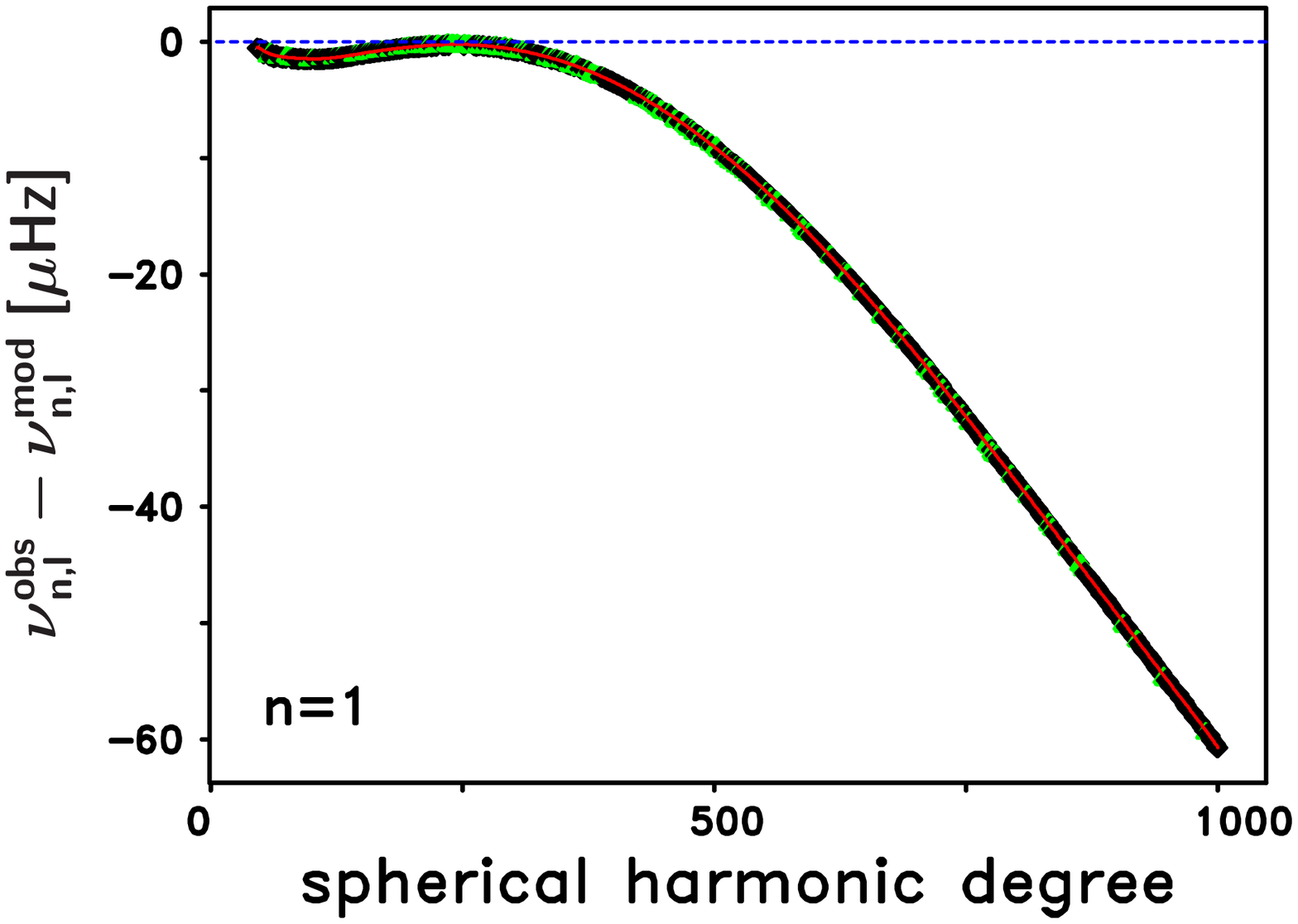}
\plotone{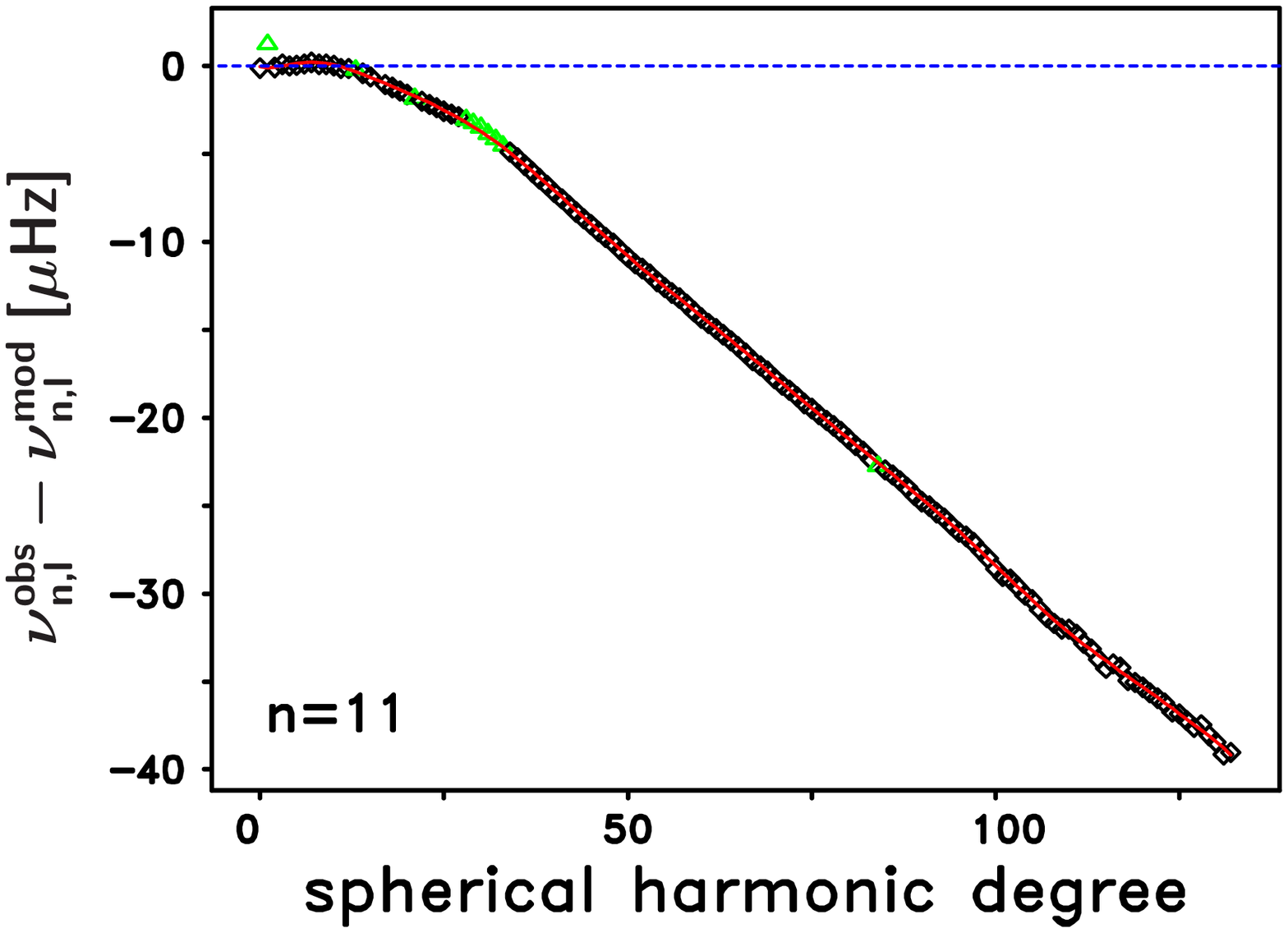} 
\caption{(left) Degree dependence of the differences between the observed
frequencies, $\nu_{n,l}^{\rm obs}$, which were computed using Method~2 on the
$m$-averaged spectral set $\cal{S}$2010\_66a, and the theoretical model
frequencies, $\nu_{n,l}^{\rm mod}$, in the sense of Sun minus model, for the
$n=1$ ridge. The frequency differences, $\nu_{n,l}^{\rm obs}-\nu_{n,l}^{\rm
mod}$, are shown as the black diamonds. The red line is for the Chebyshev
polynomial, $\varXi_{\varpi}^{(n)}(l)$, of degree $\varpi=24$ fitted to the
frequency differences. The 276 green triangles represent modes flagged as
outliers. We note that these frequency differences are the same that were shown
as the red curve in the upper-right panel of Figure~\ref{dfm1ms}. (right) Same
as left panel, but for the $n=11$ ridge. For this ridge the degree of the
fitted Chebyshev polynomial, $\varXi_{\varpi}^{(n)}(l)$, is $\varpi=14$, and 10
modes were flagged as outliers. In both panels the dashed blue line is for a
frequency difference of zero. Note that the vertical scales are different on
the two panels.
\label{fdiffvsl}}
\end{figure}

In the first step of our outlier correction scheme we fit, in the least-squares
sense, the Chebyshev polynomial, $\varXi_{\varpi}^{(n)}(l)$, of degree $\varpi$
to the unscaled frequency differences, $\Delta\nu_{n,l}$, as defined in
equation~(\ref{fdiffu}), along a ridge of given radial order $n$, i.e., as a
function of degree $l$. Typical examples are shown here in
Figure~\ref{fdiffvsl} for the $n=1$ and $n=11$ ridge, respectively. In both
panels of Figure~\ref{fdiffvsl} the unscaled frequency differences,
$\Delta\nu_{n,l}$, are shown as the black diamonds, while the fitted Chebyshev
polynomials, $\varXi_{\varpi}^{(n)}(l)$, are shown as the red lines. The degree
$\varpi$ of the polynomials $\varXi_{\varpi}^{(n)}(l)$ depends on the radial
order, $n$, and is determined by trial and error. In doing so, two requirements
must be considered. On the one hand, the degree $\varpi$ needs to be large
enough so that the fitted polynomial $\varXi_{\varpi}^{(n)}(l)$ provides a
reasonably good fit to the data. On the other hand, it must be small enough to
avoid unrealistic oscillations in the fitted curve. For the ridges in the range
$0\leq n \leq 29$ we found values of $\varpi$ in the range $1 \leq\varpi\leq
26$ to be useful. In particular, we found that $\varpi$ decreased as the order
of the ridge increased. Unfortunately, so far we have not succeeded in finding
any pre-set criteria which would allow the determination of the degree $\varpi$
in an automated fashion.

If for any mode along the ridge, $n$, the criterion
\begin{equation}
\left|\Delta\nu_{n,l} - \varXi_{\varpi}^{(n)}(l)\right| >
3 \,\sqrt{\left(\Delta\nu_{n,l}^{\rm obs}\right)^2 + 
\left(\Delta\varXi_{\varpi}^{(n)}(l)\right)^2},
\label{outla} 
\end{equation} 
is fulfilled, the mode $(n,l)$ is flagged as an outlier. Here,
$\Delta\nu_{n,l}$ is the unscaled frequency difference, $\Delta\nu_{n,l}^{\rm
obs}$ is the uncertainty of the observed frequency $\nu_{n,l}^{\rm obs}$,
$\varXi_{\varpi}^{(n)}(l)$ is the fitted polynomial, and
$\Delta\varXi_{\varpi}^{(n)}(l)$ is the uncertainty thereof. While
$\Delta\nu_{n,l}^{\rm obs}$ is calculated by default within the WMLTP fitting
code as described in Appendix~\ref{secm1}, $\Delta\varXi_{\varpi}^{(n)}(l)$ is
computed from 
\begin{equation}
\Delta\varXi_{\varpi}^{(n)}(l)=\sqrt{\mbox{var}\,\varXi_{\varpi}^{(n)}(l)}\,\,
t_{1-\gamma/2,\lambda-\varpi-1}, \label{unufit} 
\end{equation} 
where $t_{1-\gamma/2,\lambda-\varpi-1}$ is the $100\,(1-\gamma/2)$ percentage
point of Student's t-distribution with $\lambda-\varpi-1$ degrees of freedom,
$\lambda$ being the number of data points involved in the fit of
$\varXi_{\varpi}^{(n)}(l)$ to the frequency differences, $\Delta\nu_{n,l}$. As
mentioned in Appendix~\ref{secm1}, for $1\,\sigma$ confidence intervals
$\gamma=0.31731$. If we denote the coefficients of the Chebyshev polynomial
$\varXi_{\varpi}^{(n)}(l)$ by $x_i$, $i=0,\ldots,\varpi$, the variance of
$\varXi_{\varpi}^{(n)}(l)$ is given by 
\begin{equation} 
\mbox{var}\,\varXi_{\varpi}^{(n)}(l) =
\frac{2\,S}{\lambda-\varpi-1} \sum_{i=0}^{\varpi} \sum_{j=0}^{\varpi}
\frac{\partial\varXi_{\varpi}^{(n)}}{\partial
x_i}\frac{\partial\varXi_{\varpi}^{(n)}}{\partial x_j} H_{ij}^{-1}
\end{equation} 
\citep{Nag11}. Here, $S$ is the sum of squares, and $H_{ij}^{-1}$ denotes the
elements of the inverse of the Hessian Matrix.

Once an optimum value of $\varpi$ has been found for a given ridge, the
outliers are flagged by means of the criterion~(\ref{outla}), and a new
polynomial $\varXi_{\varpi}^{(n)}(l)$ is fit to the remaining points using the
same value of $\varpi$. By means of this refitted polynomial an updated set of
outliers is now identified, and the entire process is repeated  until all of
the outliers have been identified for that ridge. 

For further analyses we need to generate a set of emulated frequencies 
\begin{equation} 
\nu_{n,l}^{\rm em1} = \nu_{n,l}^{\rm mod} + \varXi_{\varpi}^{(n)}(l),
\label{nuem1} 
\end{equation} 
which correspond to observed frequencies that would fall exactly onto the curve
$\varXi_{\varpi}^{(n)}(l)$, for a given ridge. Because the theoretical model
frequencies, $\nu_{n,l}^{\rm mod}$, are error-free, the uncertainty of
$\nu_{n,l}^{\rm em1}$ is given by 
\begin{equation} 
\Delta\nu_{n,l}^{\rm em1} = \Delta\varXi_{\varpi}^{(n)}(l),
\label{unuem1} 
\end{equation} 
with $\Delta\varXi_{\varpi}^{(n)}(l)$ being given in equation~(\ref{unufit}).

%Fig. 23

\subsubsection{Step 2}

\begin{figure}
\afterpage{
\epsscale{1.000}
\plotone{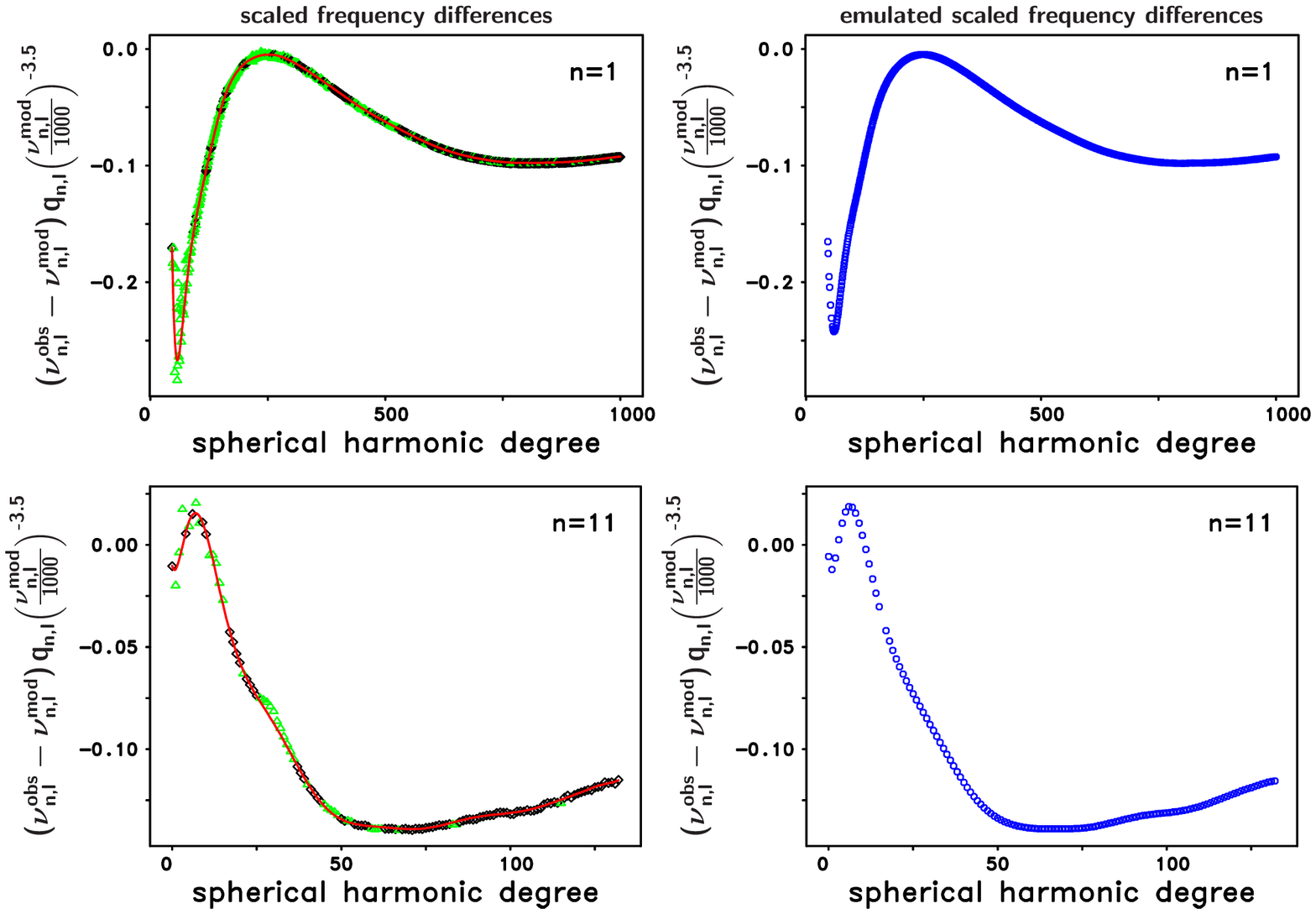}
\figcaption{
(upper-left) Degree dependence of the scaled frequency differences,
$\delta\nu_{n,l}$, as defined in equation~(\ref{fdiffs}), for the $n=1$ ridge.
The black diamonds are for the scaled frequency differences, while the red line
is for the Chebyshev polynomial, $\varPi_{\pi}^{(n)}(l)$, of degree $\pi=23$
that has been fitted to the scaled frequency differences. In this panel all of
the 435 outliers that were identified in step~1 (276), and in steps 2 and 3 (an
additional 159) are shown as the green triangles. (lower-left) Same as
upper-left panel, but for the $n=11$ ridge. For this ridge the degree of the
fitted Chebyshev polynomial, $\varPi_{\pi}^{(n)}(l)$, is $\pi=14$, and the 39
outliers that were detected in step~1 (10), and in steps 2 and 3 (an additional
29) are shown as the green triangles. (upper-right) Degree dependence of the
emulated scaled frequency differences, $\delta\nu_{n,l}^{\rm em}$, as defined
in equation~(\ref{emsres}). A comparison of the panels in the upper row
indicates that the trendline of the emulated scaled frequency differences,
$\delta\nu_{n,l}^{\rm em}$, as shown in the upper-right panel, is in excellent
agreement with the trendline of the curve fitted to the scaled frequency
differences, $\delta\nu_{n,l}$, as shown in the upper-left panel. Therefore we
can presume that the sharp variation of the scaled frequency differences,
$\delta\nu_{n,l}$, with degree at around $l\approx 60$ is a real feature.
(lower-right) Same as upper-right panel, but for the $n=11$ ridge. The march of
the fitted polynomial as shown in the lower-left panel is in excellent
agreement with the march of the emulated scaled frequency differences,
$\delta\nu_{n,l}^{\rm em}$, as shown in the lower-right panel. Hence, we can
regard the sharp variations with degree of the scaled frequency differences,
$\delta\nu_{n,l}$, at about $l\approx 1$ and $l\approx 7$, respectively, as
real features.
\label{fitresvsl}}
}
\end{figure}

In the second step of our outlier correction scheme we fit the Chebyshev
polynomial, $\varPi_{\pi}^{(n)}(l)$, of degree $\pi$ to the scaled frequency
differences, $\delta\nu_{n,l}$, as defined in equation~(\ref{fdiffs}), along a
ridge of given radial order, $n$, i.e., as a function of degree $l$. We note
that the outliers that were identified during step~1 are not used in the
fitting of the $\varPi_{\pi}^{(n)}(l)$ polynomials to the scaled frequency
differences, $\delta\nu_{n,l}$. Typical examples of fits to the scaled
frequency differences are shown here in the left-hand panels of
Figure~\ref{fitresvsl} for the $n=1$ and $n=11$ ridge, respectively. In both of
these left-hand panels of Figure~\ref{fitresvsl} the scaled frequency
differences, $\delta\nu_{n,l}$, are shown as the black diamonds, and the fitted
Chebyshev polynomials, $\varPi_{\pi}^{(n)}(l)$, are shown as the red lines.
The degree $\pi$ of the polynomials $\varPi_{\pi}^{(n)}(l)$ depends on the
radial order, $n$, and is determined by trial and error. For the ridges in the
range $0\leq n \leq 29$ we found values of $\pi$ in the range $1 \leq\pi\leq
29$ to be useful. Rather than simply decreasing with increasing $n$, as was
the case for $\varpi$, $\pi$ increased from 19 at $n=0$ to 29 at $n=4$ before
decreasing to 1 at $n=29$. As with the degree $\varpi$, we have not yet been
able to find any approach that would allow us to determine the degree $\pi$ of
the polynomials $\varPi_{\pi}^{(n)}(l)$ in an automated manner.
 
Next in this second step, we determine, for each ridge of radial order, $n$,
the fitting residuals, $r_l^{(n)}$, of the fit of $\varPi_{\pi}^{(n)}(l)$ to
the scaled frequency differences, $\delta\nu_{n,l}$, i.e.,
\begin{equation}
r_l^{(n)} = \delta\nu_{n,l} - \varPi_{\pi}^{(n)}(l).
\label{fres}
\end{equation}
We flag those modes $(n,l)$ as outliers for which the criterion
\begin{equation}
\big|r_l^{(n)}\big| > 3\,\,\mbox{std}\big(r_l^{(n)}\big)
\label{outlb}
\end{equation}
is fulfilled. Here, $\mbox{std}\big(r_l^{(n)}\big)$ denotes the standard
deviation of the fitting residuals, $r_l^{(n)}$, for the ridge of radial order,
$n$. For each ridge, $n$, the fit of the polynomial, $\varPi_{\pi}^{(n)}(l)$,
to the scaled frequency differences, $\delta\nu_{n,l}$, is repeated until all
outliers have been identified by means of the criterion (\ref{outlb}). 

For further analyses we will need to generate yet another set of emulated
frequencies, viz.
\begin{equation}
\nu_{n,l}^{\rm em2} = \nu_{n,l}^{\rm mod} + \frac{1}{q_{n,l}} 
\left(\frac{\nu_{n,l}^{\rm mod}}{1000}\right)^{3.5} \varPi_{\pi}^{(n)}(l),
\label{nuem2}
\end{equation}
which correspond to observed frequencies that would fall exactly onto the curve
$\varPi_{\pi}^{(n)}(l)$ that has been fitted to the scaled frequency
differences, $\delta\nu_{n,l}$, for the ridge of radial order, $n$. Because the
theoretical model frequencies, $\nu_{n,l}^{\rm mod}$, are error-free, the
uncertainty of $\nu_{n,l}^{\rm em2}$ is given by
\begin{equation}
\Delta\nu_{n,l}^{\rm em2} = \frac{1}{q_{n,l}} 
\left(\frac{\nu_{n,l}^{\rm mod}}{1000}\right)^{3.5} \Delta\varPi_{\pi}^{(n)}(l),
\label{unuem2}
\end{equation}
with $\Delta\varPi_{\pi}^{(n)}(l)$ being computed from
equation~(\ref{unufit}) with $\varXi_{\varpi}^{(n)}(l)$ being replaced with
$\varPi_{\pi}^{(n)}(l)$.

%Fig. 24

In the fit of the Chebyshev polynomial $\varXi_{\varpi}^{(n)}(l)$ to the
unscaled frequency differences, $\Delta\nu_{n,l}$, along a given ridge (cf.
Section~\ref{ocsstp1}) as well as in the fit of the Chebyshev polynomial
$\varPi_{\pi}^{(n)}(l)$ to the scaled frequency differences, $\delta\nu_{n,l}$,
along a given ridge, outliers can only be detected in a reliable manner if the
rough trendline of the fitted curve is known as a function of degree a priori.
While this basic requirement is fulfilled for the unscaled frequency
differences, $\Delta\nu_{n,l}$, because their trendline along a given ridge is
only a slowly varying function of degree (cf. Figure~\ref{fdiffvsl}), it is not
fulfilled for the scaled frequency differences, $\delta\nu_{n,l}$, as is
demonstrated here in the left-hand panels of Figure~\ref{fitresvsl}, for both
the $n=1$ and $n=11$ ridge. For the $n=1$ ridge a sharp variation of the
trendline with degree is prominent at around $l\approx 60$, as is shown in the
upper-left panel of Figure~\ref{fitresvsl}, while for the $n=11$ ridge similar
sharp variations do occur at about $l\approx 1$ and $l\approx 7$, respectively,
as is shown in the lower-left panel of Figure~\ref{fitresvsl}. Such sharp
variations of the trendline with degree raise the question whether or not they
are real features. However, this is important to know because suspicious trends
may reflect some essential variations in the solar structure that are not yet
known.

In dealing with this issue we found it useful to introduce the emulated scaled
frequency differences, $\delta\nu_{n,l}^{\rm em}$, given by
\begin{equation}
\delta\nu_{n,l}^{\rm em} = \left(\nu_{n,l}^{\rm em1} - \nu_{n,l}^{\rm mod}\right) \, q_{n,l}
            \left(\frac{\nu_{n,l}^{\rm mod}}{1000}\right)^{-3.5},
\label{emsres}
\end{equation}
which differ from the original scaled frequency differences, $\delta\nu_{n,l}$,
merely by the replacement of the observed frequencies, $\nu_{n,l}^{\rm obs}$,
in equation~(\ref{fdiffs}) with the emulated frequencies, $\nu_{n,l}^{\rm
em1}$, defined in equation~(\ref{nuem1}). While for some ridges the degree
dependence of the trendline of the curve fitted to the original scaled
frequencies, $\delta\nu_{n,l}$, cannot be determined in a unique manner mainly
because of lurking outliers in the observed frequencies, $\nu_{n,l}^{\rm obs}$,
the degree dependence of the trendline of the emulated scaled frequency
differences, $\delta\nu_{n,l}^{\rm em}$, is given uniquely. Primarily, this is
because the march of the observed frequencies, $\nu_{n,l}^{\rm obs}$, is only a
slowly varying function of degree without any sharp variations so that the
march of the emulated frequencies, $\nu_{n,l}^{\rm em1}$, with degree is given
unambiguously for each ridge by means of a straightforward fit of the Chebyshev
polynomial $\varXi_{\varpi}^{(n)}(l)$ to the unscaled frequency differences,
$\Delta\nu_{n,l}$ (cf. Section~\ref{ocsstp1}).

On these grounds, the degree dependence of the trendline of the emulated scaled
frequency differences, $\delta\nu_{n,l}^{\rm em}$, can be considered, for a
given ridge, as a reliable template for the trendline of the curve fitted, as a
function of degree, to the original scaled frequency differences,
$\delta\nu_{n,l}$. Therefore, by a careful identification of lurking outliers
in the observed frequencies, $\nu_{n,l}^{\rm obs}$, as well as a careful
selection of the curve (i.e., Chebyshev polynomial $\varPi_{\pi}^{(n)}(l)$ of
degree $\pi$) that is fitted to the original scaled frequency differences,
$\delta\nu_{n,l}$, along a given ridge, the degree dependence of the trendline
of this fitted curve can be made similar to that of the emulated scaled
frequency differences, $\delta\nu_{n,l}^{\rm em}$. This is demonstrated here in
Figure~\ref{fitresvsl} in which we show the degree dependence of the emulated
scaled frequency differences, $\delta\nu_{n,l}^{\rm em}$, for the $n=1$ ridge
in the upper-right panel, and for the $n=11$ ridge in the lower-right panel,
respectively. A close comparison of the left-hand panels and the corresponding
right-hand panels of Figure~\ref{fitresvsl} clearly indicates that for both the
$n=1$ and $n=11$ ridge the trendlines of the curves fitted to the original
scaled frequency differences, $\delta\nu_{n,l}$, as a function of degree are in
excellent agreement with the corresponding trendlines of the emulated scaled
frequency differences, $\delta\nu_{n,l}^{\rm em}$. We therefore can safely
presume that the sharp variations with degree to be seen in the degree
dependence of the trendlines of the original scaled frequency differences,
$\delta\nu_{n,l}$, of both the $n=1$ and $n=11$ ridge are real features. We
note that the same analysis is also useful for ridges other than the $n=1$ and
$n=11$ ridge. 

At the end of the second step of our outlier correction scheme we have the
following three diverse sets of frequencies and associated uncertainties at our
disposal: the set of observed frequencies $\nu_{n,l}^{\rm obs}$, and the
uncertainties $\Delta\nu_{n,l}^{\rm obs}$ thereof, as well as the two sets of
emulated frequencies $\nu_{n,l}^{\rm em1}$ and $\nu_{n,l}^{\rm em2}$,
respectively, the uncertainties of which are given by $\Delta\nu_{n,l}^{\rm
em1}$ and $\Delta\nu_{n,l}^{\rm em2}$, respectively.

\subsubsection{Step 3}

In the third step of our outlier correction scheme we compute from these three
sets of frequencies and associated uncertainties a total of three sets of
normalized frequency differences, viz.
\begin{align}
\Delta\nu_{n,l}^{\rm (1)} &= \frac{\nu_{n,l}^{\rm obs} - \nu_{n,l}^{\rm em1}}
{\sqrt{\left(\Delta\nu_{n,l}^{\rm obs}\right)^2 + \left(\Delta\nu_{n,l}^{\rm em1}\right)^2}},
\label{set3nud1} \\
\Delta\nu_{n,l}^{\rm (2)} &= \frac{\nu_{n,l}^{\rm obs} - \nu_{n,l}^{\rm em2}}
{\sqrt{\left(\Delta\nu_{n,l}^{\rm obs}\right)^2 + \left(\Delta\nu_{n,l}^{\rm em2}\right)^2}},
\label{set3nud2} \\
\Delta\nu_{n,l}^{\rm (3)} &= \frac{\nu_{n,l}^{\rm em1} - \nu_{n,l}^{\rm em2}}
{\sqrt{\left(\Delta\nu_{n,l}^{\rm em1}\right)^2 + \left(\Delta\nu_{n,l}^{\rm em2}\right)^2}}.
\label{set3nud3}
\end{align}
We flag those modes $(n,l)$ as outliers for which at least one of the following
criteria is fulfilled:
\begin{equation}
\big|\Delta\nu_{n,l}^{\rm (1)}\big| > 3, \quad 
\big|\Delta\nu_{n,l}^{\rm (2)}\big| > 3, \quad
\big|\Delta\nu_{n,l}^{\rm (3)}\big| > 3.
\label{outlc}
\end{equation}

\subsubsection{Step 4\label{ocs4}}

The fourth step of our outlier correction scheme actually consists of two
sub-steps. As we already have pointed out in Section~\ref{thfp}, custom-made
seed tables are required as input to the WMLTP method. This is because the
fitting profile, as given by equations~(\ref{profm2a}) through (\ref{profm2f}),
depends on the initially not very well known variation of frequency, amplitude,
and linewidth with degree, $l$. Therefore, in the first sub-step we update
those seed tables by performing an additional iteration of our fixed-point
iteration, as described in Section~\ref{thfp}. In the second sub-step we then
refit, by means of the WMLTP method, all of the modes that have been flagged
as outliers in steps~1 through 3 of our outlier correction scheme, thereby
employing the updated seed tables as input to the WMLTP method to get the
refitted frequencies, $\nu_{n,l}^{\rm new}$.

As we have demonstrated in Section~\ref{thfp}, the fixed-point iteration that
is used for the compilation of the custom-made seed tables which are required
as input to the WMLTP method converges on a solution rather rapidly. Therefore,
it is to be expected that the refitted frequencies, $\nu_{n,l}^{\rm new}$,
generated in the second sub-step deviate only slightly from the observed
frequencies, $\nu_{n,l}^{\rm obs}$. In fact, in most of our applications we
have found that $\nu_{n,l}^{\rm new}\approx \nu_{n,l}^{\rm obs}$. As a result,
the fourth step of our outlier correction scheme may be skipped provided the
preceding fixed-point iteration has been carried out carefully. In this case,
the remaining fifth step of our outlier correction scheme is performed by using
$\nu_{n,l}^{\rm new} = \nu_{n,l}^{\rm obs}$.

\subsubsection{Step 5\label{ocs5}}

In the final step of our outlier correction scheme, we update all of
the modes that have been flagged as being outliers in steps~1 through 3. In
some cases we only update the frequencies of those outliers and in other cases
we update only the associated uncertainties, and in still other cases we update
both the frequencies and their uncertainties at the same time. In addition, as
we shall demonstrate in Section~\ref{ocs6}, we also encountered a few cases for
which neither their frequency nor their uncertainty have been altered by the
procedure. 

Before making the updates, we first check whether the refitted frequencies
$\nu_{n,l}^{\rm new}$, actually constitute an improvement over the originally
observed frequencies, $\nu_{n,l}^{\rm obs}$, by calculating the distances
$\delta_{\rm old}$ and $\delta_{\rm new}$, respectively, given by
\begin{align}
\delta_{\rm old}^2 &= \left(\nu_{n,l}^{\rm obs}-\nu_{n,l}^{\rm em1}\right)^2 +
                     \left(\nu_{n,l}^{\rm obs}-\nu_{n,l}^{\rm em2}\right)^2,
\label{distold1} \\
\delta_{\rm new}^2 &= \left(\nu_{n,l}^{\rm new}-\nu_{n,l}^{\rm em1}\right)^2 +
                     \left(\nu_{n,l}^{\rm new}-\nu_{n,l}^{\rm em2}\right)^2,
\label{distold2}
\end{align}
using the emulated frequencies $\nu_{n,l}^{\rm em1}$ and $\nu_{n,l}^{\rm em2}$,
respectively, as reference values. If $\delta_{\rm new}^2 < \delta_{\rm
old}^2$, we replace the original observed frequency, $\nu_{n,l}^{\rm obs}$,
with the refitted frequency, $\nu_{n,l}^{\rm new}$, for that mode, i.e.,
\begin{equation}
\nu_{n,l}^{\rm obs,new} = \nu_{n,l}^{\rm new},
\label{nuupda}
\end{equation}
and we set for the associated uncertainty
\begin{equation}
\Delta\nu_{n,l}^{\rm obs,new} = \max\left(\Delta\nu_{n,l}^{\rm obs},
      \Delta\nu_{n,l}^{\rm new},\Delta\nu_{n,l}^{\rm em1},
      \Delta\nu_{n,l}^{\rm em2},\left|\nu_{n,l}^{\rm em1}-\nu_{n,l}^{\rm new}\right|/3,
      \left|\nu_{n,l}^{\rm em2}-\nu_{n,l}^{\rm new}\right|/3\right).
\label{dnuupda}
\end{equation}
Here, the first four terms on the right-hand side represent all of the various
uncertainties available for the mode $(n,l)$ at this point, while the last two
terms make sure that the refitted frequency is off by no more than $3\,\sigma$
from the curve that was fit to the unscaled frequency differences,
$\Delta\nu_{n,l}$, as well as from the curve that was fit to the scaled
frequency differences, $\delta\nu_{n,l}$, in the first and second step of the
outlier correction scheme, respectively. If $\delta_{\rm new}^2 \geq
\delta_{\rm old}^2$, we do not adjust the original observed frequency,
$\nu_{n,l}^{\rm obs}$, but only replace the associated uncertainty with
\begin{equation}
\Delta\nu_{n,l}^{\rm obs,new} = \max\left(\Delta\nu_{n,l}^{\rm obs},
      \Delta\nu_{n,l}^{\rm em1},
      \Delta\nu_{n,l}^{\rm em2},\left|\nu_{n,l}^{\rm em1}-\nu_{n,l}^{\rm obs}\right|/3,
      \left|\nu_{n,l}^{\rm em2}-\nu_{n,l}^{\rm obs}\right|/3\right).
\label{dnuupdb}
\end{equation}
Here, as compared to equation~(\ref{dnuupda}) the term $\Delta\nu_{n,l}^{\rm
new}$ has been dropped, and in the last two terms $\nu_{n,l}^{\rm new}$ has
been replaced with $\nu_{n,l}^{\rm obs}$.

\subsubsection{Summary of outlying cases\label{ocs6}}

By means of our outlier correction procedure we detected a total of 2,483
outliers in the entire set of 6,366 modes that we originally inverted (i.e., a
total of 39\,\% of the fitted modes were flagged as being outlying cases). For
1,156 of them we found improved frequencies, $\nu_{n,l}^{\rm new}$, by
refitting the corresponding modes, and by then replacing the original observed
frequencies, $\nu_{n,l}^{\rm obs}$, with $\nu_{n,l}^{\rm new}$ whenever
$\delta_{\rm new}^2 < \delta_{\rm old}^2$, while for the remaining 1,327
outliers it turned out that the corresponding refitted frequencies were not an
improvement over the original observed frequencies, $\nu_{n,l}^{\rm obs}$,
because of $\delta_{\rm new}^2 \geq \delta_{\rm old}^2$, and hence we left the
original observed frequencies unchanged.

\begin{figure}
\epsscale{0.60}
\plotone{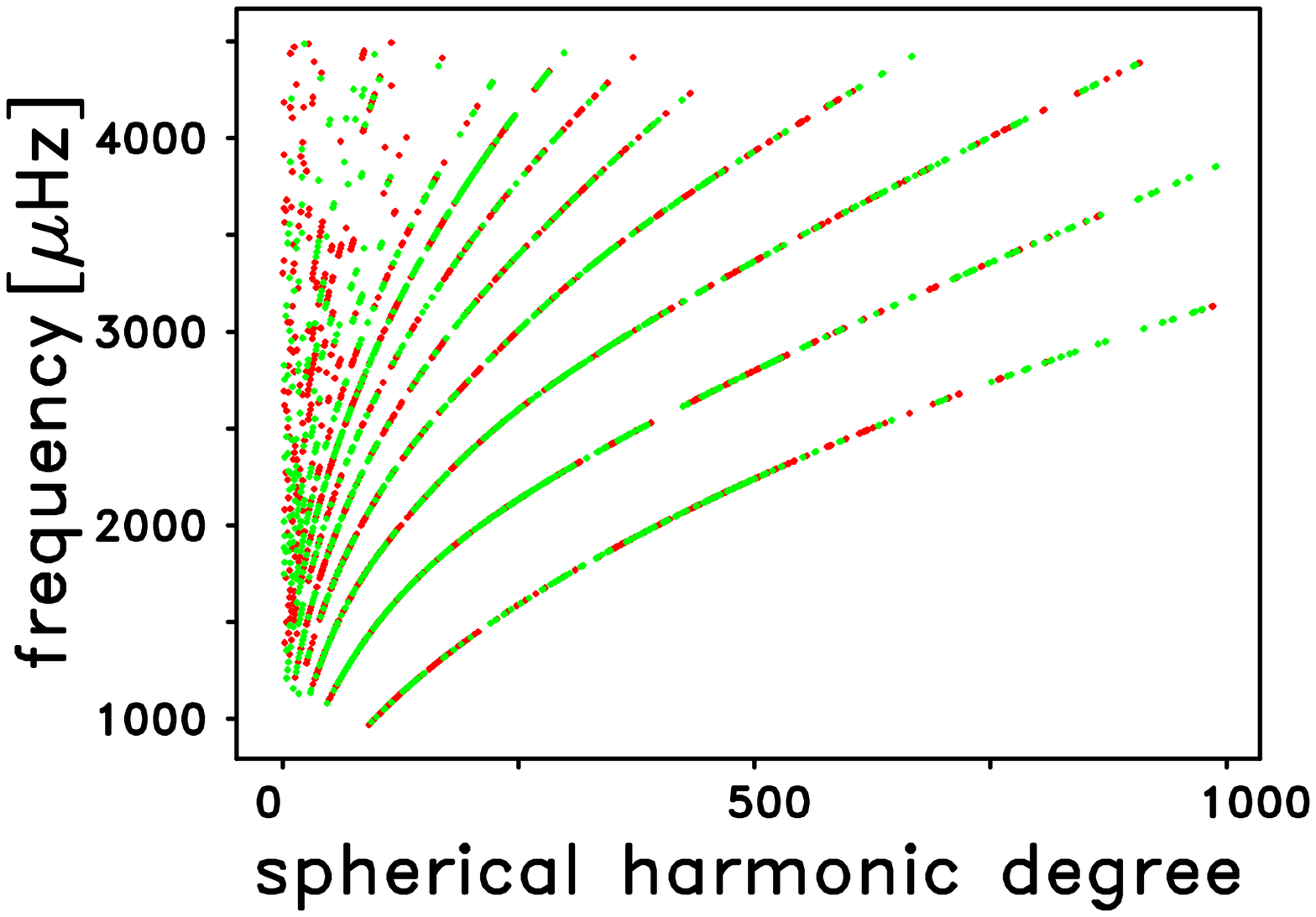}
\caption{Location of the outlying cases in the $l$-$\nu$ plane that have been
detected by means of the outlier correction scheme as described in
Section~\ref{ocs}. The red diamonds are for the 870 outliers for which both
the frequencies and their uncertainties were adjusted and for the 286
additional outliers for which only the frequencies were adjusted. The green
diamonds are for the 1,129 outliers for which only the frequency uncertainties
were adjusted and for the remaining 198 cases for which neither the frequencies
nor their uncertainties needed to be adjusted.
\label{lnuoutl}}
\end{figure}

In Figure~\ref{lnuoutl} we illustrate the location of the outliers in the
$l$-$\nu$ plane that were detected by means of our outlier correction
procedure. The 1,156 outliers for which we were able to compute an improved
refitted frequency are shown as the red diamonds, while the 1,327 outliers for
which our refitted frequency was not an improvement over our original, observed
frequency, are shown as the green diamonds. When we examined the spherical
harmonic degree distribution of both sets of outliers, we found that both sets
were primarily located at low and moderate degrees rather than at high degrees.
In fact, we found that 75\,\% of the 2,483 outliers were located at or below
$l=350$.

%Fig. 25

Of the 1,156 cases for which we were able to compute an improved refitted
frequency, we found far more cases (e.g., 1,031) for which the refitted
frequency was lower than the original frequency. That left only 125 cases for
which the refitted frequency was higher than was the original frequency.
Furthermore, for only 143 of the 1,156 cases were we able to find a normalized
frequency adjustment that was greater than $3\,\sigma$ in absolute magnitude.

By means of either equation~(\ref{dnuupda}) or (\ref{dnuupdb}) we tried to find
improved estimates for the frequency uncertainties of the 2,483 outliers.
Such adjustment of the frequency uncertainties is in a sense
equivalent to the allowance of systematic errors in the observed frequencies,
$\nu_{n,l}^{\rm obs}$, and $\nu_{n,l}^{\rm obs,new}$, respectively. A summary
of the statistics that resulted when we carefully evaluated the frequency
uncertainties of all 2,483 outliers using equations~(\ref{dnuupda}) and
(\ref{dnuupdb}) is given in Table~\ref{tab7}. In the first column of
Table~\ref{tab7} the eight terms are listed that could trigger the modification
of the frequency uncertainty of an outlying case according to either
equation~(\ref{dnuupda}) or (\ref{dnuupdb}). In the second column of
Table~\ref{tab7} we list the number of cases for which one of the terms
included in the right-hand side of equation~(\ref{dnuupda}) triggered the
adjustment of the frequency uncertainty. Interestingly, we found that
$\Delta\nu_{n,l}^{\rm obs,new}=\Delta\nu_{n,l}^{\rm obs}$ for 286 cases, as
shown in the first row of column~2 in Table~\ref{tab7}, and therefore we did
not alter the original frequency uncertainty, $\Delta\nu_{n,l}^{\rm obs}$, of
those cases. For the remaining 870 cases we increased the original frequency
uncertainty because one of the terms listed below $\Delta\nu_{n,l}^{\rm obs}$
in the first column of Table~\ref{tab7} turned out to be larger in magnitude
than $\Delta\nu_{n,l}^{\rm obs}$. In the third column of Table~\ref{tab7} we
list the number of cases that we evaluated using equation~(\ref{dnuupdb}). For
1,129 of these 1,327 cases, we ended up increasing the original frequency
uncertainties; however, for the remaining 198 cases the original frequency
uncertainty was large enough that it did not need to be replaced with one of
the other terms in equation~(\ref{dnuupdb}). In summary, considering all of the
2,483 outliers we found an average ratio of the final and original
uncertainties of $1.974\pm 1.237$. When we included only the 1,999 cases for
which the frequency uncertainty actually increased in the computation of the
average ratio of the new to the original uncertainties, we found that the
average ratio was increased slightly to $2.210\pm 1.271$.

The 198 outliers listed at the top of the third column in Table~\ref{tab7} are
those cases for which neither the frequencies nor the frequency uncertainties
needed to be changed. Therefore, the question arises as to why those 198 cases
were flagged as being outliers in the first case. This can happen 
because we are invoking two
criteria for the identification of outliers which do not depend in any way upon
the frequency uncertainty, $\Delta\nu_{n,l}^{\rm obs}$. These criteria are
given in equation~(\ref{outlb}) and in the third branch of
equation~(\ref{outlc}), respectively. As a result, if a mode is initially
flagged as being an outlier by one of those two criteria, but later its
refitted frequency is not found to be an improvement over the original
frequency of that mode, $\nu_{n,l}^{\rm obs}$, and also if the original
frequency uncertainty of that mode, $\Delta\nu_{n,l}^{\rm obs}$, is large
enough in magnitude that equation~(\ref{dnuupdb}) sets $\Delta\nu_{n,l}^{\rm
obs,new}=\Delta\nu_{n,l}^{\rm obs}$, this mode ends up with both its original
frequency and its original uncertainty intact. We note, however, that such a
mode is not expected to hamper the convergence of the inversion code because of
its exceptionally large frequency uncertainty, $\Delta\nu_{n,l}^{\rm obs}$,
that has been returned by the WMLTP code for this mode. 

In spite of the above discussion about the number of outliers that our
procedure identified, we wish to stress that, in the majority of the 6,366
original fits for which $\nu\leq 4500\,\,\mu$Hz, we did not find it necessary
to alter our original fitted frequencies and/or uncertainties thereof at all.
We illustrate this important point further in Table~\mbox{\ref{tab8}} in which
we have summarized the various outcomes of our five-step procedure. In the
second row of Table~\mbox{\ref{tab8}} we note that we did not have to alter the
frequencies of 5,210, or 81.8\,\%, of our 6,366 original fits. This
overwhelming majority of our cases was comprised of the 3,883 cases that were
never identified as being outliers by our procedure, the 1,129 cases for which
only the frequency uncertainties needed to be altered, and the 198 cases for
which neither the frequencies nor their uncertainties needed to be adjusted.
Also, in the third row of Table~\mbox{\ref{tab8}} we note that a total of
4,367, or 68.6\,\%, of our original cases did not need to have their
uncertainties altered. This number was comprised of the 3,883 cases that were
never identified as being outliers, the 286 cases for which only the
frequencies needed to be altered, and the 198 cases for which neither the
frequencies nor the uncertainties needed to be altered. In the remaining rows
of Table~\mbox{\ref{tab8}} we summarize the 2,483 outlying cases and their
outcomes that resulted from the various steps of our procedure in
diminishing order as percentages of our total of 6,366 original cases.

While Table~\ref{tab8} shows that our outlier identification and correction
procedure ended up requiring an increase in a total of 1,999 of the original
frequency uncertainties, we note that these increases were found to be needed
because the original uncertainties that the WMLTP code computed were too small.
In this regard, it is important to note that the formal uncertainties delivered
by the WMLTP code only represent the statistical errors which are not corrected
for any systematic effects.  Therefore, our outlier correction scheme is useful
not only for detecting systematic errors in the sets of fitted frequencies but
also for detecting frequency uncertainties that are unduly small. We finally
note that it is both the correction of the fitted frequencies for systematic
errors as well as the adequate adjustment of the frequency uncertainties by
means of our outlier correction scheme that finally allowed the inversion shown
in Figure~\ref{spinv} to converge properly on a solution.

\subsection{Inversion results\label{invres}}

Accurate measurements of the high-degree solar oscillation modes are
particularly important for helioseismic diagnostics of the near-surface layers
of the Sun. These layers are believed to play a critical role in the formation
of the magnetic network, sunspots and active regions, and thus are a key to
understanding the mechanisms of solar activity and variability. To illustrate
the potential of the new method of measurements of the high-degree mode
frequencies we performed a standard inversion procedure for the solar
sound-speed profile, which is based on the mode sensitivity kernels derived
from the variational principle, and the OLA inversion method. The sensitivity
kernels are calculated from the adiabatic eigenfunctions, and the surface
non-adiabatic effects are removed by assuming that these can be described as a
function of frequency, weighted with the mode inertia
(cf. equation~(\ref{fdiffs})). The inversion method provides locally averaged
estimates of the sound-speed variations at various target positions along the
solar radius. The central locations and characteristic width (``spread'') of
the averaging kernels of these estimates provide a quantitative measure of the
resolving power of the observed frequency set. The mathematical details of the
inversion procedure are described by \cite{Kos99}. 

\begin{figure}
\epsscale{0.80}
\plotone{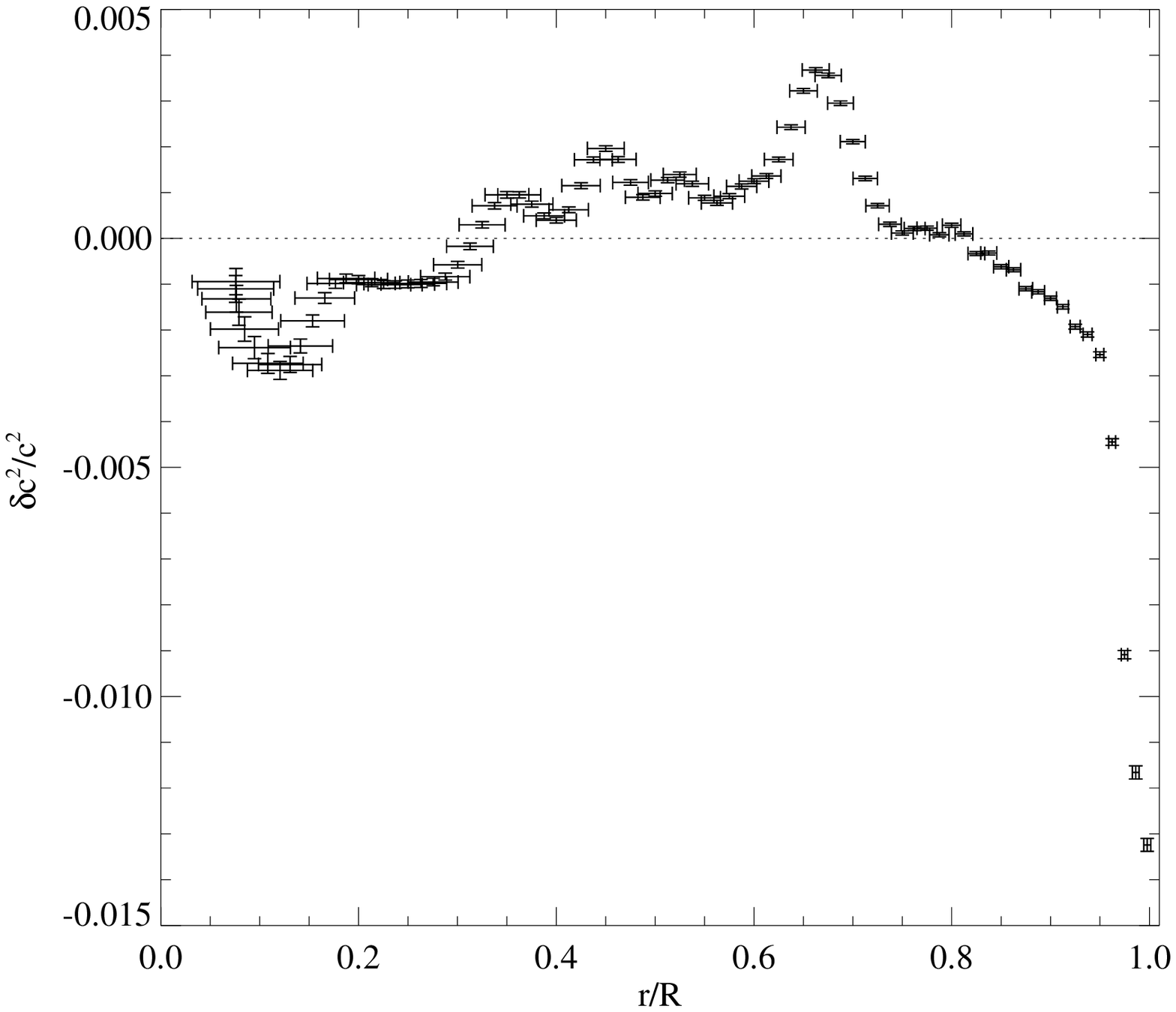}
\caption{The relative squared sound-speed deviations from the Standard Solar
Model~S of \cite{jcd96} as a function of fractional radius that we obtained by
a structural inversion of the 6,366 frequencies and their uncertainties that
resulted from the application of the outlier correction procedure to our
original frequency table that has been computed using Method~2 on the
$m$-averaged spectral set $\cal{S}$2010\_66a, and covered the frequency range
of 969 to $\,4500\,\,\mu$Hz. The horizontal bars represent the width
(``spread'') of the localized averaging kernels, providing a characteristic of
the spatial resolution; the vertical bars are the formal error estimates.
\label{spinv}}
\end{figure}

The inversion results illustrated in Figure~\ref{spinv} show the relative
deviations of the squared sound speed in the Sun from the standard solar
Model~S \citep{jcd96}. The results reproduce a well-known peak at the base of
the convection zone, which is associated with the tachocline -- a narrow area of
the strong rotational shear at the bottom of the convection zone
\citep{Kos96b,Ell99}. In the lower part of the convection zone, the sound-speed
deviation is almost zero due to the adiabatic stratification of the radial
structure. The sound-speed profile gradually decreases relative to the model in
the bulk of the convection zone, which may indicate a deviation from the
adiabatic stratification of the model (prescribed by the mixing-length theory
of convection), and then shows a sharp decrease in the upper 5\,\% of the solar
radius ($\sim 35$ Mm). Such decrease in the sound speed was previously detected
by inversions, however, its structure was not resolved. The new results show
that the structure of the near-surface layer is now well-resolved. The
substantial deviation from the model coincides with the subsurface rotational
shear layer \citep[called ``leptocline'' by][]{God01}, and indicates that the
structure of the solar convection zone and the heat transport properties may be
significantly different from the predictions of the mixing-length theory. Of
course, this result is only a starting point of systematic investigations of
the global structure and dynamics of this layer, and also of the variations
with the solar cycle.

%Fig. 26

\section{Concluding remarks\label{concl}}

Global helioseismology has proven to be an extremely powerful tool for the
investigation of the internal structure and dynamical motions of the Sun and,
thus, should not be abandoned in favor of local helioseismology alone.
Numerical inversions of low- and medium-degree solar oscillation frequencies
have confirmed that the solar structure is in remarkably good agreement with
predictions of the standard solar model. However, global helioseismic studies
have not yet provided reliable information about the very interesting
near-surface region of the Sun because high-degree modes were not measured
accurately enough due to serious problems in the understanding of the structure
of a ridge of oscillatory power and various distortion effects of the measuring
device.

In order to provide a more-accurate characterization of the high-degree ridges
of power, we have presented in this paper a sophisticated mathematical method
for the fitting of $m$-averaged power spectra which employs a theoretical
profile that contains multiple peaks which represent each mode of interest and
also its surrounding temporal and spatial sidelobes. This method includes both
spatial leakage matrices which weight each of the spatial sidelobes differently
according to the details of the instrument which provided the data from which
the power spectra were created, and it also includes the temporal window
function of each observational time series to allow for the data gaps that are
usually present in the observations. 

In Section~\ref{dagh} above, we outlined several different approaches to the
calculation of $m$-averaged spectra, and we also described the need for the
correction of the frequency splitting coefficients which are used in the
generation of such spectra for the effects of modal distortions that are
introduced by solar latitudinal differential rotation. In Section~\ref{pahds}
we presented mathematical details of the computation of the spatial leakage
matrices which are required for the accurate fitting of both isolated modal
peaks and ridges of power, and we presented the mathematical details of the
method by which such leakage matrices are corrected for the effects of
differential rotation. In Section~\ref{secm2} we presented the  mathematical
details of this new method, which we have named the WMLTP method, and which we
have also referred to in the text as our Method~2.

In Section~\ref{sfmp} we presented the results of a series of studies of the
sensitivity of the fitted frequencies that this method produces to various
details of the method of generating the $m$-averaged power spectra and of the
approximation that is made to the shape of the various leakage matrix peaks. In
Section~\ref{s6sub1} we showed that the weighting of the individual tesseral
and sectoral power spectra at each degree prior to the computation of the
$m$-averaged spectrum had only a very minor influence on the resulting
frequencies. On the other hand, in Section~\ref{s6sub2}, we showed that the use
of frequency-splitting coefficients that were computed in a non-$n$-averaged
manner and which were also corrected for the distorting effects of differential
latitudinal rotation definitely did result in substantial changes to the fitted
frequencies. In Section~\ref{s6sub3} we showed that the majority of these
systematic frequency changes were due to the use of such non-$n$-averaged
frequency-splitting coefficients in place of coefficients that were computed in
an $n$-averaged manner. That is, we demonstrated the importance of using a
narrow frequency range that was centered on each individual ridge when
cross-correlating the non-zonal power spectra with the zonal spectrum at each
degree. In  Section~\ref{slkm} we showed that an increase of 18\,\% in the
width of the Gaussian function that we employ in Method~2 as an approximation
of the peaks in the effective leakage matrices did not result in any systematic
frequency changes.

In Section~\ref{SR} we demonstrated the substantial improvements that the WMLTP
method makes in the frequencies, linewidths, amplitudes, and in the
uncertainties of all of these quantities in comparison with the same quantities
computed using our earlier Method~1, which only employed a single Lorentizan
profile in place of the more complicated multiple-peak profile of Method~2. In
particular, we have pointed out that the use of a multiple-peak profile in the
WMLTP method reduces the formal frequency uncertainties by a factor of $15.2\pm
0.3$ on average, and it also reduced the formal linewidth uncertainties by a
factor of $29.8\pm 0.9$ on average. We have also demonstrated that the
frequencies, linewidths, and amplitudes that we computed using the
multiple-peak profile do not exhibit any sharp discontinuities along the
various ridges as did the same quantities that were obtained using the
single-peak profile. These discontinuities occurred where the individual peaks
became blended into broad ridges of power at the higher degrees. We have also
compared the relative smoothness of both the frequencies and linewidths which
were computed with the WMLTP method from various sets of power spectra that
were obtained with the MDI instrument. We have also displayed the sets of
12,359 frequencies, linewidths, amplitudes, asymmetries, and power levels that
we generated by applying the WMLTP method to the 66-day long 2010 MDI Dynamics
Run power spectra. 

In Section~\ref{ocs} we presented the details of a new procedure that we
developed to improve both the numerical stability and reliability of the
structural inversion that we presented in Section~\ref{invres} and of similar
structural inversions that we expect to compute with similar Method~2 frequency
tables in the future. This procedure both identifies and corrects outliers in
both the frequencies and their associated uncertainties. We applied it to the
subset of 6,366 Method~2 frequencies that we had originally inverted and we
obtained a much smoother inverted sound speed deviation profile in the process.

In conclusion, we have shown that the development of the WMLTP method of power
spectral fitting provides a powerful new tool for the provision of accurate and
reliable estimation of low-, intermediate, and high-degree mode parameters. We
demonstrated this claim impressively in Section~\ref{invres} with the
presentation of a new structural inversion corresponding to the beginning of
Solar Cycle~24 that we computed using our cleaned table of MDI 2010 Dynamics
Run frequencies. This new inversion resolves for the first time the seismic
properties of the upper convective boundary layer, and shows a substantial and
surprisingly sharp deviation from the adiabatic sound-speed profile of the
Standard Solar Model~S of \cite{jcd96}, which is likely due to the near-surface
turbulence effects. Moreover, this is also evidence that the predictions of the
mixing-length theory may be significantly different from both the actual
structure of the convection zone and the heat transport properties in the
subsurface layers of the Sun. By comparing upcoming structural inversions
similar to that presented here in Section~\ref{invres} with realistic 3D MHD
numerical simulations of the upper solar convection zone and the subsurface
shear layer we hope to make significant progress in the understanding of the
structure and dynamics of the outer layers of the Sun. In this regard, we note
that the detailed measurements of the linewidths of the $f$- and $p$-modes
could also provide useful constraints on the properties of near-surface
convection \citep[cf.][]{jcd89,Rab99}.

Because for degrees $l\gg 1$ $m$-averaging of the originally generated zonal,
sectoral, and tesseral power spectra significantly improves the signal-to-noise
ratio, the WMLTP method is particularly well suited for the fitting of
$m$-averaged power spectra derived from time series as short as three days,
say. This facilitates the detailed investigation of temporal frequency shifts
with the phase of the solar activity cycle \citep[see,
e.g.,][]{Ron94,Jef98,Rho02,Rho03,Ros03,Rab08a,Tri10,Jai11,Rho11}. On the other
hand, the generation of $m$-averaged power spectra requires knowledge of
suitable frequency splitting coefficients (cf. Sect.~\ref{gmavg}). Therefore,
as we mentioned in Section~\ref{cdsldr}, we have also begun to develop our MPTS
fitting method which operates directly upon all $2l+1$ un-averaged power
spectra for each degree. The MPTS method allows us to determine all $2l+1$ sets
of modal parameters of a given $(n,l)$ multiplet. Once we have determined all
$2l+1$ of the frequencies of that multiplet, we can then compute both the
average frequency and the frequency-splitting coefficients for that multiplet.
In essence, the MPTS method will allow us to generate sets of non-$n$-averaged
frequency splitting coefficients without recourse to the cross-correlation
method that we described in Section~\ref{gmavg}. Details of the MPTS method
will be given in the upcoming second paper of our series of three papers.

Unfortunately, as we mentioned earlier the MPTS method cannot be applied to
power spectra that are computed from time series that are as short as three
days in duration, nor can it fit peaks having frequencies greater than about
$5000\,\,\mu$Hz because of the low signal-to-noise ratios of the underlying power
spectra. As a result of these two limitations, both the WMLTP and MPTS fitting
methods are complementary to each other. Hence, we hope that, once we have been
able to bring the current version of the MPTS code up to the level of the rev6
version of the WMLTP code, the parallel application of both fitting methods
will lead to new information concerning temporal variations in the Sun's
sub-surface structure and dynamics.


%% Included in this acknowledgments section are examples of the
%% AASTeX hypertext markup commands. Use \url without the optional [HREF]
%% argument when you want to print the url directly in the text. Otherwise,
%% use either \url or \anchor, with the HREF as the first argument and the
%% text to be printed in the second.

\acknowledgments

In this work we utilized data from the Solar Oscillations
Investigation\,/\,Michelson Doppler Imager (SOI/MDI) on board the Solar and
Heliospheric Observatory (SOHO), and we have made use of NASA’s Astrophysics
Data System. SOHO is a project of international cooperation between ESA and
NASA. The SOI/MDI project is supported by NASA grant NAG5-10483 to Stanford
University. The portion of the research that was conducted at the University of
Southern California was supported in part by NASA Grants NNX08A24G, NAG5-13510,
NAG5-11582, NAG5-11001, NAG5-8545, NAG5-8021, NAG5-6104, and NAGW-13, by
Stanford University Sub-Awards 14405890-126967, 1503169-33789-A, and 29056-C,
by Stanford University Sub-Contract Number 6914, and by USC's Office of
Undergraduate Programs. Part of this work is the result of research performed
at the Jet Propulsion Laboratory of the California Institute of Technology
under a contract with the National Aeronautics and Space Administration. We
thank the anonymous referee for his valuable contributions to improve the
presentation of this work. J.R. is grateful to R.~Bulirsch, P.~Rentrop, and
B.~Vexler of the Techni\-sche Universit\"at M\"unchen for their generous
support and hospitality, and to K.~Schittkowski of the University of Bayreuth
for providing the source code of his NLPQL optimization technique.

\appendix
\section{The single-peak, averaged-spectrum method \label{secm1}}

The Single-Peak, Averaged-Spectrum, or SPAS, Method, which is also referred to as
Method~1, was our initial, first-generation fitting method. It was developed in
the late-1980s through the mid-1990s in order to fit the peaks in the MDI power
spectra that were generated as part of that experiment's Structure Program
\citep{Scher95}. This work led to the publication of a set of tables of $f$-
and $p$-mode frequencies and their associated errors which has become a major
reference in the helioseismic literature for the MDI Medium-$l$ Program
\citep{Rho97}. We also have applied Method~1 to the analysis of power spectra
which were derived from MDI's Full-Disk (FD) Dynamics Program \citep{Scher95}.
This led to the publication of the first frequencies to be computed for the
high-degree modes from the MDI FD spectra \citep{Rho98a,Rho98b}.

In Method~1 a single, symmetric Lorentzian profile plus a linear background
term is used as the model profile to represent an oscillation peak, i.e.,
\begin{eqnarray} 
M(\nu,\mbox{\boldmath $p$}) &=& \frac{A}{1+x^2}
    +a+b\,\nu ,\,\,\,
x = \frac{2\,(\nu-\nu_0)}{w},\label{mm1}\\
\mbox{\boldmath $p$} &=& (A,\nu_0,w,a,b)^{\mbox{\footnotesize T}},
\label{pm1}
\end{eqnarray}
with $\nu$ being frequency. The five fit parameters are collected in the vector
$\mbox{\boldmath $p$}$, and include, respectively, the mode amplitude $A$, the
mode frequency $\nu_0$, the mode linewidth $w$, and the background noise
parameters $a$ and $b$. The vector $\mbox{\boldmath $p$}$ is determined by
fitting the model profile (\ref{mm1}) in the least-squares sense to the
$m$-averaged spectrum in a fitting range centered about the peak of interest
that is selected by using the heuristic approach as described below. The
resulting unconstrained least-squares problem is solved by using the FORTRAN
routine {\tt lmder1} from the MINPACK project \citep{Mor80}. This routine is
designed to minimize the sum of squares of $\lambda_1$ nonlinear functions in
$\lambda_2$ variables ($\lambda_2\leq\lambda_1$) by a modification of the
Levenberg-Marquardt algorithm \citep{Lev44,Mar63}.

The variance of the $i$th fit parameter $p_i\in\mbox{\boldmath $p$}$ is
calculated from
\begin{equation}
\mbox{var}\,p_i = \frac{2\,S}{\lambda_1 - \lambda_2}\,H^{-1}_{ii},
\end{equation}
where $\lambda_1$ is the number of data points (frequency bins), $\lambda_2$ is
the number of fit parameters, $H^{-1}_{ii}$ is the $i$th diagonal element of
the inverse of the Hessian matrix $H$, and $S$ is the sum of squares of the fit
residuals \citep{Nag11}. Both $H$ and $S$ are calculated at the solution of the
least-squares problem. In the neighborhood of the solution the Hessian matrix
can be adequately approximated by
\begin{equation}
H = 2\, J^T J,
\end{equation}
with $J$ being the Jacobian matrix, thereby avoiding the need to compute or
approximate second derivatives of the objective function \citep{Gill81}. The
$100\,(1-\gamma)$ percent confidence interval on the $i$th fit parameter $p_i$
is given by
\begin{equation}
p_i\pm\sqrt{\mbox{var}\,p_i}\,\, t_{1-\gamma/2,\lambda_1-\lambda_2},
\end{equation}
where $t_{1-\gamma/2,\lambda_1-\lambda_2}$ is the $100\,(1-\gamma/2)$
percentage point of Student's t-distribution with $\lambda_1-\lambda_2$ degrees
of freedom \citep{Wol67,Nag11}. For $1\,\sigma$ confidence intervals
$\gamma=0.31731$, and for $\lambda_1-\lambda_2\gg 1$ we have
$t_{1-\gamma/2,\lambda_1-\lambda_2}\approx 1$.

We have implemented into Method~1 an option which allows to use either a narrow
fitting range or else a wide fitting range. We switch between the narrow and
the wide fitting range based upon the ratio, $r$, between the linewidth and the
frequency change with respect to degree, namely
\begin{equation}
r=\frac{w}{\Delta\nu/\Delta l}.
\end{equation}
We estimate this ratio using linewidths and frequencies from seed tables that
have been derived from previously computed modal parameters. We use the narrow
fitting range when $r < r_{\rm crit}$ and the wide fitting range for $r \ge
r_{\rm crit}$, namely when the modes are well separated, or not, from the
spatial leaks of nearby degrees. Because there is no sharp transition between
modes and ridges of power, the choice of $r_{\rm crit}$ is somewhat arbitrary.
By visual inspection of the fits obtained with Method~1 we found the values
\begin{equation}
r_{\rm crit} = \left\{ \begin{array}{l@{\quad}l}
0.65               & \mbox{for~~$n=0$,} \\[1.0mm]
0.80               & \mbox{for~~$n>0$,}
               \end{array} \right.
\label{rcrit}
\end{equation}
to be useful. For the $n=0$ ridge we had to select a separate value of $r_{\rm
crit}$ because for this ridge individual modal peaks blend into ridges of power
for smaller ratios, $r$, than for the other ridges. The values of $r_{\rm
crit}$ given in equation~(\ref{rcrit}) correspond to $l\approx 278$ for $n=0$,
$l\approx 208$ for $n=1$, and $l\approx 15$ for $n=29$.

In the left panels of Figure~\ref{F2} we show typical fits obtained with
Method~1 when applied to the $m$-averaged spectral set $\cal{S}$2010\_66a. The
fits obtained by using a narrow fitting range are shown as the green lines,
while the blue lines are for fits that employed the wide fitting range. We note
that the use of both the narrow and wide fitting ranges for $(n,l)=(2,70)$ and
$(2,200)$ in the top-left and middle-left panel of this Figure required the
cancellation of the criterion~(\ref{rcrit}), which would have enforced the use
of the narrow fitting range for $(2,70)$ and the use of the wide fitting range
for $(2,200)$. From the colored tick marks in these two panels it becomes
evident that the frequencies determined with the wide fitting range differ
significantly from those obtained with the narrow fitting range. This is also
demonstrated in Table~\ref{tab6}. We stress that while this disagreement is
obvious in the cases of those modes for which the peaks are well resolved, the
very same problems are occurring in the cases of broad ridges of power even
though the observed peaks do not show evidence of the individual spatial leaks
as is shown here in the bottom-left panel of Figure~\ref{F2}. 

This so-called ``frequency pulling" of the measured ridge-fit frequencies away
from the true solar frequencies was first addressed by \citet{Lib88}, and is
mainly caused by the asymmetry of the power distribution of the spatial leaks
with respect to the target mode frequency \citep{Kor04}. We first tried to
resolve this problem by developing a frequency-correction scheme in which we
fitted those modes for which the target peak and the spatial leaks are
well separated with both the narrow fitting range and the wide fitting range in
Method~1 as is shown here in the top-left and the middle-left panel of
Figure~\ref{F2}. In order to be able to do so we had to override
criterion~(\ref{rcrit}), of course. By means of a multiple linear regression
we then tried to correct the frequencies obtained with the wide fitting range
for the effects of mode blending using the differences between the narrow- and
wide-fit frequencies. However, we eventually had to abandon this approach
because the range of degrees for which the multiple linear regression model can
be directly computed, namely, those degrees for which the modal peaks are not
blended into ridges of power, is not large enough to allow the regression model
to be safely extrapolated up to degrees of 1000 and higher where it is needed
\citep{Rho01}.

\section{Estimates of smoothness of fitted parameters\label{esfp}}

On physical grounds it can be assumed that any modal parameter, e.g., frequency
or linewidth, is a smooth function of spherical harmonic degree $l$ along a
ridge of radial order $n$. We suggest to measure the smoothness of any modal
parameter along a ridge by the so-called normalized point-to-point scatter,
$\varSigma(\upsilon)$, defined by
\begin{equation}
\varSigma^2(\upsilon) = \frac{1}{2\,(l_{\rm max}-l_{\rm min})}
\sum\limits_{i=l_{\rm min}}^{l_{\rm max}-1} (\upsilon_{i+1}-\upsilon_{i})^2,
\label{A3}
\end{equation}
where $\upsilon$ is the variable/parameter whose smoothness is estimated, and
$l_{\rm min}$ and $l_{\rm max}$ are, respectively, the minimum and maximum
degree $l$ of the portion of interest of the ridge. Because $\varSigma(\upsilon)$ is
rather sensitive to missing cases $\upsilon_{i}$ we recommend that any gaps
that might exist in the range from $l_{\rm min}$ to $l_{\rm max}$ are filled by
interpolation. We note that the value of $\varSigma(\upsilon)$ is biased
towards any slope of the variable $\upsilon$ that might exist in the range from
$l_{\rm min}$ to $l_{\rm max}$ because such slope causes the differences
$\upsilon_{i+1}-\upsilon_{i}$ in equation~(\ref{A3}) to be systematically
different from zero. In general, however, this bias is concealed by the scatter
of the differences $\upsilon_{i+1}-\upsilon_{i}$ due to noise.

\clearpage

%Tab. 1
\begin{deluxetable}{c|c|c|c|c}
\tablecaption{
Epochs of observing runs used in this work. 
\label{tab1}}
\tablewidth{0pt}
\tablehead{
 run & starting and end date & duration & Duration         & gap-filled \\
     &                       & (days)   & (60-sec samples) &\,\,duty cycle
}
\startdata
 $\cal{R}$1996\_61 & May 24 - Jul 23, 1996 & 60.75 &  87,480 & 97.30\,\% \\
 $\cal{R}$2001\_90 & Feb 24 - May 28, 2001 & 90    & 129,600 & 97.02\,\% \\
 $\cal{R}$2010\_66 & May 7 - Jul 11, 2010  & 66    &  95,040 & 93.87\,\% \\
 $\cal{R}$2010\_03 & May 7 - May 9, 2010   & 3     &   4,320 & 85.52\,\%
\enddata
\tablecomments{
Column~1 lists the naming convention we assigned to each observing run.
Columns~2 through 5 contain the starting and ending dates, the durations in
days, the durations in 60-second samples, and the gap-filled duty cycles of the
four runs, respectively. Each observing run was obtained as part of the MDI
Full-Disk Program
\citep{Scher95}. 
}
\end{deluxetable}

%Tab. 2
\begin{deluxetable}{c|c|c|c}
\tablecaption{
Sets of $m$-averaged spectra used in this work.
\label{tab2}}
\tablewidth{0pt}
\tablehead{
\addlinespace[-1.5mm]
observing & $m$-averaged & $m$-averaging & frequency splitting \\
run       & spectral set &               & coefficients }
\startdata
$\cal{R}$1996\_61 & $\cal{S}$1996\_61  & unweighted   & corrected, $n$-averaged     \\
$\cal{R}$2001\_90 & $\cal{S}$2001\_90  & unweighted   & corrected, $n$-averaged     \\
$\cal{R}$2010\_66 & $\cal{S}$2010\_66a & unweighted   & corrected, $n$-averaged     \\
$\cal{R}$2010\_66 & $\cal{S}$2010\_66b & weighted     & corrected, $n$-averaged     \\
$\cal{R}$2010\_66 & $\cal{S}$2010\_66c & weighted     & raw, non-$n$-averaged       \\
$\cal{R}$2010\_66 & $\cal{S}$2010\_66d & weighted     & corrected, non-$n$-averaged \\
$\cal{R}$2010\_03 & $\cal{S}$2010\_03  & unweighted   & corrected, $n$-averaged
\enddata
\tablecomments{
Column~1 lists the observing run using the naming convention introduced in
Table~\ref{tab1}. Column~2 lists the naming convention we assigned to each
$m$-averaged spectral set that we computed from the un-averaged power spectra
obtained from the observing run listed in the same row of column~1.
Column~3 indicates whether the averaged spectra were
computed using the weights described in Section~\ref{gmavg} or were computed in
an unweighted manner. Column~4 describes the type of frequency splitting
coefficients that were employed in the averaging step. 
}
\end{deluxetable}

%Tab. 3
\begin{deluxetable}{c|cccc}
\tablecaption{
Performance of the fixed-point iteration for the construction of accurate seed
tables using the example of the frequency, $\nu$. 
\label{tab3}}
\tablewidth{0pt}
\tablehead{
 difference &  \multicolumn{1}{c}{$n_c$} & \multicolumn{1}{c}{$n_{\rm out}$} & 
               \multicolumn{1}{c}{avg}  &  \multicolumn{1}{c}{std}
}
\startdata
$\displaystyle\left(\frac{\Delta\nu}{\Delta l}\right)_{n,l}^{\rm (seed,2)}-
 \left(\frac{\Delta\nu}{\Delta l}\right)_{n,l}^{\rm (seed,1)}$ &
 12,408 &  22 & $-2.456\cdot 10^{-3}$ &  $1.236\cdot 10^{-1}$ \\[5.0mm]
$\displaystyle\left(\frac{\Delta\nu}{\Delta l}\right)_{n,l}^{\rm (seed,3)}-
 \left(\frac{\Delta\nu}{\Delta l}\right)_{n,l}^{\rm (seed,2)}$ &
 12,417 &  21 & $+4.640\cdot 10^{-4}$ &  $8.988\cdot 10^{-2}$ \\[5.0mm]
$\displaystyle\frac{\nu_{n,l}^{\rm (fit,2)}-\nu_{n,l}^{\rm (fit,1)}}
   {\max(\Delta\nu_{n,l}^{\rm (fit,1)},\Delta\nu_{n,l}^{\rm (fit,2)})}$ &
 12,417 &  13 & $-1.993\cdot 10^{-1}$ &  $1.819\cdot 10^{0}$ \\[6.2mm]
$\displaystyle\frac{\nu_{n,l}^{\rm (fit,3)}-\nu_{n,l}^{\rm (fit,2)}}
   {\max(\Delta\nu_{n,l}^{\rm (fit,2)},\Delta\nu_{n,l}^{\rm (fit,3)})}$ &
 12,353 &  85 & $-8.816\cdot 10^{-7}$ &  $1.129\cdot 10^{-4}$
\enddata
\tablecomments{
Columns~2 through 5 contain the number of cases considered, $n_c$, the number
of rejected outliers, $n_{\rm out}$, the average, avg, and the standard
deviation, std, of the cases listed in column~1. The values of the frequency
derivative, $\left(\Delta\nu/\Delta l\right)_{n,l}^{\rm (seed,k)}$, $k=1,2,3$,
were input as seeds into the WMLTP code to generate the fitted frequencies,
$\nu_{n,l}^{\rm (fit,k)}$, $k=1,2,3$. The differences of the frequency
derivatives, $\left(\Delta\nu/\Delta l\right)_{n,l}^{\rm (seed,k)}$, that are
listed in rows~1 and 2, respectively, are unscaled, while the differences of
the fitted frequencies, $\nu_{n,l}^{\rm (fit,k)}$, that are listed in rows~3 and
4, respectively, are scaled by the maximum of the uncertainties of the
frequencies that are subtracted.
}
\end{deluxetable}

%Tab. 4
\begin{deluxetable}{c|lllll|llll}
\tablecaption{
Fitted parameters invoked in the individual passes used for determining the
fitting vector $\mbox{\boldmath $p$}$.
\label{tab4}}
\tablewidth{0pt}
\tablehead{
\multicolumn{1}{c|}{}      & \multicolumn{5}{c|}{fitted parameters} & \multicolumn{4}{|c}{fitted parameters}\\
\multicolumn{1}{c|}{}      & \multicolumn{5}{c|}{main peak \& background} & \multicolumn{4}{|c}{$n$-leaks}\\
\multicolumn{1}{c|}{pass}  & \multicolumn{5}{c|}{} & \multicolumn{4}{|c}{$i=1,\ldots,N$}
}
\startdata
\#\,1  & $A_{n,l}$ &             &           &           & $a$, $b$, $c$ &       &         &       &       \\
\#\,2  &           & $\nu_{n,l}$ & $w_{n,l}$ &           &               &       &         &       &       \\
\#\,3  & $A_{n,l}$ & $\nu_{n,l}$ & $w_{n,l}$ &           & $a$, $b$, $c$ &       &         &       &       \\
\#\,4  & $A_{n,l}$ &             &           &           & $a$, $b$, $c$ & $A_i$ &         &       &       \\
\#\,5  &           & $\nu_{n,l}$ &           &           & $a$, $b$, $c$ &       & $\nu_i$ &       &       \\
\#\,6  &           &             & $w_{n,l}$ &           & $a$, $b$, $c$ &       &         & $w_i$ &       \\
\#\,7  & $A_{n,l}$ & $\nu_{n,l}$ & $w_{n,l}$ &           & $a$, $b$, $c$ & $A_i$ & $\nu_i$ & $w_i$ &       \\
\#\,8  & $A_{n,l}$ & $\nu_{n,l}$ & $w_{n,l}$ &           & $a$, $b$, $c$ & $A_i$ & $\nu_i$ & $w_i$ &       \\
\#\,9  & $A_{n,l}$ & $\nu_{n,l}$ & $w_{n,l}$ &           & $a$, $b$, $c$ & $A_i$ & $\nu_i$ & $w_i$ &       \\
\#\,10 & $A_{n,l}$ & $\nu_{n,l}$ & $w_{n,l}$ &           & $a$, $b$, $c$ & $A_i$ & $\nu_i$ & $w_i$ &       \\
\#\,11 &           &             &           & $B_{n,l}$ &               &       &         &       & $B_i$ \\
\#\,12 & $A_{n,l}$ & $\nu_{n,l}$ & $w_{n,l}$ &           & $a$, $b$, $c$ & $A_i$ & $\nu_i$ & $w_i$ &       \\
\#\,13 & $A_{n,l}$ & $\nu_{n,l}$ & $w_{n,l}$ & $B_{n,l}$ & $a$, $b$, $c$ & $A_i$ & $\nu_i$ & $w_i$ & $B_i$ \\
\enddata
\tablecomments{
The value of a parameter that is not invoked in a specific pass is set to
either its initial guess or else to its value as determined in a previous pass.
For the initial guess of the line asymmetry parameter we use $B_{n,l}=0$ and
$B_i=0$, $i=1,\ldots,N$. The initial guesses of the remaining fit parameters
are taken from a seed table. If the symmetrical profile is used $B_{n,l}$ and
$B_i$, $i=1,\ldots,N$ are kept fixed to zero in passes \#\,11 and \#\,13. To
increase the numerical stability of the WMLTP code rather tight constraints are
applied to the fitted parameters in passes \#\,1 through \#\,7. The rationale
for passes \#\,8 through passes \#\,10 is to gradually relax these constraints
as much as possible.
}
\end{deluxetable}

%Tab. 5
\renewcommand\multirowsetup{\centering}
\newlength\LL \settowidth\LL{XXX}
\begin{deluxetable}{c@{\hspace{-0.0mm}}cc|cccc|cccc}
\tablecaption{
Statistical analysis of the magnitude of raw and scaled frequency differences
obtained from four comparisons C1 through C4 in the study of the sensitivity of
Method~2 to the details of the $m$-averaging procedure and the effective
leakage matrix, respectively. 
\label{tab5}}
\tablewidth{0pt}
\tablehead{
\addlinespace[-1.2mm]
 &     & \multicolumn{1}{c|}{} & \multicolumn{4}{c}{$\nu < 7000\,\,\mu$Hz} 
                               & \multicolumn{4}{|c}{$\nu < 4500\,\,\mu$Hz \rule[-2.5mm]{0.0mm}{5.0mm}} \\\cline{4-11}
 &  & \multicolumn{1}{c|}{} & $\mbox{avg}\pm\mbox{std}$ & $\mbox{ste}$
                               & $n_{\rm tot}$ & $n_{\rm out}$ & $\mbox{avg}\pm\mbox{std}$ & $\mbox{ste}$
                               & $n_{\rm tot}$ & \multicolumn{1}{c}{$n_{\rm out}$\rule[0.0mm]{0.0mm}{5.0mm}}
}
\startdata
\multirow{2}{\LL}{C1:} & 1 & \multicolumn{1}{|c|}{$|\Delta\nu|$}
                      & $0.265\pm 0.514$ & 0.005 & 12,353 & 55 & $0.044\pm 0.137$ & 0.002 & 6,421 &  4 \\
                      & 2 & \multicolumn{1}{|c|}{$|\Delta\nu/\sigma_{\Delta\nu}|$}
                      & $1.062\pm 1.075$ & 0.010 & 12,353 &  3 & $1.076\pm 1.177$  & 0.015 & 6,421 &  3 \\\specialrule{0.4pt}{1.0mm}{1.4mm}
\multirow{2}{\LL}{C2:} & 3 & \multicolumn{1}{|c|}{$|\Delta\nu|$}
                      & $0.403\pm 0.484$ & 0.004 & 12,158 & 24 & $0.278\pm 0.301$ & 0.004 & 6,235 &  0 \\
                      & 4 & \multicolumn{1}{|c|}{$|\Delta\nu/\sigma_{\Delta\nu}|$}
                      & $4.858\pm 6.398$ & 0.058 & 12,158 & 79 & $7.786\pm 7.689$  & 0.098 & 6,235 & 79 \\\specialrule{0.4pt}{1.0mm}{1.4mm}
\multirow{2}{\LL}{C3:} & 5 & \multicolumn{1}{|c|}{$|\Delta\nu|$}
                      & $0.328\pm 0.403$ & 0.004 & 12,159 & 38 & $ 0.302\pm 0.331$ & 0.004 & 6,236 &  0 \\
                      & 6 & \multicolumn{1}{|c|}{$|\Delta\nu/\sigma_{\Delta\nu}|$}
                      & $5.786\pm 8.403$ & 0.076 & 12,159 &  5 & $10.225\pm 9.759$ & 0.124 & 6,236 &  5 \\\specialrule{0.4pt}{1.0mm}{1.4mm}
\multirow{2}{\LL}{C4:} & 7 & \multicolumn{1}{|c|}{$|\Delta\nu|$}
                      & $0.149\pm 0.352$ & 0.003 & 12,368 & 38 & $0.060\pm 0.087$ & 0.001 & 6,411 &  2 \\
                      & 8 & \multicolumn{1}{|c|}{$|\Delta\nu/\sigma_{\Delta\nu}|$}
                      & $1.000\pm 1.247$ & 0.011 & 12,368 &  2 & $1.370\pm 1.452$ & 0.018 & 6,411 &  2
\enddata
\tablecomments{
C1 denotes the comparison of the frequencies from table $\cal{F}$2010\_66a with
the corresponding frequencies from table $\cal{F}$2010\_66b; C2 denotes the
comparison of the frequencies from table $\cal{F}$2010\_66c with the
corresponding frequencies from table $\cal{F}$2010\_66d; C3 denotes the
comparison of the frequencies from table $\cal{F}$2010\_66b with the
corresponding frequencies from table $\cal{F}$2010\_66d; and C4 denotes the
comparison of the frequencies from table $\cal{F}$2010\_66b with the
corresponding frequencies from table $\cal{F}$2010\_66b.lkm. For each of the
four comparisons C1 through C4 we show the average, avg, the standard
deviation, std, and the standard error of the average, ste, of the magnitude of
the raw frequency differences, $|\Delta\nu|$, measured in $\mu$Hz, in the first
row, and of the magnitude of the normalized frequency differences,
$|\Delta\nu/\sigma_{\Delta\nu}|$, where $\sigma_{\Delta\nu}$ denotes the formal
error of the frequency difference $\Delta\nu$, in the second row. In our
analysis we weeded out from the given $n_{\rm tot}$ frequency differences
$n_{\rm out}$ gross outliers by using the rejection criterion $|\Delta\nu| >
5\,\mu$Hz for the raw differences and $|\Delta\nu/\sigma_{\Delta\nu}| > 35$ for
the normalized differences. As shown in the last four columns, we repeated the
entire analysis for the subset of modes having $\nu<4500\,\,\mu$Hz.
}
\end{deluxetable}

%Tab. 6
\begin{deluxetable}{c|ccc|ccc}
\tablecaption{
Normalized comparison of sample Method~1 and Method~2 fit results based upon
the $m$-averaged spectral set $\cal{S}$2010\_66a for the modes $(n,l)=(2,70)$, $(2,200)$, and $(2,600)$.
\label{tab6}}
\tablewidth{0pt}
\tablehead{
\multicolumn{1}{c|}{$(n,l)$} & 
        \multicolumn{1}{|c}{$\displaystyle\frac{\nu_{\sf 1,wfr}-\nu_{\sf 1,nfr}}{\sigma_{\nu,{\sf 1,nfr}}}$} & 
        \multicolumn{1}{c}{$\displaystyle\frac{\nu_{\sf 2}-\nu_{\sf 1,nfr}}{\sigma_{\nu,{\sf 2}}}$} & 
        \multicolumn{1}{c|}{$\displaystyle\frac{\nu_{\sf 2}-\nu_{\sf 1,wfr}}{\sigma_{\nu,{\sf 2}}}$} &
        \multicolumn{1}{|c}{$\displaystyle\frac{w_{\sf 1,wfr}-w_{\sf 1,nfr}}{\sigma_{w,{\sf 1,nfr}}}$} &
        \multicolumn{1}{c}{$\displaystyle\frac{w_{\sf 2}-w_{\sf 1,nfr}}{\sigma_{w,{\sf 2}}}$} & 
        \multicolumn{1}{c}{$\displaystyle\frac{w_{\sf 2}-w_{\sf 1,wfr}}{\sigma_{w,{\sf 2}}}$ \rule[-5.0mm]{0.0mm}{4.0mm}}
}
\startdata
  (2,70) & $862.7$  & $  0.2$  & $-258.6$  & $2225.0$     & $24.9$    & $-1121.3$ \\[0.2mm]
 (2,200) & $ 10.1$  & $-10.6$  & $-210.8$  & $  10.6$     & $ 5.3$    & $ -669.3$ \\[0.2mm]
 (2,600) &          &          & $ -16.4$  &              &           & $ -261.8$ \\
\enddata
\tablecomments{
Column~2 contains the normalized frequency differences between the narrow (nfr)
and wide (wfr) fitting range versions of Method~1 for which the denominator was
narrow fitting range uncertainty. Column~3 contains the normalized frequency
differences between Method~2 and the narrow fitting range version of Method~1
for which the denominator was the Method~2 uncertainty. Column~4 contains the
normalized frequency differences between Method~2 and the wide fitting range
version of Method~1 for which the Method~2 uncertainty was the denominator.
Columns~5 through 7 contain the similar quantities for the linewidths.
}
\end{deluxetable}

%Tab. 7
\begin{deluxetable}{c|c|c}
\tablecaption{
Classification of the 2,483 outliers detected by means of the outlier
correction scheme, as described in Section~\ref{ocs}, in terms of the various
terms that could trigger the adjustment of the frequency uncertainty of the
outlying cases. 
\label{tab7}}
\tablewidth{0pt}
\tablehead{
\addlinespace[-1.5mm]
\multicolumn{1}{c}{} & \multicolumn{2}{c}{no. of cases} \\\cline{2-3}
\multicolumn{1}{c|}{term} & \multicolumn{1}{c|}{eq.~(\ref{dnuupda})} & 
                            \multicolumn{1}{c}{eq.~(\ref{dnuupdb})\phantom{\rule{0.0mm}{5.0mm}}} 
}
\startdata
$\Delta\nu_{n,l}^{\rm obs}$                     &   286                                        &   198 \\
$\Delta\nu_{n,l}^{\rm new}$                     &   103                                        &  \raisebox{+1.0mm}{\rule{10.0mm}{0.4pt}}     \\
$\Delta\nu_{n,l}^{\rm em1}$                     &    13                                        &     5 \\
$\Delta\nu_{n,l}^{\rm em2}$                     &   224                                        &   209 \\
$|\nu_{n,l}^{\rm em1}-\nu_{n,l}^{\rm new}|/3$   &   329                                        &  \raisebox{+1.0mm}{\rule{10.0mm}{0.4pt}}     \\
$|\nu_{n,l}^{\rm em1}-\nu_{n,l}^{\rm obs}|/3$   &  \raisebox{+1.0mm}{\rule{10.0mm}{0.4pt}}     &   482 \\
$|\nu_{n,l}^{\rm em2}-\nu_{n,l}^{\rm new}|/3$   &   201                                        &  \raisebox{+1.0mm}{\rule{10.0mm}{0.4pt}}     \\
$|\nu_{n,l}^{\rm em2}-\nu_{n,l}^{\rm obs}|/3$   &  \raisebox{+1.0mm}{\rule{10.0mm}{0.4pt}}     &   433 \\[2.0mm]\hline
\multicolumn{1}{r|}{total}                      & 1,156                                        & 1,327 \phantom{\rule{0.0mm}{5.5mm}}
\enddata
\tablecomments{
In the first column we list the terms as given in the right-hand side of
equations~(\ref{dnuupda}) and (\ref{dnuupdb}), respectively, that trigger the
evaluation of the frequency uncertainties. The meaning of the various terms is
given in Section~\ref{ocs}. In column~2 we list the number of cases that were
triggered by the corresponding term listed in column~1 in case of
equation~(\ref{dnuupda}). We note that this equation is for those cases for
which we were able to compute an improved refitted frequency in the manner as
described in Section~\ref{ocs4}. Column~3 contains the same quantities as
column~2, but the quantities in this column were triggered by the different
terms in the right-hand side of equation~(\ref{dnuupdb}), which applies for
those cases for which our refitted frequency was not an improvement over our
original, observed  frequency. In the last row we give the total number of
cases that were triggered by the respective equations. A horizontal bar in a
column means that the corresponding term listed in the first column does not
apply for the respective equations.
}
\end{deluxetable}

%Tab. 8
\begin{deluxetable}{ccl}
\tablecaption{
Statistical overview of the results of the application of our 5-step outlier
identification and correction procedure to the table of fits that was employed
in the structural inversion shown in Figure~\ref{spinv}. 
\label{tab8}}
\tablewidth{0pt}
\tablehead{
\addlinespace[-1.5mm]
\multicolumn{1}{c}{\#} & \multicolumn{1}{c}{\%} & \multicolumn{1}{c}{Description}
}
\startdata
6,366 & 100.0 &  Original fits for which $\nu\leq 4500\,\,\mu$Hz \\
5,210 &  81.8 &  Cases for which $\nu$ was unchanged \\
4,367 &  68.6 &  Cases for which $\Delta\nu$ was unchanged \\
3,883 &  61.0 &  Cases that were never identified as being outliers \\
2,483 &  39.0 &  Cases that were identified as being outliers \\
1,999 &  31.4 &  Outliers that resulted in an increase in $\Delta\nu$ \\
1,327 &  20.8 &  Outliers for which $\nu$ was unchanged \\
1,156 &  18.2 &  Outliers for which $\nu$ was altered \\
1,129 &  17.7 &  Outliers for which $\nu$ was unchanged, but $\Delta\nu$ was increased \\
1,031 &  16.2 &  Outliers for which $\nu$ was decreased \\
  870 &  13.7 &  Outliers for which both $\nu$ and $\Delta\nu$ were altered \\
  286 &   4.5 &  Outliers for which $\nu$ was altered but $\Delta\nu$ was unchanged \\
  198 &   3.1 &  Outliers for which neither $\nu$ nor $\Delta\nu$ were altered \\
  125 &   2.0 &  Outliers for which $\nu$ was increased
\enddata
\tablecomments{
Here, $\nu$ denotes the frequency and $\Delta\nu$ the uncertainty thereof. All
of the percentages shown in the second column were computed using the 6,366
original cases in this table.
}
\end{deluxetable}

\clearpage

\bibliography{jr201412}{}

\begin{thebibliography}{}
\expandafter\ifx\csname natexlab\endcsname\relax\def\natexlab#1{#1}\fi

\bibitem[{{Antia} \& {Chitre}(1998)}]{Antc98}
{Antia}, H.~M., \& {Chitre}, S.~M. 1998, \aap, 339, 239

\bibitem[{{Beck} {et~al.}(2002){Beck}, {Gizon}, \& {Duvall}}]{Beck02}
{Beck}, J.~G., {Gizon}, L., \& {Duvall}, Jr., T.~L. 2002, \apjl, 575, L47

\bibitem[{{Braun} {et~al.}(1987){Braun}, {Duvall}, \& {Labonte}}]{Bra87}
{Braun}, D.~C., {Duvall}, Jr., T.~L., \& {Labonte}, B.~J. 1987, \apjl, 319, L27

\bibitem[{{Braun} \& {Lindsey}(2001)}]{Bra01}
{Braun}, D.~C., \& {Lindsey}, C. 2001, \apjl, 560, L189

\bibitem[{{Brown}(1985)}]{Bro85}
{Brown}, T.~M. 1985, \nat, 317, 591

\bibitem[{{Christensen-Dalsgaard}(2002)}]{jcd02a}
{Christensen-Dalsgaard}, J. 2002, RvMP, 74, 1073

\bibitem[{{Christensen-Dalsgaard}(2003)}]{jcd03}
---. 2003, {Lecture Notes on Stellar Oscillations}, 5th edn. (Aarhus:
  University of Aarhus)

\bibitem[{{Christensen-Dalsgaard} {et~al.}(2000){Christensen-Dalsgaard},
  {D{\"a}ppen}, {Dziembowski}, \& {Guzik}}]{jcd00}
{Christensen-Dalsgaard}, J., {D{\"a}ppen}, W., {Dziembowski}, W.~A., \&
  {Guzik}, J.~A. 2000, in NATO Advanced Science Institutes (ASI) Series C, Vol.
  544, Variable stars as essential astrophysical tools, ed. C.~{Ibano\u{g}lu},
  59

\bibitem[{{Christensen-Dalsgaard} \& {Gough}(1984)}]{jcd84}
{Christensen-Dalsgaard}, J., \& {Gough}, D.~O. 1984, in Solar Seismology from
  Space, ed. R.~K. {Ulrich}, J.~{Harvey}, E.~J. {Rhodes}, Jr., \& J.~{Toomre},
  199--204

\bibitem[{{Christensen-Dalsgaard} {et~al.}(1989){Christensen-Dalsgaard},
  {Gough}, \& {Libbrecht}}]{jcd89}
{Christensen-Dalsgaard}, J., {Gough}, D.~O., \& {Libbrecht}, K.~G. 1989, \apjl,
  341, L103

\bibitem[{{Christensen-Dalsgaard} {et~al.}(1996){Christensen-Dalsgaard},
  {D\"appen}, {Ajukov}, {Anderson}, {Antia}, {Basu}, {Baturin}, {Berthomieu},
  {Chaboyer}, {Chitre}, {Cox}, {Demarque}, {Donatowicz}, {Dziembowski},
  {Gabriel}, {Gough}, {Guenther}, {Guzik}, {Harvey}, {Hill}, {Houdek},
  {Iglesias}, {Kosovichev}, {Leibacher}, {Morel}, {Proffitt}, {Provost},
  {Reiter}, {Rhodes}, {Rogers}, {Roxburgh}, {Thompson}, \& {Ulrich}}]{jcd96}
{Christensen-Dalsgaard}, J., {D\"appen}, W., {Ajukov}, S.~V., {et~al.} 1996,
  Sci, 272, 1286

\bibitem[{{Claverie} {et~al.}(1979){Claverie}, {Isaak}, {McLeod}, {van der
  Raay}, \& {Cortes}}]{Cla79}
{Claverie}, A., {Isaak}, G.~R., {McLeod}, C.~P., {van der Raay}, H.~B., \&
  {Cortes}, T.~R. 1979, \nat, 282, 591

\bibitem[{{Claverie} {et~al.}(1980){Claverie}, {Isaak}, {McLeod}, {van der
  Raay}, \& {Roca-Cortes}}]{Cla80}
{Claverie}, A., {Isaak}, G.~R., {McLeod}, C.~P., {van der Raay}, H.~B., \&
  {Roca-Cortes}, T. 1980, in Lecture Notes in Physics, Berlin Springer Verlag,
  Vol. 125, Nonradial and Nonlinear Stellar Pulsation, ed. H.~A. {Hill} \&
  W.~A. {Dziembowski}, 181--183

\bibitem[{{Deubner}(1975)}]{Deu75}
{Deubner}, F.-L. 1975, \aap, 44, 371

\bibitem[{{Deubner}(1977)}]{Deu77}
{Deubner}, F.~L. 1977, in Proceeding of the November 7-10, 1977 OSO-8 Workshop,
  ed. L.~for Atmospheric \& S.~P. (LASP), University of Colorado, Boulder,
  295--310

\bibitem[{{Deubner} {et~al.}(1979){Deubner}, {Ulrich}, \& {Rhodes}}]{Deu79}
{Deubner}, F.-L., {Ulrich}, R.~K., \& {Rhodes}, Jr., E.~J. 1979, \aap, 72, 177

\bibitem[{{Di Mauro} {et~al.}(2002){Di Mauro}, {Christensen-Dalsgaard},
  {Rabello-Soares}, \& {Basu}}]{Dim02}
{Di Mauro}, M.~P., {Christensen-Dalsgaard}, J., {Rabello-Soares}, M.~C., \&
  {Basu}, S. 2002, \aap, 384, 666

\bibitem[{{Domingo} {et~al.}(1995){Domingo}, {Fleck}, \& {Poland}}]{Dom95}
{Domingo}, V., {Fleck}, B., \& {Poland}, A.~I. 1995, \solphys, 162, 1

\bibitem[{{Duvall} {et~al.}(1991){Duvall}, {Harvey}, {Jefferies}, \&
  {Pomerantz}}]{Duv91}
{Duvall}, Jr., T.~L., {Harvey}, J.~W., {Jefferies}, S.~M., \& {Pomerantz},
  M.~A. 1991, \apj, 373, 308

\bibitem[{{Duvall} {et~al.}(1986){Duvall}, {Harvey}, \& {Pomerantz}}]{Duv86}
{Duvall}, Jr., T.~L., {Harvey}, J.~W., \& {Pomerantz}, M.~A. 1986, \nat, 321,
  500

\bibitem[{{Duvall} {et~al.}(1993){Duvall}, {Jefferies}, {Harvey}, \&
  {Pomerantz}}]{Duv93}
{Duvall}, Jr., T.~L., {Jefferies}, S.~M., {Harvey}, J.~W., \& {Pomerantz},
  M.~A. 1993, \nat, 362, 430

\bibitem[{{Elliott} \& {Gough}(1999)}]{Ell99}
{Elliott}, J.~R., \& {Gough}, D.~O. 1999, \apj, 516, 475

\bibitem[{{Fahlman} \& {Ulrych}(1982)}]{Fah82}
{Fahlman}, G.~G., \& {Ulrych}, T.~J. 1982, \mnras, 199, 53

\bibitem[{Frieden(1983)}]{Fri83}
Frieden, B.~R. 1983, Probability, statistical optics, and data testing: a
  problem solving approach (Berlin: Springer)

\bibitem[{{Giles} {et~al.}(1997){Giles}, {Duvall}, {Scherrer}, \&
  {Bogart}}]{Gil97}
{Giles}, P.~M., {Duvall}, Jr., T.~L., {Scherrer}, P.~H., \& {Bogart}, R.~S.
  1997, \nat, 390, 52

\bibitem[{Gill {et~al.}(1981)Gill, Murray, \& Wright}]{Gill81}
Gill, P.~E., Murray, W., \& Wright, M.~H. 1981, Practical optimization (London:
  Acad. Press)

\bibitem[{{Gizon} \& {Birch}(2005)}]{Giz05}
{Gizon}, L., \& {Birch}, A.~C. 2005, LRSP, 2, 6

\bibitem[{{Gizon} {et~al.}(2010){Gizon}, {Birch}, \& {Spruit}}]{Giz10}
{Gizon}, L., {Birch}, A.~C., \& {Spruit}, H.~C. 2010, \araa, 48, 289

\bibitem[{{Godier} \& {Rozelot}(2001)}]{God01}
{Godier}, S., \& {Rozelot}, J.~P. 2001, \solphys, 199, 217

\bibitem[{{Gonz{\'a}lez-Hern{\'a}ndez}(2008)}]{GH08}
{Gonz{\'a}lez-Hern{\'a}ndez}, I. 2008, Journal of Physics Conference Series,
  118, 012034

\bibitem[{{Gough}(1977)}]{Gou77}
{Gough}, D.~O. 1977, in IAU Colloq. 36: The Energy Balance and Hydrodynamics of
  the Solar Chromosphere and Corona, ed. B.~{Bonnet} \& P.~{Delache}, 3--36

\bibitem[{{Gough}(1980)}]{Gou80}
{Gough}, D.~O. 1980, in Lecture Notes in Physics, Berlin Springer Verlag, Vol.
  125, Nonradial and Nonlinear Stellar Pulsation, ed. H.~A. {Hill} \& W.~A.
  {Dziembowski}, 273--299

\bibitem[{{Gough}(1993)}]{Gou93}
{Gough}, D.~O. 1993, in Astrophysical Fluid Dynamics - Les Houches 1987, ed.
  J.-P. {Zahn} \& J.~{Zinn-Justin}, 399--560

\bibitem[{{Gough} {et~al.}(1996){Gough}, {Kosovichev}, {Toomre}, {Anderson},
  {Antia}, {Basu}, {Chaboyer}, {Chitre}, {Christensen-Dalsgaard},
  {Dziembowski}, {Eff-Darwich}, {Elliott}, {Giles}, {Goode}, {Guzik}, {Harvey},
  {Hill}, {Leibacher}, {Monteiro}, {Richard}, {Sekii}, {Shibahashi}, {Takata},
  {Thompson}, {Vauclair}, \& {Vorontsov}}]{Gou96}
{Gough}, D.~O., {Kosovichev}, A.~G., {Toomre}, J., {et~al.} 1996, Sci, 272,
  1296

\bibitem[{{Haber} {et~al.}(2002){Haber}, {Hindman}, {Toomre}, {Bogart},
  {Larsen}, \& {Hill}}]{Hab02}
{Haber}, D.~A., {Hindman}, B.~W., {Toomre}, J., {et~al.} 2002, \apj, 570, 855

\bibitem[{{Hill}(1988)}]{Hil88}
{Hill}, F. 1988, \apj, 333, 996

\bibitem[{{Hill}(1980)}]{HHill80}
{Hill}, H.~A. 1980, in Lecture Notes in Physics, Berlin Springer Verlag, Vol.
  125, Nonradial and Nonlinear Stellar Pulsation, ed. H.~A. {Hill} \& W.~A.
  {Dziembowski}, 174--180

\bibitem[{{Jain} {et~al.}(2011){Jain}, {Tripathy}, \& {Hill}}]{Jai11}
{Jain}, K., {Tripathy}, S.~C., \& {Hill}, F. 2011, \apj, 739, 6

\bibitem[{{Jefferies}(1998)}]{Jef98}
{Jefferies}, S.~M. 1998, in IAU Symposium, Vol. 185, New Eyes to See Inside the
  Sun and Stars, ed. F.-L. {Deubner}, J.~{Christensen-Dalsgaard}, \&
  D.~{Kurtz}, 415--422

\bibitem[{{Korzennik}(1990)}]{Kor90}
{Korzennik}, S.~G. 1990, PhD thesis, California Univ., Los Angeles.

\bibitem[{{Korzennik} {et~al.}(2004){Korzennik}, {Rabello-Soares}, \&
  {Schou}}]{Kor04}
{Korzennik}, S.~G., {Rabello-Soares}, M.~C., \& {Schou}, J. 2004, \apj, 602,
  481

\bibitem[{{Korzennik} {et~al.}(2008){Korzennik}, {Rabello-Soares}, \&
  {Schou}}]{Kor08}
---. 2008, JPhCS, 118, 012027

\bibitem[{{Kosovichev}(1996{\natexlab{a}})}]{Kos96b}
{Kosovichev}, A.~G. 1996{\natexlab{a}}, \apjl, 469, L61

\bibitem[{{Kosovichev}(1996{\natexlab{b}})}]{Kos96a}
---. 1996{\natexlab{b}}, \apjl, 461, L55

\bibitem[{{Kosovichev}(1999)}]{Kos99}
---. 1999, JCoAM, 109, 1

\bibitem[{{Kosovichev} {et~al.}(2000){Kosovichev}, {Duvall}, \&
  {Scherrer}}]{Kos00}
{Kosovichev}, A.~G., {Duvall}, Jr., T.~L.~., \& {Scherrer}, P.~H. 2000,
  \solphys, 192, 159

\bibitem[{{Kosovichev} {et~al.}(1998){Kosovichev}, {Schou}, {Scherrer},
  {Goode}, {Dziembowski}, {Rhodes}, \& {SOI Structure Inversion Team}}]{Kos98}
{Kosovichev}, A.~G., {Schou}, J., {Scherrer}, P.~H., {et~al.} 1998, in IAU
  Symposium, Vol. 185, New Eyes to See Inside the Sun and Stars, ed. F.-L.
  {Deubner}, J.~{Christensen-Dalsgaard}, \& D.~{Kurtz}, 157--164

\bibitem[{{Leibacher} \& {Stein}(1971)}]{Lei71}
{Leibacher}, J.~W., \& {Stein}, R.~F. 1971, ApL, 7, 191

\bibitem[{{Levenberg}(1944)}]{Lev44}
{Levenberg}, K. 1944, QApMa, 2, 164

\bibitem[{{Libbrecht} \& {Kaufman}(1988)}]{Lib88}
{Libbrecht}, K.~G., \& {Kaufman}, J.~M. 1988, \apj, 324, 1172

\bibitem[{{Lindsey} \& {Braun}(2000)}]{Lin00}
{Lindsey}, C., \& {Braun}, D.~C. 2000, \solphys, 192, 261

\bibitem[{{Lindsey} \& {Braun}(2004)}]{Lin04}
---. 2004, \apjs, 155, 209

\bibitem[{{Marquardt}(1963)}]{Mar63}
{Marquardt}, D.~W. 1963, J. Soc. Ind. Appl. Math., 11, 431

\bibitem[{{Mor\'{e}} {et~al.}(1980){Mor\'{e}}, {Garbow}, \&
  {Hillstrom}}]{Mor80}
{Mor\'{e}}, J.~J., {Garbow}, B.~S., \& {Hillstrom}, K.~E. 1980, {User Guide for
  MINPACK-1}, Tech. rep., Argonne National Laboratory Report ANL-80-74

\bibitem[{{NAG Fortran Library, Mark 23}(2011)}]{Nag11}
{NAG Fortran Library, Mark 23}. 2011, {NAG Library Chapter Introduction: e04 -
  Minimizing or Maximizing a Function}, The Numerical Algorithms Group (NAG),
  Oxford, UK, http:/$\!$/www.nag.co.uk/numeric/fl/manual/pdf/E04/e04\_intro.pdf

\bibitem[{{Nigam} \& {Kosovichev}(1998)}]{Nig98}
{Nigam}, R., \& {Kosovichev}, A.~G. 1998, \apjl, 505, L51

\bibitem[{{Nigam} {et~al.}(1998){Nigam}, {Kosovichev}, {Scherrer}, \&
  {Schou}}]{Nigetal98}
{Nigam}, R., {Kosovichev}, A.~G., {Scherrer}, P.~H., \& {Schou}, J. 1998,
  \apjl, 495, L115

\bibitem[{{Rabello-Soares}(2011)}]{Rab11}
{Rabello-Soares}, M.~C. 2011, Journal of Physics Conference Series, 271, 012026

\bibitem[{{Rabello-Soares} {et~al.}(2000){Rabello-Soares}, {Basu},
  {Christensen-Dalsgaard}, \& {Di Mauro}}]{Rab00}
{Rabello-Soares}, M.~C., {Basu}, S., {Christensen-Dalsgaard}, J., \& {Di
  Mauro}, M.~P. 2000, \solphys, 193, 345

\bibitem[{{Rabello-Soares} {et~al.}(1999){Rabello-Soares}, {Houdek}, \&
  {Christensen-Dalsgaard}}]{Rab99}
{Rabello-Soares}, M.~C., {Houdek}, G., \& {Christensen-Dalsgaard}, J. 1999, in
  Astronomical Society of the Pacific Conference Series, Vol. 173, Theory and
  Tests of Convection in Stellar Structure: First Granada Workshop, ed.
  A.~{Gimenez}, E.~F. {Guinan}, \& B.~{Montesinos}, 301

\bibitem[{{Rabello-Soares} {et~al.}(2001){Rabello-Soares}, {Korzennik}, \&
  {Schou}}]{Rab01}
{Rabello-Soares}, M.~C., {Korzennik}, S.~G., \& {Schou}, J. 2001, in ESA
  Special Publication, Vol. 464, SOHO 10/GONG 2000 Workshop: Helio- and
  Asteroseismology at the Dawn of the Millennium, ed. A.~{Wilson} \& P.~L.
  {Pall{\'e}}, 129--136

\bibitem[{{Rabello-Soares} {et~al.}(2008{\natexlab{a}}){Rabello-Soares},
  {Korzennik}, \& {Schou}}]{Rab08}
{Rabello-Soares}, M.~C., {Korzennik}, S.~G., \& {Schou}, J. 2008{\natexlab{a}},
  \solphys, 251, 197

\bibitem[{{Rabello-Soares} {et~al.}(2008{\natexlab{b}}){Rabello-Soares},
  {Korzennik}, \& {Schou}}]{Rab08a}
---. 2008{\natexlab{b}}, Advances in Space Research, 41, 861

\bibitem[{{Reiter}(2007)}]{Rei07}
{Reiter}, J. 2007, AN, 328, 245

\bibitem[{{Reiter} {et~al.}(2003){Reiter}, {Kosovichev}, {Rhodes}, \&
  {Schou}}]{Rei03}
{Reiter}, J., {Kosovichev}, A.~G., {Rhodes}, Jr., E.~J., \& {Schou}, J. 2003,
  in ESA Special Publication, Vol. 517, GONG+ 2002. Local and Global
  Helioseismology: the Present and Future, ed. H.~{Sawaya-Lacoste}, 369--372

\bibitem[{{Rhodes} {et~al.}(1997){Rhodes}, {Kosovichev}, {Schou}, {Scherrer},
  \& {Reiter}}]{Rho97}
{Rhodes}, Jr., E.~J., {Kosovichev}, A.~G., {Schou}, J., {Scherrer}, P.~H., \&
  {Reiter}, J. 1997, \solphys, 175, 287

\bibitem[{{Rhodes} {et~al.}(1998{\natexlab{a}}){Rhodes}, {Reiter},
  {Kosovichev}, {Schou}, \& {Scherrer}}]{Rho98a}
{Rhodes}, Jr., E.~J., {Reiter}, J., {Kosovichev}, A.~G., {Schou}, J., \&
  {Scherrer}, P.~H. 1998{\natexlab{a}}, in ESA Special Publication, Vol. 418,
  Structure and Dynamics of the Interior of the Sun and Sun-like Stars, ed.
  S.~G. {Korzennik}, 73--82

\bibitem[{{Rhodes} {et~al.}(1998{\natexlab{b}}){Rhodes}, {Reiter},
  {Kosovichev}, {Schou}, {Scherrer}, {Rose}, {Irish}, \& {Jones}}]{Rho98b}
{Rhodes}, Jr., E.~J., {Reiter}, J., {Kosovichev}, A.~G., {et~al.}
  1998{\natexlab{b}}, in ESA Special Publication, Vol. 418, Structure and
  Dynamics of the Interior of the Sun and Sun-like Stars, ed. S.~G.
  {Korzennik}, 311--316

\bibitem[{{Rhodes} {et~al.}(2002){Rhodes}, {Reiter}, \& {Schou}}]{Rho02}
{Rhodes}, Jr., E.~J., {Reiter}, J., \& {Schou}, J. 2002, in ESA Special
  Publication, Vol. 508, From Solar Min to Max: Half a Solar Cycle with SOHO,
  ed. A.~{Wilson}, 37--40

\bibitem[{{Rhodes} {et~al.}(2003){Rhodes}, {Reiter}, \& {Schou}}]{Rho03}
{Rhodes}, Jr., E.~J., {Reiter}, J., \& {Schou}, J. 2003, in ESA Special
  Publication, Vol. 517, GONG+ 2002. Local and Global Helioseismology: the
  Present and Future, ed. H.~{Sawaya-Lacoste}, 173--182

\bibitem[{{Rhodes} {et~al.}(2001){Rhodes}, {Reiter}, {Schou}, {Kosovichev}, \&
  {Scherrer}}]{Rho01}
{Rhodes}, Jr., E.~J., {Reiter}, J., {Schou}, J., {Kosovichev}, A.~G., \&
  {Scherrer}, P.~H. 2001, \apj, 561, 1127

\bibitem[{{Rhodes} {et~al.}(1976){Rhodes}, {Ulrich}, \& {Simon}}]{Rho76b}
{Rhodes}, Jr., E.~J., {Ulrich}, R.~K., \& {Simon}, G.~W. 1976, in Bulletin of
  the American Astronomical Society, Vol.~8, 533

\bibitem[{{Rhodes} {et~al.}(1977){Rhodes}, {Ulrich}, \& {Simon}}]{Rho77}
{Rhodes}, Jr., E.~J., {Ulrich}, R.~K., \& {Simon}, G.~W. 1977, \apj, 218, 901

\bibitem[{{Rhodes} {et~al.}(2011){Rhodes}, {Reiter}, {Schou}, {Larson},
  {Scherrer}, {Brooks}, {McFaddin}, {Miller}, {Rodriguez}, \& {Yoo}}]{Rho11}
{Rhodes}, Jr., E.~J., {Reiter}, J., {Schou}, J., {et~al.} 2011, JPhCS, 271,
  012029

\bibitem[{{Ronan} {et~al.}(1994){Ronan}, {Cadora}, \& {Labonte}}]{Ron94}
{Ronan}, R.~S., {Cadora}, K., \& {Labonte}, B.~J. 1994, \solphys, 150, 389

\bibitem[{{Rose} {et~al.}(2003){Rose}, {Rhodes}, {Reiter}, \&
  {Rudnisky}}]{Ros03}
{Rose}, P., {Rhodes}, Jr., E.~J., {Reiter}, J., \& {Rudnisky}, W. 2003, in ESA
  Special Publication, Vol. 517, GONG+ 2002. Local and Global Helioseismology:
  the Present and Future, ed. H.~{Sawaya-Lacoste}, 373--376

\bibitem[{{Roth} \& {Stix}(2008)}]{Rth08}
{Roth}, M., \& {Stix}, M. 2008, \solphys, 251, 77

\bibitem[{{Schad} {et~al.}(2011){Schad}, {Timmer}, \& {Roth}}]{Schad11}
{Schad}, A., {Timmer}, J., \& {Roth}, M. 2011, \apj, 734, 97

\bibitem[{{Schad} {et~al.}(2013){Schad}, {Timmer}, \& {Roth}}]{Schad13}
---. 2013, \apjl, 778, L38

\bibitem[{{Scherrer} {et~al.}(1995){Scherrer}, {Bogart}, {Bush}, {Hoeksema},
  {Kosovichev}, {Schou}, {Rosenberg}, {Springer}, {Tarbell}, {Title},
  {Wolfson}, {Zayer}, \& {MDI Engineering Team}}]{Scher95}
{Scherrer}, P.~H., {Bogart}, R.~S., {Bush}, R.~I., {et~al.} 1995, \solphys,
  162, 129

\bibitem[{{Schittkowski}(1986)}]{Schitt85}
{Schittkowski}, K. 1986, Annals of Operations Research, 5, 485

\bibitem[{{Schou}(1992)}]{Schou92}
{Schou}, J. 1992, PhD thesis, Aarhus University, Aarhus, Denmark

\bibitem[{{Schou} \& {Bogart}(1998)}]{Schou98a}
{Schou}, J., \& {Bogart}, R.~S. 1998, \apjl, 504, L131

\bibitem[{{Schou} {et~al.}(1998){Schou}, {Antia}, {Basu}, {Bogart}, {Bush},
  {Chitre}, {Christensen-Dalsgaard}, {di Mauro}, {Dziembowski}, {Eff-Darwich},
  {Gough}, {Haber}, {Hoeksema}, {Howe}, {Korzennik}, {Kosovichev}, {Larsen},
  {Pijpers}, {Scherrer}, {Sekii}, {Tarbell}, {Title}, {Thompson}, \&
  {Toomre}}]{Schou98}
{Schou}, J., {Antia}, H.~M., {Basu}, S., {et~al.} 1998, \apj, 505, 390

\bibitem[{{Snodgrass} \& {Ulrich}(1990)}]{Sno90}
{Snodgrass}, H.~B., \& {Ulrich}, R.~K. 1990, \apj, 351, 309

\bibitem[{{Thompson} {et~al.}(1996){Thompson}, {Toomre}, {Anderson}, {Antia},
  {Berthomieu}, {Burtonclay}, {Chitre}, {Christensen-Dalsgaard}, {Corbard}, {De
  Rosa}, {Genovese}, {Gough}, {Haber}, {Harvey}, {Hill}, {Howe}, {Korzennik},
  {Kosovichev}, {Leibacher}, {Pijpers}, {Provost}, {Rhodes}, {Schou}, {Sekii},
  {Stark}, \& {Wilson}}]{Thom96}
{Thompson}, M.~J., {Toomre}, J., {Anderson}, E.~R., {et~al.} 1996, Sci, 272,
  1300

\bibitem[{{Tomczyk}(1988)}]{Tom88}
{Tomczyk}, S. 1988, PhD thesis, California Univ., Los Angeles

\bibitem[{{Tripathy} {et~al.}(2010){Tripathy}, {Jain}, {Hill}, \&
  {Leibacher}}]{Tri10}
{Tripathy}, S.~C., {Jain}, K., {Hill}, F., \& {Leibacher}, J.~W. 2010, \apjl,
  711, L84

\bibitem[{{Ulrich}(1970)}]{Ulr70}
{Ulrich}, R.~K. 1970, \apj, 162, 993

\bibitem[{{Ulrich} \& {Rhodes}(1977)}]{Ulr77}
{Ulrich}, R.~K., \& {Rhodes}, Jr., E.~J. 1977, \apj, 218, 521

\bibitem[{{Ulrich} {et~al.}(1979){Ulrich}, {Rhodes}, \& {Deubner}}]{Ulr79}
{Ulrich}, R.~K., {Rhodes}, Jr., E.~J., \& {Deubner}, F.-L. 1979, \apj, 227, 638

\bibitem[{{Vorontsov}(2011)}]{Vor11}
{Vorontsov}, S.~V. 2011, \mnras, 418, 1146

\bibitem[{Wolberg(1967)}]{Wol67}
Wolberg, J.~R. 1967, Prediction analysis (Toronto: Nostrand)

\bibitem[{{Woodard}(1989)}]{Woo89}
{Woodard}, M.~F. 1989, \apj, 347, 1176

\bibitem[{{Woodard}(2000)}]{Woo00}
---. 2000, \solphys, 197, 11

\bibitem[{{Zhao}(2004)}]{Zha04}
{Zhao}, J. 2004, PhD thesis, Stanford university

\bibitem[{{Zhao}(2008)}]{Zha08}
---. 2008, Advances in Space Research, 41, 838

\end{thebibliography}

\end{document}